\documentclass[preprint,3p,12pt,letterpaper, compress]{elsarticle}
\usepackage[breaklinks=true,colorlinks=true,linkcolor=blue,anchorcolor=blue,citecolor=blue,filecolor=blue,menucolor=blue,pagecolor=blue,urlcolor=blue]{hyperref}

\usepackage{graphicx}
\usepackage{float}
\usepackage{subfig}
\usepackage{epsfig}
\usepackage{xfrac}
\usepackage{psfrag}
\usepackage{amssymb}
\usepackage{amsmath}
\usepackage{framed}
\usepackage{array} 
\usepackage{multirow}
\usepackage{epstopdf}
\usepackage{mathbbol}
\usepackage{dutchcal}
\usepackage{tablefootnote}
\usepackage[flushleft]{threeparttable}
\usepackage{adjustbox}
\usepackage{siunitx}
\usepackage[T1]{fontenc}
\usepackage{xcolor}
\usepackage{lmodern}
\usepackage{listings}
\usepackage{float}
\usepackage{times}
\usepackage{relsize}

\usepackage[linesnumbered]{algorithm2e}

\DeclareMathOperator*{\argmax}{arg\,max}

\usepackage{tabularx,booktabs}
\newcolumntype{Y}{>{\centering\arraybackslash}X}
\usepackage{enumitem}

\usepackage[framemethod=tikz]{mdframed}

\journal{Journal of the Mechanics and Physics of Solids} 

\begin{document}
	
	\begin{frontmatter}
		
		\title{Thermodynamically consistent concurrent material and structure
			optimization of elastoplastic multiphase hierarchical systems}

		\author[civil,TUD]{Tarun Gangwar\corref{correspondingauthor}}
		\ead{gangw007@umn.edu}
		\author[TUD]{Dominik Schillinger}
		\ead{dominik.schillinger@tu-darmstadt.de}
		
		\address[civil]{Department of Civil, Environmental, and Geo-$\,$Engineering, University of Minnesota, Twin Cities, USA}
		\address[TUD]{Institute of Mechanics, Computational Mechanics Group, Technical University of Darmstadt, Germany}
		
		\cortext[correspondingauthor]{\textbf{Corresponding Author}: Tarun Gangwar, Department of Civil, Environmental, and Geo-$\,$Engineering, University of Minnesota, 500 Pillsbury Drive S.E., Minneapolis, MN 55455, USA;}
		
\begin{abstract}
	
The concept of concurrent material and structure optimization aims at alleviating the computational discovery of optimum microstructure configurations in multiphase hierarchical systems, whose macroscale behavior is governed by their microstructure composition that can evolve over multiple length scales from a few micrometers to centimeters.  It is based on the split of the multiscale optimization problem into two nested sub-problems, one at the macroscale (structure) and the other at the microscales (material). In this paper, we establish a novel formulation of concurrent material and structure optimization for multiphase hierarchical systems with elastoplastic constituents at the material scales. Exploiting the thermomechanical foundations of elastoplasticity, we reformulate the material optimization problem based on the maximum plastic dissipation principle such that it assumes the format of an elastoplastic constitutive law and can be efficiently solved via modified return mapping algorithms. We integrate continuum micromechanics based estimates of the stiffness and the yield criterion into the formulation, which opens the door to a computationally feasible treatment of the material optimization problem. To demonstrate the accuracy and robustness of our framework, we define new benchmark tests with several material scales that, for the first time, become computationally feasible. We argue that our formulation naturally extends to multiscale optimization under further path-dependent effects such as viscoplasticity or multiscale fracture and damage.


\end{abstract}


\end{frontmatter}

\newpage
\tableofcontents 
\newpage	

\section{Introduction}


Multiphase hierarchical systems apply the concept of microheterogeneity repetitively across a hierarchy of well-separated length scales: composite microstructures at a smaller scale form the base constituents for new microstructures at the next larger scale. This principle constitutes the backbone of virtually all biological materials, enabling them to combine various functional properties at different length scales with favorable mechanical properties at the macroscale through evolutionary mechanisms \cite{wegst2015bioinspired,zheng2014ultralight,fratzl2007nature,ritchie2009plasticity,bhushan2009biomimetics,egan2015role}. In other words, biological materials adapt their ``\textit{form}'' (or shape/structure) against the dynamic external environment and improve the ``\textit{microstructure architecture},'' fulfilling the local needs imposed by physiological, phylogenetic, and reproductive constraints \cite{wolf1986law,gibson2012hierarchical,gao2003materials}. A rational understanding of microstructure interdependencies across hierarchical scales on the macroscale properties helps pave the way forward to many engineering applications involving biological materials such as the genetic tailoring of crops \cite{brule2016hierarchies,mccann2014plants}, bone remodeling \cite{rodrigues1999global,blanchard2016patient}, and the fabrication of bioinspired engineering materials \cite{wegst2015bioinspired,holstov2015hygromorphic}.


Multiscale modeling of hierarchical materials in conjunction with structural optimization methods
constitutes a promising pathway to elucidate optimum microstructure configurations in multiphase
hierarchical systems. In this context, recently developed concurrent multiscale analysis and topology optimization methods \cite{xia2014concurrent,xia2015multiscale,rodrigues2002hierarchical,coelho2008hierarchical,nakshatrala2013nonlinear,da2017concurrent,zhang2018multiscale} naturally fit to the dual optimization of structure (``\textit{form}'') and material (``\textit{microstructure architecture}''). The idea is to decompose the multiscale problem into two nested sub-problems, one at the macroscale (structure) and the other at the microscale (material).
At each macroscale material point, 
the microscale sub-problem provides a locally optimal material response and can be interpreted as the reformulation of a material constitutive law for the macroscale structure optimization problem. The material optimization problem is typically solved within an $\text{FE}^2 $ type computational homogenization framework \cite{feyel2000fe2,fish2013practical} at each macroscale Gauss point. Other approaches not based on scale separation or periodicity such as \cite{nguyen2019multiscale} also seem possible.
We note that throughout the article, we will use the term \textit{multiphase hierarchical system} for the combined representation of the \textit{multiphase hierarchical material} and the \textit{macrostructure} domain that habitats it. 




The base constituents in multiphase hierarchical systems often exhibit elastoplastic material properties resulting in a path-dependent macroscale mechanical response and dissipation-driven self-adapting mechanisms. In the case of path-dependent problems, efficient methods for the computational optimization of multiphase hierarchical systems are still in its infancy \cite{da2019topology}. 
In the context of fracture resistance and damage, recent contributions have proposed structure optimization methods optimizing the inclusion characteristics in matrix-inclusion type multiphase materials for path-dependent objective functions \cite{xia2018topology,li2021simp,kato2010material,kato2013multiphase,hilchenbach2015optimization}. 
The corresponding optimization formulations, however, remain in the format of a monoscale design, where all the morphological design parameters of the material are represented at the structure scale. In this case, the structure scale discretization is dictated by the smallest length scale of constituents, making these methods computationally prohibitive for multiphase hierarchical systems with several well-separated length scales. To the best of our knowledge, no work has been reported so far in the literature on a decomposed concurrent material and structure optimization formulation for path-dependent problems involving elastoplastic multiphase hierarchical systems.





The first major roadblock in the context of elastoplastic behavior across hierarchical scales is the non-trivial problem decomposition into material and structure optimization subproblems. The current state of concurrent material and structure optimization methods focuses only on end-compliance type optimization problems with an overall linear elastic response at both the material and structure levels. The variational structure of the displacement-based formulation of end-compliance optimization corresponds to a saddle point problem with respect to the admissible set for design variables and the space of kinematically admissible displacements \cite{lipton1994saddle}. With pointwise definitions of material design variables, the saddle point property enables a natural decomposition into the material and structure optimization subproblems \cite{jog1994topology}. The equivalent variational structure for combined analysis and optimization of path-dependent problems considering the complete deformation process is yet to be investigated. 

The second critical roadblock is the computational cost for multiscale analysis through computational homogenization that for multiphase hierarchical systems grows exponentially with each scale characterization \cite{yuan2009multiple,le2015computational,liu2016self,bessa2017framework}. Adding the topology optimization at the structure level results in even higher computational cost, since it requires solving many multiscale problems for different realizations of the structure topology during a typical optimization algorithm. This drawback limits existing approaches to small two-scales problems, even in the simplest case of hierarchical materials with linear elastic constituents. 
Continuum micromechanics provides a rigorous framework to derive analytical estimates of macroscale elastoplastic properties \cite{zaoui2002continuum,suquet2014continuum,morin2017micromechanics} and has been successfully applied to describe many multiphase hierarchical systems such as plant, wood, bone, and cementitious materials \cite{gangwar2019microimaging,gangwar2020multiscale,hofstetter2005development,hellmich2004can,fritsch2009ductile,pichler2011upscaling}. In the context of concurrent material and structure optimization, we recently integrated continuum micromechanics based homogenized estimates in end-compliance optimization problems, 
which rendered our framework computationally tractable for multiphase hierarchical systems with several material length scales \cite{gangwar2021concurrent}. 

In this article, we establish, for the first time, a thermodynamically consistent formulation of concurrent material and structure optimization, including suitable sub-problem formulations for multiphase hierarchical systems with elastoplastic constituents at the material scales. 
The structure optimization problem addresses the macroscale density distribution, while the material optimization problem at each material point seeks the optimized macroscale response with respect to microscale (material) design variables. In particular, we reformulate the material optimization problem based on the maximum plastic dissipation principle such that it assumes the format of an elastoplastic constitutive law that can be efficiently solved via modified return mapping algorithms. We express the homogenized stiffness and yield criterion as a function of material design variables within a continuum micromechanics framework, which enables the computationally tractable treatment of our optimization formulation.



Our article is organized as follows: In Section~2, we briefly review the relevant thermomechanical principles of elastoplasticity along with multiscaling concepts in continuum micromechanics, which form the basis of our further developments. In Section~3, we formulate the path-dependent stiffness maximization problem, decomposing material and structure optimization sub-problems for elastoplastic multiphase hierarchical systems. We then describe its discretization within the framework of the finite element method. 
In Section~4, we develop an algorithmic procedure for the material optimization problem based on the maximum plastic dissipation principle. In Section~5, we consolidate all our developments in an algorithmic framework and provide pertinent implementation details. Finally, we verify our framework with benchmark problems in Section~6. 

\section{A brief review of fundamental concepts}




\subsection{Thermomechanical formulation of elastoplasticity }
\label{sec:ch7_71}


We start by briefly reviewing the basic principles of elastoplasticity from a thermodynamics viewpoint, including the mechanical work identity, the notion of plastic dissipation from the second law of thermodynamics, and the derivation of the classical elastoplastic constitutive equations reflecting on the principle of maximum plastic dissipation. We primarily follow the exposition of \textsc{Simo \& Hughes} \cite{simo2006computational}. 

\subsubsection{The mechanical work identity}
\label{sec:ch7_711}

We consider a  macroscale initial boundary value problem defined on a domain $ \Omega $ and restrict our attention to a time interval $ [0,T] $. The position of a material point in the domain $ \Omega $ is denoted by $ \boldsymbol{x} $. The macroscale density at a material point $ \boldsymbol{x} $ is denoted as $ \rho (\boldsymbol{x}) $. The domain $\Omega $ is subjected to a traction $ \boldsymbol{\bar{t}}(t) $ at the Neumann boundary $ \Gamma_N $ and the prescribed displacements $ \boldsymbol{\bar{u}}^E(t) $ at the Dirichlet boundary $ \Gamma_D $ with a body force $\boldsymbol{b}(\boldsymbol{x},t)$, where $ t \in [0,T] $. Then, the macroscale displacement field $\boldsymbol{\bar{u}} (\boldsymbol{x},t)$ at a material point $ \boldsymbol{x}  $ and at time $ t \in [0,T]  $ is a mapping $ \boldsymbol{\bar{u}}: \Omega \times [0,T] \rightarrow \mathbb{R}^3 $. We define the corresponding velocity and strain fields at $ (\boldsymbol{x},t) \in \Omega \times  [0,T]  $ as
\begin{equation}
	\boldsymbol{v}(\boldsymbol{x},t): = \frac{\partial \boldsymbol{\bar{u}} (\boldsymbol{x},t)}{\partial t} \;\; \text{and} \;\; \boldsymbol{E}(\boldsymbol{x},t): = \frac{\partial \boldsymbol{\bar{u}} (\boldsymbol{x},t)}{\partial\boldsymbol{x} }.
	\label{eq:ch6_eq2}
\end{equation}


With the kinematically admissible velocity field $\boldsymbol{v}(\boldsymbol{x},t)$ and the macroscale stress field $ \boldsymbol{\Sigma}(\boldsymbol{x},t) $, the \textit{mechanical work identity} is
\begin{equation}
	\frac{d}{dt} T(\boldsymbol{v}) + P_{int}(\boldsymbol{\Sigma}, \boldsymbol{v}) = P_{ext}(\boldsymbol{v})\;\; \forall t \in [0,T],
	\label{eq:ch7_eq6} 
\end{equation}
where
\begin{equation}
	\begin{split}
		& \text{kinetic energy}\;\;\; T(\boldsymbol{v}) = \frac{1}{2} \int_{\Omega}\rho |\boldsymbol{v}|^2 \;d\Omega,    \\
		&\text{stress power}\;\; \;\;\;\;  P_{int}(\boldsymbol{\Sigma}, \boldsymbol{v}) = \int_{\Omega} \boldsymbol{\Sigma} : \frac{ \partial \boldsymbol{v}}{ \partial \boldsymbol{x}} \; d\Omega,   \\
		&\text{external power}\;\;  P_{ext}(\boldsymbol{v}) = \int_{\Omega}  \boldsymbol{b}  (\boldsymbol{x},t) \cdot \boldsymbol{v} \;d\Omega \; +  \int_{\Gamma_N} \boldsymbol{\bar{t}} (t) \cdot \boldsymbol{v} \;ds.
	\end{split}
	\label{eq:ch7_eq7}
\end{equation}
This result represents the energy conservation principle or the first law of thermodynamics. We note that this form is independent of the specific nature of the constitutive equations.\\

\noindent\textbf{Remark 1:} We emphasize that we chose $ (\bar{\square}) $ notation for the macroscale displacement $\boldsymbol{\bar{u}}$, and the given boundary conditions $ \boldsymbol{\bar{u}}^E$ and $\boldsymbol{\bar{t}}$. Later in this article, the displacement solution and boundary conditions drive the optimization algorithm, and, therefore, this choice directly relates with the introduced notations for the sought solutions for the optimized multiscale configurations. 



\subsubsection{Constitutive relations from the second law of thermodynamics}

The constitutive relations between the macroscale stress $ \boldsymbol{\Sigma}(\boldsymbol{x},t) $ and the macroscale displacement field $ \boldsymbol{\bar{u}} (\boldsymbol{x},t) $ (through the macroscale strains $ \boldsymbol{E}(\boldsymbol{x},t) $) close the global governing equations stated in \eqref{eq:ch7_eq6} and \eqref{eq:ch7_eq7}. The second law of thermodynamics governs the form of these constitutive relations, and we summarize the important results in the following. First, we decompose the macroscale strain tensor $ \boldsymbol{E} $ into an elastic and plastic part assuming small strains, denoted by $ \boldsymbol{E}^e $ and $ \boldsymbol{E}^p $, as
\begin{equation}
	\boldsymbol{E} = \boldsymbol{E}^e + \boldsymbol{E}^p. 
	\label{eq:ch6_eq8}
\end{equation}
We introduce the notion of \textit{internal potential energy} and \textit{dissipation} within the context of elastoplasticity. 
We define the internal energy of the system as
\begin{equation}
	V_{int}  = \int_{\Omega} \Psi(\boldsymbol{E}^{e})\; d \Omega,
	\label{eq:ch7_eq7a}
\end{equation}
where $\Psi(\boldsymbol{E}^{e}) $ is the \textit{Helmholtz free energy} density defined in terms of the stored elastic energy function $ W $ and the contributions from \textit{hardening} effects. In this presentation, we consider the case of perfect plasticity, which implies that the contribution from hardening is zero and $ \Psi = W $.

Next, we look at the difference between the stress power $ P_{int}(\boldsymbol{\Sigma}, \boldsymbol{v}) $ and the rate of change of the internal energy $V_{int} $, which we denote by $\mathcal{D}^{mech}$. Assuming isothermal conditions, the \textit{Clausius-Duhem} version of the second law of thermodynamics \cite{truesdell2004non,tadmor2012continuum} follows as
\begin{equation}
	\mathcal{D}^{mech}:= P_{int}(\boldsymbol{\Sigma}, \boldsymbol{v}) - \frac{d}{dt} V_{int} \geq 0 \;\; \forall t \in [0,T]. 
	\label{eq:ch7_eq9}
\end{equation}
We identify $\mathcal{D}^{mech} $ as the total instantaneous \textit{mechanical dissipation} in the domain $\Omega $ at time $ t \in [0,T] $, which is always non-negative. 

Inserting the definitions of $ P_{int}(\boldsymbol{\Sigma}, \boldsymbol{v}) $ and  $V_{int} $ from \eqref{eq:ch7_eq7} and \eqref{eq:ch7_eq7a} and using the strain decomposition \eqref{eq:ch6_eq8}, we arrive at
\begin{equation}
\mathcal{D}^{mech}= \int_{\Omega} \Bigg[   \Bigg(\boldsymbol{\Sigma} - \frac{\partial \Psi (\boldsymbol{E}^{e})}{\partial \boldsymbol{E}^{e}}\Bigg) : \dot{\boldsymbol{E}^e}   + \boldsymbol{\Sigma}:\dot{\boldsymbol{E}^{p}} \Bigg] \; d\Omega \geq 0,
	\label{eq:ch7_eq10}
\end{equation}
where $ (\dot{\square}) $ denotes the material time derivative of a quantity. The principle of thermodynamic determinism requires that (\ref{eq:ch7_eq10}) remains valid for any kinematic process defined by $\boldsymbol{E}^{e} $. This condition implies that 
\begin{equation}
	\boldsymbol{\Sigma} = \frac{\partial \Psi (\boldsymbol{E}^{e})}{\partial \boldsymbol{E}^{e}}\;\; \text{and}\;\; \boldsymbol{\Sigma}:\dot{\boldsymbol{E}^{p}}  \geq 0 . 
	\label{eq:ch7_eq11}
\end{equation}
The first equation is a typical elastic constitutive relation, where the stress is defined as the derivative of the free energy function with respect to the elastic part of the strain tensor. The second equation (\ref{eq:ch7_eq11}) represents the irreversible nature of an elastoplastic process implying that the dissipation energy is always non-negative. This relation constrains the possible stress states a material can undergo and indicates that the stress depends on the rate of the plastic part of the strain tensor.

\subsubsection{The principle of maximum plastic dissipation }
\label{sec:sec_213}


The principle of maximum plastic dissipation is a cornerstone of the mathematical formulation of plasticity. In the following, we derive the material constitutive equations for perfect plasticity from the viewpoint of this principle. We first assume a yield criterion $ \mathfrak{F}(\boldsymbol{\tau}) $, with $\boldsymbol{\tau}\in\mathbb{S}$ denoting any possible stress state, where $\mathbb{S}$ is the space of second-order symmetric tensors. Its zero isosurface is the usually convex yield surface that encloses 
the space of admissible stresses 
\begin{equation}
	\mathbb{E}_{\Sigma}:= \Big\{  \boldsymbol{\tau} \in \mathbb{S} \;|\; \mathfrak{F}(\boldsymbol{\tau}) \leq 0\Big\}.
	\label{eq:ch7_eq3}
\end{equation}
For a given plastic strain $\boldsymbol{E}^{p}\in \mathbb{S}$, we define the plastic dissipation $\mathcal{D}^{p} $ at a material point for perfect plasticity as
\begin{equation}
	\mathcal{D}^{p}[\boldsymbol{\tau}; \dot{\boldsymbol{E}}^p] : = \boldsymbol{\tau}:\dot{\boldsymbol{E}}^{p}.
	\label{eq:ch7_eq1}
\end{equation}
where $\boldsymbol{\tau}  \in \mathbb{E}_{\Sigma} $ now denotes an admissible stress state. 

In the local form, the principle of maximum plastic dissipation states that, for a given plastic strain $\boldsymbol{E}^{p} \in \mathbb{S}$, the plastic dissipation $ \mathcal{D}^{p} $ attains its maximum for the actual stress tensor $ \boldsymbol{\Sigma} $ among all possible stresses $\boldsymbol{\tau} \in \mathbb{E}_{\Sigma} $. Mathematically, the principle is
\begin{equation}
	\mathcal{D}^{p}[\boldsymbol{\Sigma}; \dot{\boldsymbol{E}}^p] = \max_{ \boldsymbol{\tau} \in \mathbb{E}_{\Sigma}} \Big\{ \mathcal{D}^{p}[\boldsymbol{\tau}; \dot{\boldsymbol{E}}^p] \Big\}. 
	\label{eq:ch7_eq2}
\end{equation}

The classical formulation of plasticity (associative flow rule, loading/unloading conditions) directly follows from this principle. To this end, we first transform the maximization principle into a minimization problem by changing the sign of the objective function. Next, we transform the constraint minimization problem into an unconstrained problem by introducing the cone of Lagrange multipliers
\begin{equation}
	\mathbb{K} := \Big\{  \delta \in L^2(\Omega) \;|\; \delta \geq 0\Big\},
	\label{eq:ch7_eq3cone} 
\end{equation}
where $L^2$ denotes the space of all square integrable functions. The corresponding Lagrangian functional $\mathcal{L}^{p}: \mathbb{S} \times \mathbb{K} \times \mathbb{S} \rightarrow \mathbb{R}$ is then
\begin{equation}
	\mathcal{L}^{p}( \boldsymbol{\tau}, \delta; \dot{\boldsymbol{E}}^p) := - \boldsymbol{\tau}:\dot{\boldsymbol{E}}^{p} + \delta\; \mathfrak{F}(\boldsymbol{\tau}). 
	\label{eq:ch7_eq4} 
\end{equation}
The solution to (\ref{eq:ch7_eq2}) is then given by a point $(\boldsymbol{\Sigma},\gamma) \in \mathbb{S} \times \mathbb{K}$ satisfying the Karush-Kuhn-Tucker optimality conditions for the Lagrangian functional
\eqref{eq:ch7_eq4}. The conditions are
\begin{equation}
	\begin{split}
		& \left.\frac{\partial \; \mathcal{L}^{p}( \boldsymbol{\tau}, \gamma; \dot{\boldsymbol{E}}^p) }{\partial \; \boldsymbol{\tau}} \right\vert_{\boldsymbol{\Sigma},\gamma} = 	- \dot{\boldsymbol{E}}^p + \gamma\; \left.\frac{\partial \mathfrak{F}^{}_{}(\boldsymbol{\tau})}{\partial \boldsymbol{\tau}}\right\vert_{\boldsymbol{\Sigma}} = 0,\\ &
		\gamma \geq 0, \;\; \mathfrak{F}(\boldsymbol{\Sigma}) \leq 0, \;\; \text{and}\;\; \gamma\; \mathfrak{F}(\boldsymbol{\Sigma}) = 0.
	\end{split}	
	\label{eq:ch7_eq5} 
\end{equation}

The first equation in \eqref{eq:ch7_eq5} is the \textit{associated flow rule}, often also called the normality of the flow rule. The second and third equations in \eqref{eq:ch7_eq5} are the classical \textit{loading/unloading conditions}. The only requirement for these equations to hold uniquely is the convexity of the elastic range $ \mathbb{E}_{\Sigma} $. A sufficient condition for this requirement is the convexity of the yield criterion function $ \mathfrak{F}(\boldsymbol{\tau}) $. We will exploit these aspects later on for devising solution strategies for our optimization framework.

\subsection{Multiscaling concepts in continuum micromechanics}
\label{sec:sec22}


Continuum micromechanics forms a rigorous foundation for the (semi-)analytical estimation of homogenized stiffness and strength properties of materials with hierarchical microstructures. Here, we state the key results that are relevant in the context of this article. For a detailed review, interested readers are referred to the presentations given in \cite{zaoui2002continuum,suquet1997effective}. 
 
\subsubsection{Estimation of homogenized elastic properties}
\label{sec:sec221}


The goal of continuum micromechanics is to estimate the homogenized response of a representative volume element (RVE) filled with microheterogeneous material. For the existence of such an RVE, a minimal requirement is that the characteristic length, $ d $, of the considered inhomogeneities and deformation mechanisms is much smaller than the size, $ l $, of the RVE. Moreover, $l$ must be much smaller than the characteristic length scale of the variation in the loading on the macroscale structure, $L$. Therefore, a proper scale separation implies
\begin{equation}
	d \ll l \ll L.
	\label{eq:ch2_eq1}
\end{equation}

In each phase $r $ of the RVE, the average microscopic stress $\boldsymbol{\sigma}_r $, the average microscopic strain $\boldsymbol{\varepsilon}_r $, and the phase stiffness $\mathbb{c}_r$ are linked as: $ \boldsymbol{\sigma}_r = \mathbb{c}_r:\boldsymbol{\varepsilon}_r $. The kinematic compatibility for the homogeneous strain boundary conditions for the RVE relates the macroscale strain tensor $\boldsymbol{E}$ with the volume average of microscopic strains $\boldsymbol{\varepsilon}_r $. Similarly, the equilibrated microscopic stresses $\boldsymbol{\sigma}_r $ and the macroscale stress tensor $\boldsymbol{\Sigma}$ fulfill the volume average relation following the homogeneous stress boundary conditions. With $\phi_r$ as the volume fraction of the phase $ r $, these relations are
\begin{equation}
	\boldsymbol{E} = \sum_{r} \phi_r \boldsymbol{\varepsilon}_r \;\;\;\; \text{and} \;\;\;\; \boldsymbol{\Sigma} = \sum_{r} \phi_r \boldsymbol{\sigma}_r.
	\label{eq:eqs39g}
\end{equation}
A link between the macroscale strain $\boldsymbol{E}$ and the average microscopic strain $\boldsymbol{\varepsilon}_r $ of phase $r$ is established with a fourth order concentration tensor $\mathbb{A}_r$ as
\begin{equation}
	\boldsymbol{\varepsilon} _r = \mathbb{A}_r :\boldsymbol{E}.
	\label{eq:eqs39h}
\end{equation}

A comparison of the macroscale constitutive relation $ \boldsymbol{\Sigma} = \mathbb{C}:\boldsymbol{E} $ with \eqref{eq:eqs39g} and \eqref{eq:eqs39h} yields the homogenized stiffness $\mathbb{C} $ in terms of the volume fraction, stiffness, and concentration tensor of constituent phases as
\begin{equation}
	\mathbb{C} = \sum_{r} \phi_r \mathbb{c}_r: \mathbb{A}_r.
	\label{eq:eqs39i}
\end{equation}
It is clear from (\ref{eq:eqs39i}) that the estimation of the concentration tensors $ \mathbb{A} $ entails the homogenized stiffness $ \mathbb{C} $. The simplest choice for  $ \mathbb{A} $ is to assume a uniform strain state throughout the RVE, that is  $\mathbb{A} = \mathbb{I} $, where $ \mathbb{I} $ is a fourth-order symmetric unit tensor. This choice leads to the well-established Voigt mixture rule for homogenized stiffness. However, the Voigt rule does not consider any other statistical information beyond the volume fraction of phases. We note that the Voigt rule is often applied for ``homogenization/interpolation'' between a solid material and voids in conjunction with relaxation to ill-defined 0-1 type problems in topology optimization \cite{bendsoe1999material,allaire1999optimal}.

The estimation of the concentration tensor $\mathbb{A}_r$ based on Eshelby's matrix-inclusion solutions can incorporate the volume fraction, the shape of phases, and their interaction with each other. Eshelby's matrix-inclusion problem relates strains in an ellipsoidal inclusion perfectly bonded with the surrounded homogeneous infinite elastic matrix to the applied homogeneous strains at infinity. Following \cite{zaoui2002continuum}, the estimation of $\mathbb{A}_r$ from the matrix-inclusion problem entails the homogenized stiffness expression as
\begin{equation}
	\begin{split}
		\mathbb{C} =  \sum_{r} \phi_r  \mathbb{c}_r : [\mathbb{I}  + \mathbb{P}_r^0:  (\mathbb{c}_r - \mathbb{C}^0)]^{-1}: \Big[\sum_{s} \phi_s [\mathbb{I} + \mathbb{P}_s^0:(\mathbb{c}_s-\mathbb{C}^0)]^{-1}\Big]^{-1}\;.
	\end{split}
	\label{eq:eqs39l} 
\end{equation} 
Here, the Hill tensor $\mathbb{P}_r^0$ characterizes the morphology of the inclusion phase $r$ and its interaction with the surrounding reference matrix with stiffness tensor $\mathbb{C}^0$. The Hill tensor $\mathbb{P}_r^0$ depends on the morphology, that is, the shape and orientation of the inclusion phase as well as the stiffness tensor of the reference matrix. The analytical expressions for $\mathbb{P}_r^0$ can be found in \cite{laws1977determination,laws1985note,masson2008new}. With \eqref{eq:eqs39l}, the homogenized stiffness of the RVE can be expressed as a function of constituent phase characteristics. 

\subsubsection{Estimation of homogenized elastic limit strength}
\label{sec:sec222}




A macroscale RVE reaches the elastic limit state when any one of the constituents in the RVE yields. Let us focus on the weakest constituent phase, denoted by index $ r = w $. We assume that its elastic limit behavior is described by the yield criterion
\begin{equation}
	\boldsymbol{\mathfrak{f}}_{w} (\boldsymbol{\sigma}^{*}_w) \leq 0,
	\label{eqs31:eq13}
\end{equation}
where $\boldsymbol{\sigma}^{*}_w $ is the effective stress distribution in the weak phase $ w$. Moreover, we assume that it is the only constituent that exhibits inelastic behavior. 
The effective stress or ``stress peaks'' in phase $ w $ can then be estimated with the second-order moment of the stress field in this phase, which is the quadratic stress average $ \overline{\overline{\boldsymbol{\sigma}}}_w $ over the phase volume $ V_w $ \cite{suquet1997effective}, expressed as
\begin{equation}
	\boldsymbol{\sigma}^{*}_w = \overline{\overline{\boldsymbol{\sigma}}}_w = \langle \boldsymbol{\sigma}:\boldsymbol{\sigma}   \rangle^{1/2}_w =  \Big( \frac{1}{V_w} \int_{V_w}^{} \frac{1}{2} \boldsymbol{\sigma}:\boldsymbol{\sigma} \; dV \Big)^{1/2}\;.
	\label{eqs31:eq15}
\end{equation}


Next we assume that $ \boldsymbol{\mathfrak{f}}_w $ is a scalar deviatoric stress-based yield criterion such as the von Mises criterion with known yield strength $ \sigma^{Y}_{w}$, bulk modulus $\mu_{w} $ and effective volume function $\bar{\phi}_w$ for the weak phase $ w $. Following \cite{suquet1997effective}, with the admissible stress $\boldsymbol{\tau}$ and homogenized stiffness $\mathbb{C} $, the weak phase criterion $ \boldsymbol{\mathfrak{f}}_w $ translates to the macroscopic yield criterion $ \mathfrak{F} $ as
\begin{equation}
	\mathfrak{F} (\boldsymbol{\tau}) = \sqrt{\boldsymbol{\tau}:[\mathbb{C}]^{-1}:\frac{\partial \mathbb{C}}{\partial \mu_{w}}:[\mathbb{C}]^{-1}:\boldsymbol{\tau}} \; - \;  \sqrt{\frac{\bar{\phi}_w}{3}} \; \frac{\sigma^{Y}_{w}}{\mu_{w}} \leq 0. 
	\label{eq:ch6_eq15}
\end{equation}
We emphasize that (\ref{eq:ch6_eq15}) is of the form of $ \mathfrak{F} = \sqrt{\boldsymbol{\tau}:\mathbb{M}:\boldsymbol{\tau}}  - R \leq 0 $ that represents the general quadratic form of classical rate-independent plasticity models. In the context of this article, it implies that the elastic domain defined by (\ref{eq:ch6_eq15}) satisfies two critical geometric properties. These properties are (1) the \textit{convexity} of the elastic domain and (2) the \textit{degree-one homogeneity} of the yield criterion. These properties are very important for developing solution algorithms for our optimization formulation, and we will recall them later in subsequent sections. 


\begin{figure*}[t!]
	\centering
	\includegraphics[width=0.85\textwidth]{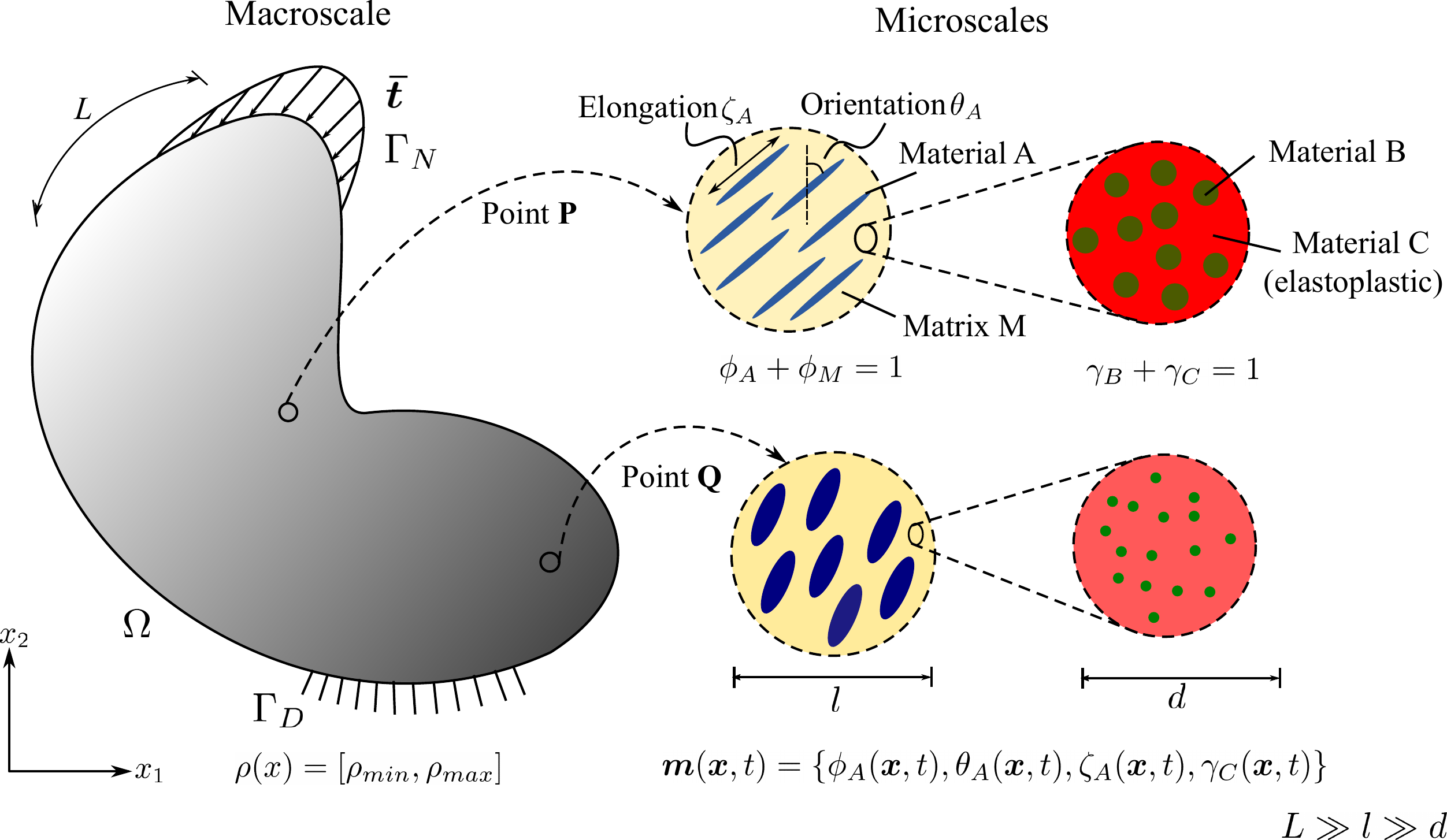}	
	\caption{Sketch of a representative elastoplastic multiphase hierarchical system.}
	\label{Figs:ch7_fig1}
\end{figure*}

\section{A framework for path-dependent concurrent material and structure optimization}

In this section, we formulate a concurrent material and structure optimization method maximizing the 
path-dependent stiffness for elastoplastic multiphase hierarchical systems. We then focus on the finite element discretization of this formulation. 


\subsection{Thermodynamically consistent 
formulation}




We start by looking at a representative problem illustrated in Fig.~\ref{Figs:ch7_fig1}. We assume a fixed reference domain $ \Omega $ subjected to a traction $ \boldsymbol{\bar{t}}(t) $ at the Neumann boundary $ \Gamma_N $ and the prescribed displacements $ \boldsymbol{\bar{u}}^E(t) $ at the Dirichlet boundary $ \Gamma_D $ with a body force $\boldsymbol{b}(\boldsymbol{x},t)$, where $ t \in [0,T] $. 

\subsubsection{Global optimization problem with micromechanical design variables }

%
%

We introduce the definition of macroscale density $ \rho (\boldsymbol{x}) $ and microstructure characterization $ \boldsymbol{m}(\boldsymbol{x},t) $. 
We assume that the macroscale density $ \rho (\boldsymbol{x}) $ is fixed with respect to time, while $\boldsymbol{m}(\boldsymbol{x},t) $ is a function of loading history representing a local adaption of microstructure with time. The set $ \boldsymbol{m}(\boldsymbol{x},t) $ contains the geometric and mechanical characterization of phases that span multiple well-separated microscales, consisting of volume fraction, material properties, shape, and orientation of the different phases in the hierarchical system. The homogenized material constitutive relations defined by the plastic dissipation $\mathcal{D}^{p} $ and the Helmholtz free energy $ \Psi $ in Section~\ref{sec:ch7_71} 
depend on the macroscale density $ \rho (\boldsymbol{x}) $ and the microstructure characterization field $ \boldsymbol{m}(\boldsymbol{x},t) $. Therefore, the design vector is $ [\rho(\boldsymbol{x}), \boldsymbol{m}(\boldsymbol{x},t)]^{T}$. 

A typical objective is to maximize the structural stiffness for the path-dependent nonlinear structure designs. It translates as the maximization of the total mechanical work expended in the course of a deformation process \cite{fritzen2016topology}. Assuming a quasi-static case with no inertial effects, the total mechanical work $ f_w $ in the considered time interval $[0,T] $ follows directly from the mechanical work identity (\ref{eq:ch7_eq6}) and the definition of stress power \eqref{eq:ch7_eq9} as
\begin{equation}
	\begin{split}
		f_w =\int_{0}^{T} \Big[\mathcal{D}^{mech} + \frac{d}{dt} V_{int}\Big] dt = \int_{0}^{T} P_{ext}(\boldsymbol{v}) dt. 
	\end{split}
	\label{eq:ch7_eq12}
\end{equation}
Utilizing the definitions of $\mathcal{D}^{mech} $, $ V_{int} $ and $P_{ext}(\boldsymbol{v})$ from Section~\ref{sec:ch7_71}, we arrive at
\begin{equation}
	\begin{split}
		f_w	= &\int_{0}^{T} \Big[ \int_{\Omega} \Big\{\mathcal{D}^{p}[\boldsymbol{\Sigma}; \dot{\boldsymbol{E}}^p] +  \dot{\Psi}(\boldsymbol{E}^{e}) \Big\} d \Omega  \Big] \;dt \\ = & \int_{0}^{T} \Big[ \int_{\Omega}  \boldsymbol{b}(\boldsymbol{x},t) \cdot \boldsymbol{v} \;d\Omega \; +  \int_{\Gamma_N} \boldsymbol{\bar{t}}(t) \cdot \boldsymbol{v} \;ds \Big] dt.
	\end{split}
	\label{eq:ch7_eq13}
\end{equation}
We note that the (pseudo-)time $ t $ 
represents the loading history. 

Augmenting 
$\mathcal{D}^{p} $ and $ \Psi $ with $ \rho (\boldsymbol{x}) $ and $ \boldsymbol{m}(\boldsymbol{x},t) $, we set up our optimization problem using the introduced definition of total mechanical work $ f_w $ from \eqref{eq:ch7_eq13} as
\begin{equation}
	\scalebox{0.96}{$
		\begin{split}
			\max_{ \substack{\rho(\boldsymbol{x}) \in \mathcal{A}_{\textit{ad}} \\  \boldsymbol{m}(\boldsymbol{x},t) \in E_{\textit{ad}}  }} f_w = & \max_{ \substack{\rho(\boldsymbol{x}) \in \mathcal{A}_{\textit{ad}} \\  \boldsymbol{m}(\boldsymbol{x},t) \in E_{\textit{ad}}  }} \int_{0}^{T}  \int_{\Omega} \Big\{\mathcal{D}^{p}[\rho(\boldsymbol{x}), \boldsymbol{m}(\boldsymbol{x},t),\boldsymbol{\Sigma}; \dot{\boldsymbol{E}}^p] + \dot{\Psi} [\rho(\boldsymbol{x}), \boldsymbol{m}(\boldsymbol{x},t);\boldsymbol{E}^{e}] \Big\} d \Omega \; dt \\ 
			= & \max_{ \substack{\rho(\boldsymbol{x}) \in \mathcal{A}_{\textit{ad}} \\  \boldsymbol{m}(\boldsymbol{x},t) \in E_{\textit{ad}}  }} \int_{0}^{T} \Big[ \int_{\Omega}  \boldsymbol{b}(\boldsymbol{x},t) \cdot \boldsymbol{v} \;d\Omega \; +  \int_{\Gamma_N} \boldsymbol{\bar{t}}(t) \cdot \boldsymbol{v} \;ds \Big] dt\;. 
		\end{split}
		$}
	\label{eq:ch7_eq16}
\end{equation}
$ \mathcal{A}_{\textit{ad}}  $ and $ E_{\textit{ad}} $ define the set of admissible design variables at the macro- and microscales, respectively, with possible design constraints. The second part of this equation is the continuous version of the objective function proposed in \cite{fritzen2016topology}. On the one hand, the velocity field $ \boldsymbol{v}(\boldsymbol{x},t) $ and, thus, the displacement field $\boldsymbol{\bar{u}} (\boldsymbol{x},t)$ implicitly depend on $ \rho (\boldsymbol{x}) $ and $ \boldsymbol{m}(\boldsymbol{x},t) $. On the other hand, the first part of (\ref{eq:ch7_eq16}) is an explicit expression in terms of the design variables $ \rho (\boldsymbol{x}) $ and $ \boldsymbol{m}(\boldsymbol{x},t)$ with known macroscale strains $ \boldsymbol{E} $ and $\boldsymbol{E}^p$ that arise from the solution of the global equilibrium equations. The first part of \eqref{eq:ch7_eq16} is the basis for the development and mathematical analysis of our optimization formulation. Later on in this article, we will come back to the second part of \eqref{eq:ch7_eq16} to motivate sensitivity calculations. 


\subsubsection{Definition of the sets of admissible design variables}

The admissible set $ \mathcal{A}_{\textit{ad}} $ seeks a limit on the total material mass $ M_{\textit{req}} $ available for design.  Mathematically, it can be defined as
\begin{equation}
	\begin{split}
		\mathcal{A}_{\textit{ad}} = \Big\{   \rho(\boldsymbol{x}) \; | \; \rho(\boldsymbol{x}) = [\rho_{\textit{min}}, \rho_{\textit{max}}], \;\int_{\Omega} \rho(\boldsymbol{x}) d \Omega \leq M^{}_{\textit{req}}, \; \boldsymbol{x} \in \Omega    \Big\},
	\end{split}
	\label{eq:ch7_eq17}
\end{equation}
where $ \rho_{\textit{min}} $ and  $ \rho_{\textit{max}} $ are the bounds on the macroscale material density $ \rho $. 

Without loss of generality, the definition of the admissible set $E_{\textit{ad}} $ is illustrated via the representative multiphase hierarchical system shown in Fig.~\ref{Figs:ch7_fig1}. We observe a well-separated three-scale hierarchical system with three base constituent materials denoted as Material A, B, and C with densities $ \rho_A$, $\rho_B$, and $ \rho_C $, respectively. Material A and B are linear elastic, and Material C exhibits a perfectly elastoplastic response. At a material point P, the volume fraction of Material B and C at the lowermost scale are $ \gamma_B $ and $ \gamma_C $ such that $ \gamma_B + \gamma_C = 1  $. Material B forms spherical inclusions in the matrix of Material C at this scale. The homogenized material from this scale forms the matrix M that hosts the inclusions of Material A with the orientation $ \theta_A  $ and elongation $\zeta_A$ at the mesoscale. The density of the matrix M is simply $ \rho_M = (\gamma_B \rho_B + \gamma_C \rho_C)$. The volume fraction of Material A and matrix M are $ \phi_A $ and $ \phi_M $ with $ \phi_A + \phi_M = 1 $. 
 The microstructure characterization field set $ \boldsymbol{m}(\boldsymbol{x},t) $ is thus $ \{ {\phi}_A(\boldsymbol{x},t), {\theta}_A (\boldsymbol{x},t) ,\zeta_A (\boldsymbol{x},t), {\gamma}_C (\boldsymbol{x},t) \} $. We note that this set can be arbitrarily extended if necessary.
 
 \allowdisplaybreaks
 
With these definitions, the microscale design admissible set $ E_{\textit{ad}} $ for the representative multiphase hierarchical system shown in Fig.~\ref{Figs:ch7_fig1} follows as
 \begin{align}
 			E_{\textit{ad}} =  \Big\{ \boldsymbol{m}(\boldsymbol{x},t) \; | & \; \nonumber \rho(\boldsymbol{x}) = \rho_A \phi_A (\boldsymbol{x},t) + \rho_M (\boldsymbol{x},t) (1 - \phi_A (\boldsymbol{x},t)),\; \\ \nonumber & 0 < \phi^{\textit{min}}_A < \phi_A(\boldsymbol{x},t) < \phi^{\textit{max}}_A \leq 1, \;  \label{eq:ch7_eq18} \\  & \rho_M (\boldsymbol{x},t) = \rho_B  (1-\boldsymbol{\gamma}_C(\boldsymbol{x},t))   + \rho_C \; \gamma_C(\boldsymbol{x},t), \;\\ \nonumber & 0 < \gamma^{\textit{min}}_C < \gamma_C(\boldsymbol{x},t) < \gamma^{\textit{max}}_C \leq 1,\\ \nonumber &
 			\theta_A (\boldsymbol{x},t) \in [-\pi/2,\pi/2],\; \\ &\boldsymbol{\zeta}_A (\boldsymbol{x},t) \in [1,\zeta^{\textit{max}}],\; \forall (\boldsymbol{x},t) \in \Omega \times  [0,T]  \Big\}. \nonumber
 \end{align}
Here, the volume fraction of Material A is bounded by $ \phi_A^{\textit{min}}  $ and $ \phi_A^{\textit{max}} $, and the volume fraction of Material C is bounded by $ \gamma_C^{\textit{min}} $ and $ \gamma_C^{\textit{max}} $ at their respected scales. Furthermore, the elongation ratio of the inclusions of Material A is bounded by $ \zeta^{\textit{max}} $. We note that the bounds are constant and do not depend on the loading history or the material position in the domain. We emphasize again that the multiscale configuration of Fig.~\ref{Figs:ch7_fig1} is used for illustration, but the underlying represenation is easily generalized to cover any other multiphase hierarchical system.

\subsubsection{Decomposition into material and structure optimization problems}

%
%



The definition of the admissible set $ E_{\textit{ad}} $ at a material point $\boldsymbol{x} $ only depends on the macroscale density $ \boldsymbol{\rho}(\boldsymbol{x}) $ of this point. It implies that the definition of $ E_{\textit{ad}} $ is pointwise, and we can thus rewrite the statement (\ref{eq:ch7_eq16}) as
\begin{equation}
	\begin{split}
		\max_{ \substack{\rho(\boldsymbol{x}) \in \mathcal{A}_{\textit{ad}} }} \;\; \max_{\boldsymbol{m}(\boldsymbol{x},t) \in E_{\textit{ad}} (\rho(\boldsymbol{x})) }  \int_{0}^{T}  \int_{\Omega} \Big\{\mathcal{D}^{p}[\rho(\boldsymbol{x}), \boldsymbol{m}(\boldsymbol{x},t), \boldsymbol{\Sigma}; \dot{\boldsymbol{E}}^p] + \dot{\Psi} [\rho(\boldsymbol{x}), \boldsymbol{m}(\boldsymbol{x},t);\boldsymbol{E}^{e}] \Big\} d \Omega  \;dt.
	\end{split}
	\label{eq:ch7_eq29}
\end{equation}
We also assume that the macroscale density $ \boldsymbol{\rho}(\boldsymbol{x}) $ is fixed with respect to the loading history. Therefore, we are allowed to swap the integral and maximization operations in (\ref{eq:ch7_eq29}), hence
\begin{equation}
	\begin{split}
		\max_{ \substack{\rho(\boldsymbol{x}) \in \mathcal{A}_{\textit{ad}} }} \;\;   \int_{0}^{T} \int_{\Omega} \max_{\boldsymbol{m}(\boldsymbol{x},t) \in E_{\textit{ad}} (\rho(\boldsymbol{x})) } \Big\{\mathcal{D}^{p}[\rho(\boldsymbol{x}), \boldsymbol{m}(\boldsymbol{x},t), \boldsymbol{\Sigma}; \dot{\boldsymbol{E}}^p] + \dot{\Psi} [\rho(\boldsymbol{x}), \boldsymbol{m}(\boldsymbol{x},t);\boldsymbol{E}^{e}] \Big\} d \Omega \;dt.
	\end{split}
	\label{eq:ch7_eq30}
\end{equation}

The statement (\ref{eq:ch7_eq30}) allows us to decompose the optimization problem into two sub-problems. The outer ``structure'' optimization problem is
\begin{equation}
	\begin{split}
		& \max_{ \substack{\rho(\boldsymbol{x}) \in \mathcal{A}_{\textit{ad}}  }} \int_{0}^{T} \int_{\Omega} \Big\{\mathcal{D}^{p}[\rho(\boldsymbol{x}), \boldsymbol{\bar{m}}(\boldsymbol{x},t),\boldsymbol{\Sigma} ; \dot{\boldsymbol{E}}^p] + \dot{\Psi} [\rho(\boldsymbol{x}), \boldsymbol{\bar{m}}(\boldsymbol{x},t);\boldsymbol{E}^{e}] \Big\} d \Omega  \;dt\\
	  = & \max_{ \substack{\rho(\boldsymbol{x}) \in \mathcal{A}_{\textit{ad}}  }} \int_{0}^{T} \Big[ \int_{\Omega}  \boldsymbol{b}(\boldsymbol{x},t) \cdot \boldsymbol{v} \;d\Omega \; +  \int_{\Gamma_N} \boldsymbol{\bar{t}}(t) \cdot \boldsymbol{v} \;ds \Big] dt.
	\end{split}
	\label{eq:ch7_eq31}
\end{equation}
We also explicitly write the equivalent statement in terms of the total external work done in the deformation process from \eqref{eq:ch7_eq16}, which we will exploit in the subsequent sections for discretization and sensitivity calculations. In this statement, $\boldsymbol{\bar{m}}(\boldsymbol{x},t) $ optimizes the following sub-problem or ``material'' optimization problem: 
\begin{equation}
	\begin{split}
		&\max_{\boldsymbol{m}(\boldsymbol{x},t) \in E_{\textit{ad}}(\rho(\boldsymbol{x}))} \Big\{\mathcal{D}^{p}[ \boldsymbol{m}(\boldsymbol{x},t), \boldsymbol{\Sigma}; \dot{\boldsymbol{E}}^p] + \dot{\Psi} [ \boldsymbol{m}(\boldsymbol{x},t);\boldsymbol{E}^{e}] \Big\}\;\; 
		 = \\
		&\max_{\boldsymbol{m}(\boldsymbol{x},t) \in E_{\textit{ad}}(\rho(\boldsymbol{x}))} \Big\{  \max_{ \boldsymbol{\tau} \in \mathbb{E}_{\Sigma}} \{ \boldsymbol{\tau}:\dot{\boldsymbol{E}}^p \} + \dot{\Psi} [ \boldsymbol{m}(\boldsymbol{x},t);\boldsymbol{E}^{e}] \Big\}\;\;\;\; \forall (\boldsymbol{x},t) \in \Omega \times  [0,T].
	\end{split}
	\label{eq:ch7_eq32}
\end{equation}
The macroscale density $ \rho(\boldsymbol{x}) $ dictates the construction of the admissible space $ E_{ad} $, and, therefore, we take it out from the definitions of $\mathcal{D}^{p} $ and $\Psi $ in (\ref{eq:ch7_eq32}) and consider it in $ E_{ad} $. In the second line, we rewrite the definition of the plastic dissipation $\mathcal{D}^{p}$ with the help of the principle of maximum plastic dissipation discussed in Section~\ref{sec:sec_213}. 

The constitutive relations in \eqref{eq:ch7_eq32} defined through the Helmholtz free energy $\Psi $ and the plastic dissipation $\mathcal{D}^{p} $ remain to be discussed. For linearized elasticity, the stored elastic energy function $ W $ takes a quadratic form in the elastic part of the strain tensor $ \boldsymbol{E}^{e} = \boldsymbol{E}^{} -  \boldsymbol{E}^{p}  $. The homogenized elasticity tensor $ \mathbb{C} $ is a function of the microstructure characterization field set $ \boldsymbol{m}(\boldsymbol{x},t) \in E_{ad}(\rho(\boldsymbol{x})) $. For the perfect plasticity case, that is $ \Psi = W $, the quadratic form follows as 
\begin{equation}
	\begin{split}
		\Psi [\boldsymbol{m}(\boldsymbol{x},t);\boldsymbol{E}^{e}] = \; \frac{1}{2}(\boldsymbol{E}^{} - \boldsymbol{E}^{p} ): \mathbb{C} (\boldsymbol{m}(\boldsymbol{x},t) ):(\boldsymbol{E}^{} - \boldsymbol{E}^{p}),
	\end{split}
	\label{eq:ch7_eq13a}
\end{equation}
The elastic constitutive equation from \eqref{eq:ch7_eq11} entails the following stress-strain relationship: 
\begin{equation}
	\boldsymbol{\Sigma} = \frac{\partial \Psi (\boldsymbol{E}^{e})}{\partial \boldsymbol{E}^{e}}\; =   \mathbb{C}(\boldsymbol{m}(\boldsymbol{x},t)): (\boldsymbol{E}^{} - \boldsymbol{E}^{p}). 
	\label{eq:ch7_eq13d}
\end{equation}
Similarly, the elastoplastic material constitutive equations through the maximum plastic dissipation principle stated in (\ref{eq:ch7_eq2}) 
is augmented to include $ \rho (\boldsymbol{x}) $ and $ \boldsymbol{m}(\boldsymbol{x},t) $ as
\begin{equation}
	\mathcal{D}^{p}[ \boldsymbol{m}(\boldsymbol{x},t), \boldsymbol{\Sigma}; \dot{\boldsymbol{E}}^p] = \max_{ \boldsymbol{\tau} \in \mathbb{E}_{\Sigma}} \Big\{ \boldsymbol{\tau}:\dot{\boldsymbol{E}}^p \Big\},
\end{equation}
where the definition of the elastic closure $\mathbb{E}_{\Sigma}$ in terms of $ \rho(\boldsymbol{x}) $ and $\boldsymbol{\bar{m}}(\boldsymbol{x},t)$ is 
\begin{equation}
	\mathbb{E}_{\Sigma}:= \Big\{  \boldsymbol{\tau} \in \mathbb{S}\;|\; \boldsymbol{m}(\boldsymbol{x},t) \in E_{ad}(\rho(\boldsymbol{x})),\; \mathfrak{F}(\boldsymbol{\tau},\boldsymbol{m}(\boldsymbol{x},t)) \leq 0\Big\}. 
\label{eq:ch7_eq13b}
\end{equation}

The homogenized stiffness $ \mathbb{C} (\boldsymbol{m}(\boldsymbol{x},t) ) $ and the homogenized yield criterion $ \mathfrak{F}(\boldsymbol{\tau},\boldsymbol{m}(\boldsymbol{x},t))$ can be estimated as a function of microstructure variables for instance via the continuum micromechanics principles outlined in Section~\ref{sec:sec22}.

\subsubsection{Interpretation as an inelastic constitutive law}

The combination of (\ref{eq:ch7_eq31}) and  (\ref{eq:ch7_eq32}) constitutes the concurrent material and structure optimization formulation. The maximization problem in (\ref{eq:ch7_eq31}) seeks the optimal material distribution $ \rho(\boldsymbol{x}) $ in the domain $\Omega$. For a given material distribution $ \rho(\boldsymbol{x}) $, the optimization problem \eqref{eq:ch7_eq32} finds the optimal microstructure configuration maximizing the stress/deformation power for the known macroscale strains at each material point $\boldsymbol{x}$. Both statements are coupled through the macroscale strains and, therefore, through the displacement field solution $\boldsymbol{\bar{u}} (\boldsymbol{x},t)$ that satisfies the global equilibrium equations. This interdependency makes the global equilibrium a constitutively nonlinear problem analogous to a typical initial boundary value problem with an inelastic constitutive law. Therefore, we propose to interpret the material optimization problem as a reformulated elastoplastic constitutive law that provides the locally optimal material response with respect to the external loading history. The microstructure variable $ \boldsymbol{m}(\boldsymbol{x},t)$ can be thought of as an ``internal state variable'' analogous to any path-dependent history variable encountered in elastoplasticity formulations. This interpretation will be used in Section~\ref{sec:sec_4} for devising the optimization algorithm for the material optimization problem.

\subsection{Finite element discretization}
In the next step, we discretize our concurrent material and structure optimization formulation within the context of the finite element method. We use vector-matrix notation, consistent with the standard finite element discretization of the initial boundary value problem introduced in \cite{simo2006computational}, to represent the introduced quantities in the global equilibrium equations. 

\subsubsection{Model definitions in the discrete setting}


We start by dividing the time interval $ [0,T] $ into $n_{load} $ partitions and split the domain $ \Omega $ into $ N_e $ finite elements:
\begin{equation}
	[0,T] = \bigcup^{n_{load}-1}_{n=0} [t_n,t_{n+1}]\;\;\text{and}\;\;\Omega = \bigcup^{N_e}_{j=1} \Omega_{j}.
	\label{eq:ch7_eq19}
\end{equation}
Here, $\Omega_{j} $ is the domain of element $ j $, and each element is equipped with $ N_{\textit{gp}} $ Gauss quadrature points. 
We focus on a typical (quasi-)time interval $ [t_n,t_{n+1}] $ with known equilibrated state at time step $ t_n $. In the context of this work, we choose standard nodal finite elements with Lagrange basis functions that can be assembled to approximate the macroscale displacements and strains at load increment ${(n +1)}$ over the complete domain:
\begin{equation}
	\boldsymbol{u} \approx \boldsymbol{N} \boldsymbol{u}_{n+1} \;\;\; \text{and} \;\;\; \boldsymbol{E} \approx \boldsymbol{B} \boldsymbol{u}_{n+1}
	\label{eq:ch7_eq19b}
\end{equation}
where $\boldsymbol{N}$ is the (assembled) displacement interpolation operator and $\boldsymbol{B}$ is the (assembled) strain-displacement operator \cite{hughes2012finite}. In the sense of the standard Galerkin method, we use the same finite element basis functions for the representation of the solution and the test functions \cite{hughes2012finite}.\\

\noindent\textbf{Remark 2:} We emphasize that from here on, displacement-type vector quantities with subscript $n$ or $({n+1})$ such as $\boldsymbol{u}_{n+1}$ denote the vector of unknown displacement-type coefficients at the corresponding load increment. Tensor quantities with subscript $n$ or $(n+1)$ such as macroscale strains $\boldsymbol{E}_{n+1}$ or stresses $\boldsymbol{\Sigma}_{n+1}$ denote the corresponding tensor fields approximated in terms of the corresponding displacement finite element solution  at the corresponding load increment.\\


The design vector $ [\rho(\boldsymbol{x}), \boldsymbol{m}(\boldsymbol{x},t)]^{T}$ for our example multiscale configuration in Fig.~\ref{Figs:ch7_fig1} can now be defined in this discrete setting as $ [\boldsymbol{\rho}, \boldsymbol{m}]^{T}$, where
\begin{equation}
	\begin{split}
		\boldsymbol{\rho} = & \;[\rho_{1}, \rho_{2},\rho_{3}, ..., \rho_{N_e}]^{},\\
		\boldsymbol{m} = & \;[\boldsymbol{m}_{0},\boldsymbol{m}_{1}, ...,\boldsymbol{m}_{n+1},...,\boldsymbol{m}_{n_{load}-1} ], \\
		\boldsymbol{m}_{n+1} = & \; [ (m^{1,1}_{n+1},..,m^{N_{\textit{gp}},1}_{n+1}), ...,(..,m^{x,j}_{n+1}, ..), ..., (m^{1,N_e}_{n+1},..,m^{N_{\textit{gp}},N_e}_{n+1})]^{}, \\
		m^{x,j}_{n+1} = & \; [\phi^{x,j}_{A,n+1}, \theta^{x,j}_{A,n+1},  \zeta^{x,j}_{A,n+1}, \gamma^{x,j}_{C,n+1} ]_{}, \\
		\text{where}\;\;  x & \in \{1,..,N_{\textit{gp}}\},\; j \in \{1,..,N_e\}, \; n \in \{0,..,n_{load}-1\}.
	\end{split}
	\label{eq:ch7_eq20}
\end{equation}
The macroscale density $ \rho_{j} $ is assumed to be constant in each element and load increment, with $j$ being the element index. The microstructure design variable set $ \boldsymbol{m} $ is defined at each (macroscale) Gauss point and load increment with $\boldsymbol{m}_{n+1} $ as the microstructure characterization set at load increment $(n+1)$. The microstructure configuration $ m^{x,j}_{n+1} $ at a Gauss point $\boldsymbol{x}$ inside element $ j $ at load increment $(n+1)$ consists of volume fraction $\phi^{x,j}_{A,n+1}$, orientation $ \theta^{x,j}_{A,n+1} $ elongation $\zeta^{x,j}_{A,n+1} $ for Material A, and volume fraction $ \gamma^{x,j}_{C,n+1} $ of Material C. 
We again emphasize that we use this definition of the multiscale configuration in Fig.~\ref{Figs:ch7_fig1} for illustration purposes, this procedure can easily be generalized to cover any other multiphase hierarchical system. 

\begin{figure*}[b!]
	\centering
	\includegraphics[width=0.45\textwidth]{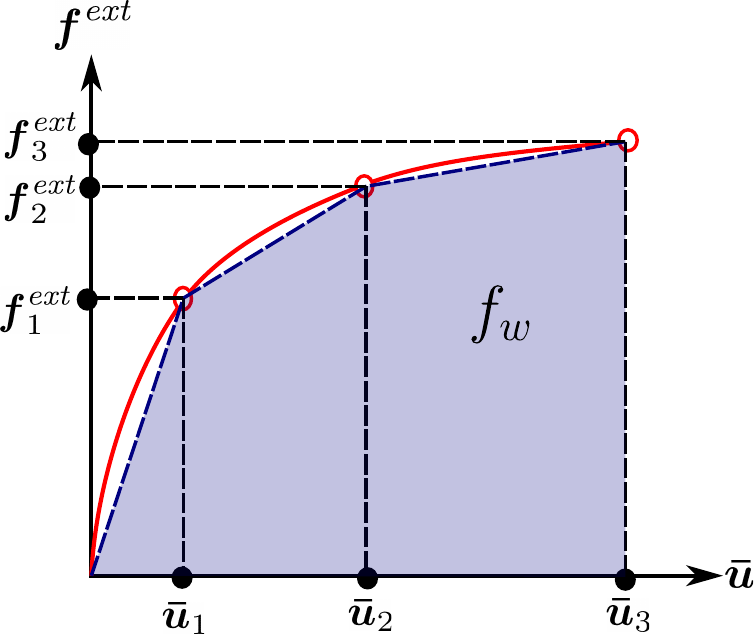}	
	\caption{Total mechanical work $f_w $ in the course of the deformation process.}
	\label{Figs:ch6_fig2}
\end{figure*}

The constitutive equation relating the macroscale stress $ \boldsymbol{\Sigma}_{n+1} $ with the macroscale strains $ \boldsymbol{E}_{n+1} $ and $ \boldsymbol{E}^{p}_{n+1} $ at the Gauss point $\boldsymbol{x}$ 
follows from \eqref{eq:ch7_eq13d} as
\begin{equation}
	\begin{split}	
		\boldsymbol{\Sigma}_{n+1} =  \mathbb{C}({m}^{x,j}_{n+1}):(\boldsymbol{E}^{}_{n+1} - \boldsymbol{E}^{p}_{n+1}),\;\; 
	\end{split}
	\label{eq:ch7_eq22}
\end{equation}
where the homogenized stiffness $\mathbb{C}(m^{x,j}_{n+1}) $ is evaluated for the microstructure configuration $ m^{x,j}_{n+1} $. To derive the incremental form of the elastoplastic constitutive equations, the discrete version of the maximum plastic dissipation principle at the Gauss point $\boldsymbol{x}$ from \eqref{eq:ch7_eq13b} is
	\begin{equation}
		 \mathcal{D}^{p}[m^{x,j}_{n+1}, \boldsymbol{\Sigma}_{n+1}; \boldsymbol{E}^{p}_{n+1}] = \max_{ \boldsymbol{\tau} \in \mathbb{E}_{\Sigma_{n+1}}} \Big\{ \boldsymbol{\tau}_{}^{}: (\boldsymbol{E}^{p}_{n+1} - \boldsymbol{E}^{p}_{n} )\Big\}, \;\; 
	\end{equation}
where the admissible stresses $\boldsymbol{\tau} $ lie in a set $\mathbb{E}_{\Sigma_{n+1}}$ defined by the  homogenized yield criterion $\mathfrak{F}(\boldsymbol{\tau},m^{x,j}_{n+1}) $ evaluated at $ m^{x,j}_{n+1} $:
	\begin{equation}
		\mathbb{E}_{\Sigma_{n+1}}:= \Big\{  \boldsymbol{\tau} \in \mathbb{S}\;|\; m^{x,j}_{n+1}\in E_{\textit{ad}}(\rho_j),\; \mathfrak{F}(\boldsymbol{\tau},m^{x,j}_{n+1}) \leq 0\Big\}. 
	\label{eq:ch7_eq23}
	\end{equation}
The macroscale plastic strain $\boldsymbol{E}^{p}_{n}$ is known from the equilibrated solution state at load step $n$. We note that we write these constitutive relations in tensor notation given its direct relation with continuum micromechanics principles stated in Section~\ref{sec:sec22}. 

\subsubsection{Discrete form of the structure optimization problem}

With the introduced definitions, we can write the discrete form of the material and structure optimization formulation (\ref{eq:ch7_eq31}) and (\ref{eq:ch7_eq32}). In this work, we employ the trapezoidal rule for the numerical evaluation of the integrals over the (quasi-)time domain, which is second-order accurate with respect to the (quasi-)time step size. 
The discrete version of the structure optimization problem \eqref{eq:ch7_eq31} then becomes
\begin{equation}
	\begin{split}
		\max_{\boldsymbol{\rho}}: & \;\; f_w (\boldsymbol{\rho}) = \frac{1}{2} \sum_{n=0}^{n_{load}-1} (\boldsymbol{f}_{n+1}^{\textit{ext}} + \boldsymbol{f}_{n}^{\textit{ext}})^{T} \;\Delta \boldsymbol{\bar{u}}_{n+1}\\ 
		\text{s.t.}: & \;\; \boldsymbol{\bar{r}}_{n+1} (\boldsymbol{\rho},\boldsymbol{\bar{u}}_{n+1},\boldsymbol{\bar{m}}_{n+1})  = 0\;\; \forall n = 0,1,...,n_{load}-1 \\
		\;\; & \;\; M(\boldsymbol{\rho})  = \sum_{j=1}^{N_e} \rho_{j} |\Omega_{j}| =   M_{\textit{req}} = M_{\textit{frac}} \times \rho_C \times|\Omega|;  \;\; \forall n = 1,2,...,n_{load}-1 \\
		& \;\; \rho_{j} \in [\rho_{\textit{min}},\rho_{\textit{max}}] , \; \forall j = 1,2,...,N_e. \\
	\end{split}
	\label{eq:ch7_eq26}
\end{equation}
In accordance with the notation introduced above in the context of a finite element discretization, $ \boldsymbol{f}_{n+1}^{\textit{ext}} $ is the external force vector, $ \boldsymbol{\bar{u}}_{n+1} $ is the converged vector of the macroscale nodal displacements, and $\Delta \boldsymbol{\bar{u}}_{n+1} := \boldsymbol{\bar{u}}_{n+1} - \boldsymbol{\bar{u}}_{n}$ is the increment of the displacement vector in load increment $(n+1)$. The force residual $\boldsymbol{\bar{r}}_{n+1} $ is calculated utilizing the optimal microstructure configuration $ \boldsymbol{\bar{m}}_{n+1}$ in load increment $(n+1)$. We note that the optimal microstructure configuration $ \boldsymbol{\bar{m}}_{n+1}$ and $ \boldsymbol{\bar{u}}_{n+1}$ are dependent on each other justifying the choice of $ (\bar{\square}) $ notations introduced in Section~\ref{sec:ch7_71}. $ M(\boldsymbol{\rho})  $ is the total mass of the occupying domain, and 
$\rho_{j} $ and $|\Omega_{j}| $ are the density and volume of element $j$. 

The total mechanical work $f_w $ is the discrete version of the second statement in \eqref{eq:ch7_eq31}, which is equivalent to the objective function proposed in \cite{fritzen2016topology}. This essentially is the area under the characteristic force-displacement curve approximated with the trapezoidal rule, as illustrated in Fig.~\ref{Figs:ch6_fig2}. 
The first condition in \eqref{eq:ch7_eq26} ensures that the global equilibrium is satisfied in all load steps. 
The second and third conditions of \eqref{eq:ch7_eq26} are the discrete definitions of the macroscale admissible design variable set $ \mathcal{A}_{\textit{ad}}$. The total available mass $M_{req} $ can be expressed in terms of the fraction $ M_{\textit{frac}} $ with respect to the mass when the densest material occupies the complete domain.

The force residual $\boldsymbol{\bar{r}}_{n+1}$ 
at load increment $(n+1)$ is defined as 
\begin{equation}
	\begin{split}
		\boldsymbol{\bar{r}}_{n+1} (\boldsymbol{\rho},\boldsymbol{\bar{u}}_{n+1},\boldsymbol{\bar{m}}_{n+1}):= \boldsymbol{f}^{\textit{ext}}_{n+1} - \boldsymbol{f}^{\textit{int}}_{n+1} = \boldsymbol{f}^{\textit{ext}}_{n+1} - \sum_{j=1}^{N_e}  \Bigg[\sum_{x=1}^{N_{\textit{gp}}} \boldsymbol{B}^{T} \boldsymbol{\Sigma}_{n+1} \; w_x \Bigg], \\ 
	\end{split}
	\label{eq:ch7_eq28}
\end{equation}
where 
$ w_x $ contains the Gauss point weight and the determinant of the Jacobian matrix for element $ j $. We observe that the microstructure design variable set $\boldsymbol{m} $ is implicitly accounted for by the residual definitions in each load increment. The macroscale stress $ \boldsymbol{\Sigma}_{n+1}$ at each Gauss point is evaluated by solving the nonlinear elastoplastic constitutive relations \eqref{eq:ch7_eq22} and \eqref{eq:ch7_eq23} with known microstructure configuration $ \bar{m}^{x,j}_{n+1} $ that solves the material optimization problem detailed in the following subsection. Therefore, the global equilibrium equation \eqref{eq:ch7_eq28} is nonlinear and requires iterative solution approaches such as the Newton-Raphson incremental procedure \cite{simo2006computational,de2011computational}. 

Structure optimization involving inelastic material models is potentially ill-posed in a force-controlled setting \cite{swan1997voigtnon,huang2008optimal,schwarz2001topology,maute1998adaptive,cho2003design}. Therefore, we only consider displacement-controlled loading in this article, incrementally applied through prescribed displacements $ \boldsymbol{\bar{u}}^{E} (t) $, see Section~\ref{sec:sec221}. In a displacement-controlled setting, $ \boldsymbol{f}_{n+1}^{\textit{ext}} $ represents the discretized form of the loading potential resulting from the non-zero displacement boundary conditions. This assumption also simplifies the sensitivity calculations of the objective function with respect to the design variables for the optimization algorithms, which we will discuss in Section~\ref{sec:sec51}. 


\subsubsection{Discrete form of the material optimization problem}

For a given material distribution $\boldsymbol{\rho} $ and the macroscale displacement solution vector $\boldsymbol{\bar{u}}_{n+1} $, 
the material optimization problem for the Gauss point $\boldsymbol{x} $ inside element $j$ for load increment $(n+1)$ follows from (\ref{eq:ch7_eq32}) as
\begin{equation}
	\bar{m}^{x,j}_{n+1} = \argmax_{	m^{x,j}_{n+1}\in E_{\textit{ad}}(\rho_j)} 
	\Big\{  \max_{ \boldsymbol{\tau} \in \mathbb{E}_{\Sigma_{n+1}}} \boldsymbol{\tau}_{}^{}: (\boldsymbol{E}^{p}_{n+1} - \boldsymbol{E}^{p}_{n}) + \Psi(\boldsymbol{E}_{n+1} - \boldsymbol{E}^{p}_{n+1})
	- \Psi(   \boldsymbol{E}_{n} - \boldsymbol{E}^{p}_{n}) \Big\}. 
	\label{eq:ch7_eq33a}
\end{equation}
The first part of this equation directly comes from the incremental form of the principle of maximum plastic dissipation outlined in \eqref{eq:ch7_eq23}. Similarly, the second part is the incremental form of the Helmholtz free energy rate defined in (\ref{eq:ch7_eq32}). We emphasize that all quantities at load increment $n$ are known, and, therefore, $\Psi( \boldsymbol{E}_{n} - \boldsymbol{E}^{p}_{n}) $ does not play any role in this maximization problem. The optimized configuration $\bar{m}^{x,j}_{n+1} $ is sought in the microscale design variable space $m^{x,j}_{n+1} = [\phi^{x,j}_{A,n+1}, \theta^{x,j}_{A,n+1}, \zeta^{x,j}_{A,n+1}, \gamma^{x,j}_{C,n+1}] $ with constraints definitions that follow from the admissible set $E_{\textit{ad}}$ \eqref{eq:ch7_eq18}. 
From \eqref{eq:ch7_eq33a}, we can rewrite the material optimization statement as
\begin{equation}
	\begin{split}
		\bar{m}^{x,j}_{n+1} = \argmax_{	m^{x,j}_{n+1}(\rho_{j})} &  
		\Big\{   \boldsymbol{\Sigma}_{n+1}^{}: (\boldsymbol{E}^{p}_{n+1} - \boldsymbol{E}^{p}_{n}   )  +    \Psi(\boldsymbol{E}_{n+1} - \boldsymbol{E}^{p}_{n+1})
		- \Psi(   \boldsymbol{E}_{n} - \boldsymbol{E}^{p}_{n})\Big\}\\
		\text{s.t.}: & \;\; \boldsymbol{\Sigma}_{n+1} =  \mathbb{C}(m^{x,j}_{n+1}): (\boldsymbol{E}^{}_{n+1} - \boldsymbol{E}^{p}_{n+1})  \\		
		& \;\;\mathfrak{F}(\boldsymbol{\Sigma}_{n+1},m^{x,j}_{n+1}) \leq 0 \\
		& \;\; \Psi (\boldsymbol{E}_{n+1}   - \boldsymbol{E}^{p}_{n+1} )  = \frac{1}{2}(\boldsymbol{E}_{n+1}   - \boldsymbol{E}^{p}_{n+1} ):\mathbb{C}(	m^{x,j}_{n+1}): (\boldsymbol{E}^{}_{n+1} - \boldsymbol{E}^{p}_{n+1}) \\
		& \;\; \rho_j = \rho_A  \phi^{x,j}_{A,n+1} + \rho_M  (1 - \phi^{x,j}_{A,n+1});\; \rho_M = \rho_B  (1-  \gamma^{x,j}_{C,n+1})  + \rho_C  \gamma^{x,j}_{C,n+1} \\
		& \;\; \phi^{x,j}_{A,n+1} \in [\phi^{\textit{min}}_A,\phi^{\textit{max}}_A]; \;\; \theta^{x,j}_{A,n+1} \in [-\pi/2,\pi/2]; \; \zeta^{x,j}_{A,n+1} \in [1,\zeta^{\textit{max}}]\\
		& \;\; \gamma^{x,j}_{C,n+1} \in [\gamma^{\textit{min}}_C,\gamma^{\textit{max}}_C] \;, 
	\end{split}
	\label{eq:ch7_eq27}
\end{equation}
including all constraints defined through the stress admissible set $\mathbb{E}_{\Sigma_{n+1}}$ and microscale design admissible set $ E_{\textit{ad}} $. The first two conditions are essentially the elastoplastic constitutive equations relating the macroscale stress with the macroscale strains via \eqref{eq:ch7_eq22} and the constraint on the macroscale stress defined through the homogenized yield criterion \eqref{eq:ch7_eq23}. The third condition is the definition of the Helmholtz free energy in terms of microscale design variable $m^{x,j}_{n+1} $. The rest of the conditions follow in a straightforward manner from the constraints definitions in $E_{ad} $. The solution of \eqref{eq:ch7_eq27} at each Gauss point in each load increment yields the optimized microstructure configuration set $ \boldsymbol{\bar{m}} $.



We emphasize that in contrast to the equivalent strain energy maximization for concurrent optimization problems involving overall linear elastic multiphase hierarchical systems \cite{xia2014concurrent,gangwar2021concurrent}, finding the solution to the material optimization problem (\ref{eq:ch7_eq27}) is not straightforward. Both the macroscale plastic strain $\boldsymbol{E}^{p}_{n+1} $ and the optimized microstructure $\bar{m}^{x,j}_{n+1}$ are unknown. 
Intuitively, the material optimization problem maximizes the area under the homogenized elastoplastic stress-strain curve for each material point. Multiple stress-strain curves are available at each load increment, defined by the the microscale design variable $m^{x,j}_{n+1}$.  This inter-dependency couples the history variable $\boldsymbol{E}^{p}_{n+1} $ with  $m^{x,j}_{n+1}$, necessitating a challenging novel algorithmic treatment to tackle this maximization problem.

\section{Algorithmic treatment of the material optimization problem }
\label{sec:sec_4}


For the algorithmic treatment, we interpret the material optimization problem as a reformulated constitutive law at each material point that provides a locally optimal mechanical response to the loading history. This interpretation allows us to treat the microscale design variable $m^{x,j}_{n+1}$ as an additional internal state variable within the context of the classical formulation of plasticity. With this interpretation, we first exploit the principle of maximum plastic dissipation to motivate a solution strategy for the material optimization problem. We then cast this strategy into an algorithmic procedure that assumes the format of a typical return map algorithm for the integration of elastoplastic constitutive equations. Finally, we leverage continuum micromechanics and the associated homogenized elastoplastic constitutive relations, which enable further simplifications that make our framework computationally feasible. 

\subsection{The principle of maximum plastic dissipation revisited}
\label{sec:sec_41}

As explained above, the first part of the material optimization problem (\ref{eq:ch7_eq33a}) 
is the incremental statement of the maximum plastic dissipation principle. This part defines the interaction between the next stress state $\boldsymbol{\Sigma}_{n+1} $ and the optimized microstructure state $\bar{m}^{x,j}_{n+1} $ through the  homogenized yield criterion $\mathfrak{F}(\boldsymbol{\tau},m^{x,j}_{n+1}) $. Focusing on this part only, we can combine both statements in a single one as 
\begin{subequations}
	\begin{equation}
		\{\boldsymbol{\Sigma}_{n+1},\hat{m}^{x,j}_{n+1}\} = \argmax_{	( \boldsymbol{\tau}, m^{x,j}_{n+1})\;\in\; \mathbb{E}_{\Sigma_{n+1}}} 
		\Big\{   \boldsymbol{\tau}_{}^{}: (\boldsymbol{E}^{p}_{n+1} - \boldsymbol{E}^{p}_{n}   ) \Big\}, \;\; \text{where}
		\label{eq:ch7_eq34a}
	\end{equation}
	\begin{equation}
		\mathbb{E}_{\Sigma_{n+1}}:= \Big\{  \boldsymbol{\tau} \in \mathbb{S},\;\; m^{x,j}_{n+1}\in E_{\textit{ad}}(\rho_j)\;|\; \mathfrak{F}(\boldsymbol{\tau},m^{x,j}_{n+1}) \leq 0\Big\}. 
		\label{eq:ch7_eq34b}
	\end{equation}
\end{subequations}
This maximization problem seeks the macroscale stress state $\boldsymbol{\Sigma}_{n+1} $ and a solution $\hat{m}^{x,j}_{n+1} $ within the modified admissible space definition $\mathbb{E}_{\Sigma_{n+1}} $. The solution $\hat{m}^{x,j}_{n+1}$ restricts the search space for the solution $\bar{m}^{x,j}_{n+1} $ of the original material optimization problem (\ref{eq:ch7_eq27}), which we will further detail in the subsequent discussion. We note that the interpretation of the microscale design variable $m^{x,j}_{n+1}$ as internal state variable naturally arises from these statements. 


We define a Lagrangian functional that converts the constraint optimization problem \eqref{eq:ch7_eq34a} into an unconstrained problem following (\ref{eq:ch7_eq4}):\begin{equation}
	\mathcal{L}_{n+1}( \boldsymbol{\tau}, m^{x,j}_{n+1},\delta; \boldsymbol{E}^{p}_{n+1}) := - \boldsymbol{\tau}_{}^{}: (\boldsymbol{E}^{p}_{n+1} - \boldsymbol{E}^{p}_{n}) + \delta\;\mathfrak{F}(\boldsymbol{\tau},m^{x,j}_{n+1}),
	\label{eq:ch7_eq35} 
\end{equation}
where $\delta $ is in the cone of Lagrange multipliers defined through \eqref{eq:ch7_eq3cone}. The solution to (\ref{eq:ch7_eq34a}) is given by a point $ (\boldsymbol{\Sigma}_{n+1},\hat{m}^{x,j}_{n+1},\Delta \gamma_{n+1}) $ that satisfies the Karush-Kuhn-Tucker optimality conditions for (\ref{eq:ch7_eq35}). The conditions entail
\begin{equation}
	\begin{split}
		& \left.\frac{\partial \; \mathcal{L}_{n+1} }{\partial \; \boldsymbol{\tau}} \right\vert_{\boldsymbol{\Sigma}_{n+1},\hat{m}^{x,j}_{n+1},\Delta \gamma_{n+1}} = 	- (\boldsymbol{E}^{p}_{n+1} - \boldsymbol{E}^{p}_{n}) + \Delta \gamma_{n+1} \left.\frac{\partial \mathfrak{F}^{}_{}(\boldsymbol{\tau},m^{x,j}_{n+1})}{\partial \boldsymbol{\tau}}\right\vert_{\boldsymbol{\Sigma}_{n+1},\hat{m}^{x,j}_{n+1}} = 0,\\ 
		&\left.\frac{\partial \; \mathcal{L}_{n+1} }{\partial \; m^{x,j}_{n+1}} \right\vert_{\boldsymbol{\Sigma}_{n+1},\hat{m}^{x,j}_{n+1},\Delta \gamma_{n+1}} = 	 \Delta \gamma_{n+1} \left.\frac{\partial \mathfrak{F}^{}_{}(\boldsymbol{\tau},m^{x,j}_{n+1})}{\partial m^{x,j}_{n+1}}\right\vert_{\boldsymbol{\Sigma}_{n+1},\hat{m}^{x,j}_{n+1}} = 0,\\
		& \Delta \gamma_{n+1} \geq 0, \;\; \mathfrak{F}(\boldsymbol{\Sigma}_{n+1},\hat{m}^{x,j}_{n+1}) \leq 0, \;\; \text{and}\;\; \Delta \gamma_{n+1}\; \mathfrak{F}(\boldsymbol{\Sigma}_{n+1},\hat{m}^{x,j}_{n+1}) = 0.
	\end{split}	
	\label{eq:ch7_eq36} 
\end{equation}
The general structure of \eqref{eq:ch7_eq36} is similar to the typical local constitutive equations for plasticity (flow rule, loading/unloading conditions) as described in \eqref{eq:ch7_eq5}. Equation $\eqref{eq:ch7_eq36}_{2} $ represents the evolution of microstructure state in a particular load increment $(n+1)$. All these equations together form a coupled nonlinear system that requires a special computational treatment. 
The solution $\hat{m}^{x,j}_{n+1}$ from \eqref{eq:ch7_eq36} provides important insights into the interaction of the plastic update and the microstructure update. 
In the following, we will can utilize these insights to design an algorithmic framework for solving the original material optimization problem \eqref{eq:ch7_eq33a}. 


\subsection{Algorithmic procedure in the form of return map algorithms}
\label{sec:sec_42}

Analogous to the \textit{elastic-plastic operator split} formulas for inelastic constitutive equations, we define a \textit{trial elastic state} by \textit{freezing} the plastic flow and microstructure evolution state during the current load increment. It implies that the macroscale plastic strain and optimal microstructure configuration state in the current load increment are known and equal to that of the previous load increment, that is $\boldsymbol{E}_{n+1}^{p} = \boldsymbol{E}_{n}^{p},\;\hat{m}^{x,j}_{n+1} = \bar{m}^{x,j}_{n}  $. With these assumptions, the trial elastic state is
\begin{equation}
	\begin{split}
		& \boldsymbol{E}_{n+1}^{p} = \boldsymbol{E}_{n}^{p} \implies \boldsymbol{E}^{e,tr}_{n+1} := \boldsymbol{E}^{}_{n+1} - \boldsymbol{E}^{p}_{n} \\
		& \boldsymbol{\Sigma}^{tr}_{n+1}:= \mathbb{C}(\bar{m}^{x,j}_{n}):(\boldsymbol{E}^{}_{n+1} - \boldsymbol{E}^{p}_{n})\\
		& \mathfrak{F}^{tr}_{n+1}:= \mathfrak{F}(\boldsymbol{\Sigma}^{tr}_{n+1},\bar{m}^{x,j}_{n}).
	\end{split}
	\label{eq:ch7_eq41} 
\end{equation}
Here, $\boldsymbol{E}^{e,tr}_{n+1} $, $\boldsymbol{\Sigma}^{tr}_{n+1}$ and $\mathfrak{F}^{tr}_{n+1}$ denote the macroscale trial elastic strain, trial elastic stress, and trial yield criterion, respectively. Further, the convexity of the homogenized yield criterion $\mathfrak{F} $ leads to the following important property \cite{simo2006computational}:
\begin{equation}
	\mathfrak{F}^{tr}_{n+1} \geq \mathfrak{F}^{}_{n+1}, \;\; \text{and}\;\; \mathfrak{F}^{}_{n+1} := \mathfrak{F}(\boldsymbol{\Sigma}^{}_{n+1},\bar{m}^{x,j}_{n+1}),
	\label{eq:ch7_eq42}
\end{equation}
where $\mathfrak{F}^{}_{n+1} $ is the homogenized yield criterion calculated at the macroscale stress $\boldsymbol{\Sigma}^{}_{n+1}$ and the optimal microstructure configuration $\bar{m}^{x,j}_{n+1} $ after load increment $(n+1)$. 

\subsubsection{Elastic trial state and plastic vs.\ microstructure updates}

The trial state \eqref{eq:ch7_eq41} and property \eqref{eq:ch7_eq42} in combination with \eqref{eq:ch7_eq36} lead to three important cases defining the restrictions posed by the solution $\hat{m}^{x,j}_{n+1}$ on the search space for the original material optimization problem (\ref{eq:ch7_eq27}). In the following, we will discuss all three cases in detail.\\

\noindent \textbf{Case 1:} If $\mathfrak{F}^{tr}_{n+1} < 0 $, then property \eqref{eq:ch7_eq42} entails $\mathfrak{F}^{}_{n+1} < 0 $, implying a \textbf{purely elastic step}. In this case, the solution $\{{\Sigma}_{n+1},\bar{m}^{x,j}_{n+1} \} $ to the original material optimization problem \eqref{eq:ch7_eq33a} automatically satisfies \eqref{eq:ch7_eq36}. To see this, one can put $\hat{m}^{x,j}_{n+1} = \bar{m}^{x,j}_{n+1}$, implying $\mathfrak{F}(\boldsymbol{\Sigma}_{n+1},\hat{m}^{x,j}_{n+1}) =  \mathfrak{F}^{}_{n+1}$, then the discrete KKT condition $ \Delta \gamma_{n+1}\; \mathfrak{F}(\boldsymbol{\Sigma}_{n+1},\hat{m}^{x,j}_{n+1}) = \Delta \gamma_{n+1}\;\mathfrak{F}^{}_{n+1} = 0 $ in \eqref{eq:ch7_eq36} results in $\Delta \gamma_{n+1} = 0 $. This means that $\eqref{eq:ch7_eq36}_{1} $ implies $\boldsymbol{E}_{n+1}^{p} = \boldsymbol{E}_{n}^{p} $, and $\eqref{eq:ch7_eq36}_{2} $ is automatically satisfied with no restrictions on the solution space of the microstructure configuration ${m}^{x,j}_{n+1}$. 
Therefore, with zero dissipation, the solution $\bar{m}^{x,j}_{n+1} $ of the material optimization problem reduces to the strain energy maximization that follows from \eqref{eq:ch7_eq27} as
\begin{equation}
	\bar{m}^{x,j}_{n+1} = \argmax_{m^{x,j}_{n+1}\in E_{\textit{ad}}(\rho_j)}
	\frac{1}{2}(\boldsymbol{E}_{n+1}   - \boldsymbol{E}^{p}_{n+1} ):\mathbb{C}(	m^{x,j}_{n+1}): (\boldsymbol{E}^{}_{n+1} - \boldsymbol{E}^{p}_{n+1}).
	\label{eq:ch7_eq43}
\end{equation}
In conclusion, the solution $\hat{m}^{x,j}_{n+1} $ does not pose any restrictions to the search space $E_{\textit{ad}}$ for the optimized microstructure configuration  $\bar{m}^{x,j}_{n+1} $  in this case.\\


\begin{figure*}[b!]
	\centering
	\subfloat[Case 2: $\Delta \gamma_{n+1} = 0\; \text{update}\; m^{x,j}_{n} \implies m^{x,j}_{n+1}  $ ]{\includegraphics[width=0.42\textwidth]{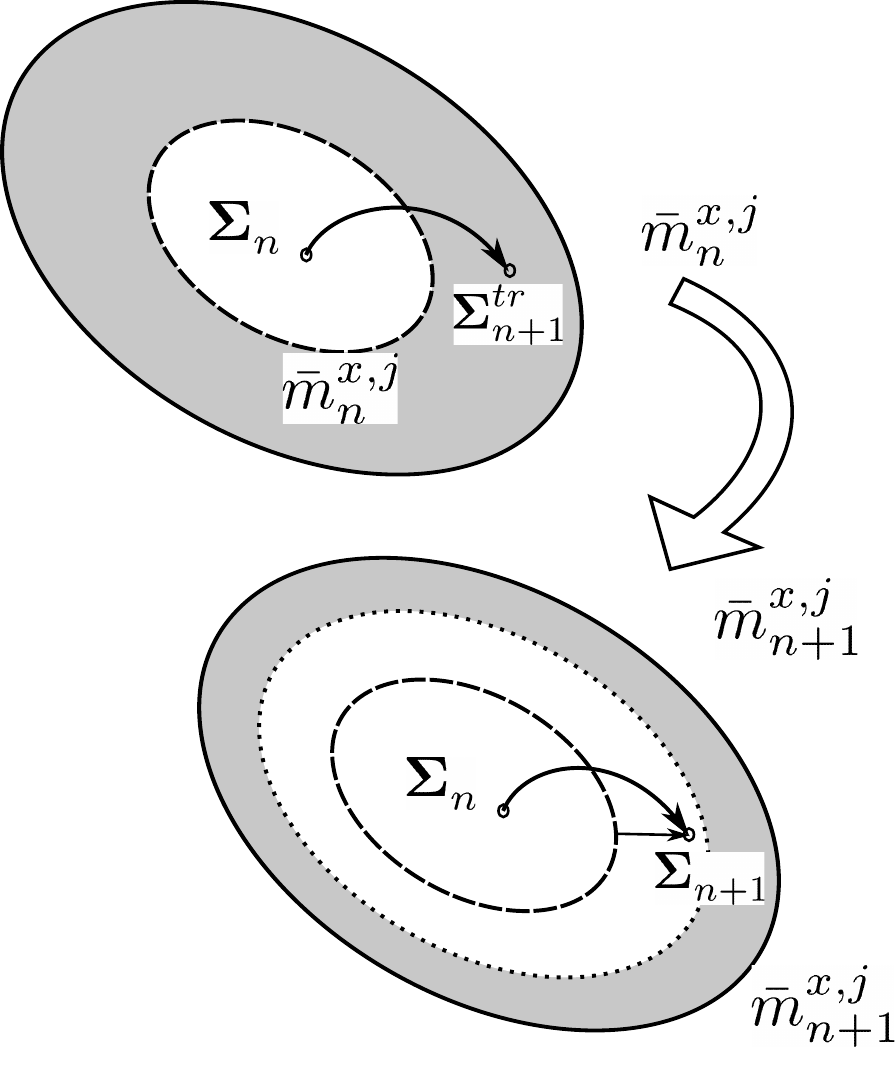}\label{Figs:ch7_fig2a}}
	\hfil
	\subfloat[Case 3: $\Delta \gamma_{n+1} > 0\; \implies \partial \mathfrak{F}^{}_{}/ \partial m^{x,j}_{n+1}$]{\includegraphics[width=0.42\textwidth]{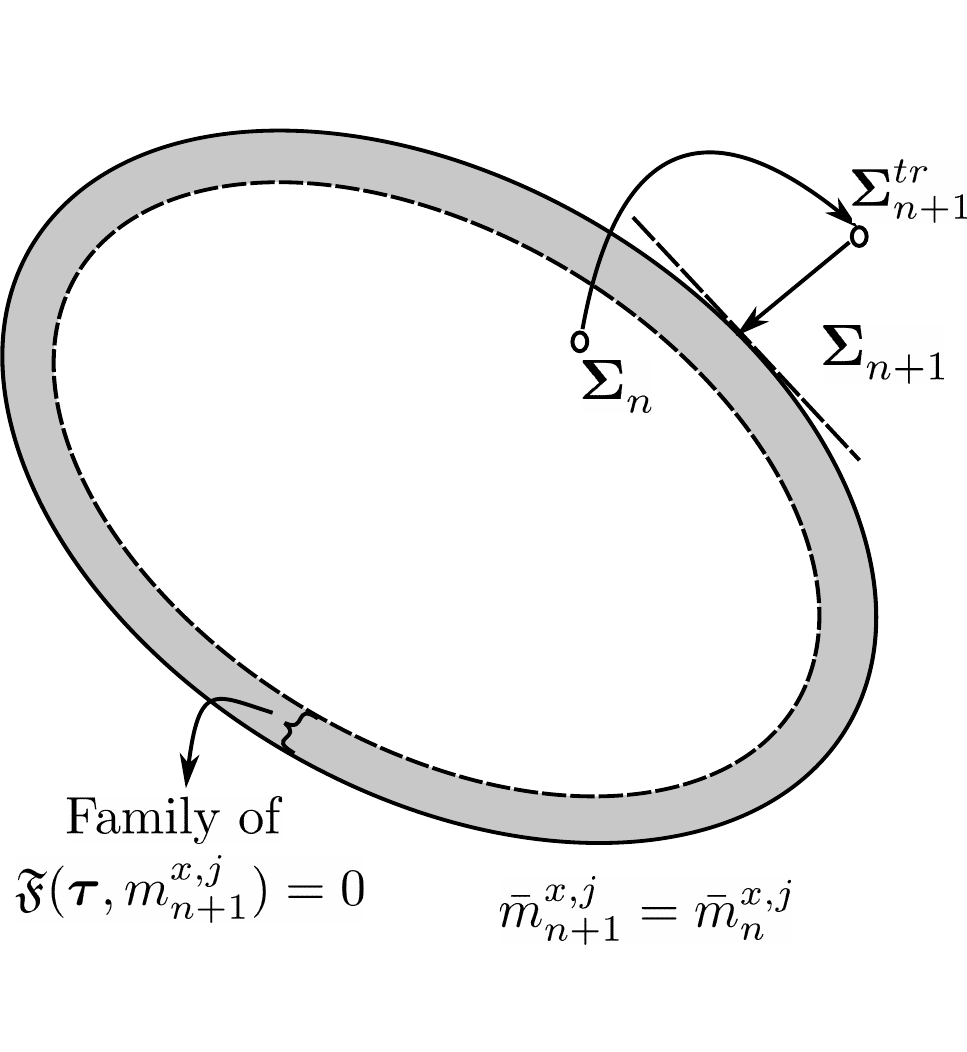}\label{Figs:ch7_fig2b}}	
	\caption{Geometric illustration of solution strategy for the material optimization problem, based on the return-map algorithm interpretation.}
	\label{Figs:ch7_fig2}
\end{figure*}

To support our discussion of the remaining two cases, Figure~\ref{Figs:ch7_fig2} presents a geometric interpretation in a typical return-mapping context. 
We note that in Fig.~\ref{Figs:ch7_fig2}, the gray region represents the family of available homogenized yield criterion envelops $\mathfrak{F}(\boldsymbol{\tau},m^{x,j}_{n+1}) = 0$ at each load increment, defined by the set of microscale design variables $m^{x,j}_{n+1}$.\\

\noindent \textbf{Case 2:} If $\mathfrak{F}^{tr}_{n+1} > 0 $, and if it is possible to find at least one microstructure configuration $\hat{m}^{x,j}_{n+1}  \in E_{ad}(\rho_j)$ such that $\mathfrak{F}^{tr(2)}_{n+1}:=\mathfrak{F}(\boldsymbol{\Sigma}^{tr}_{n+1},\hat{m}^{x,j}_{n+1}) \leq 0 $, property \eqref{eq:ch7_eq42} indicates that $ \mathfrak{F}^{tr(2)}_{n+1} \geq \mathfrak{F}^{}_{n+1} \implies \mathfrak{F}^{}_{n+1} < 0 $. The solution is possible via the microstructure update provided that new microstructure state $\bar{m}^{x,j}_{n+1} $ follows the constraint $\mathfrak{F}(\boldsymbol{\Sigma}^{tr}_{n+1},\bar{m}^{x,j}_{n+1}) \leq \mathfrak{F}^{tr(2)}_{n+1}  \leq 0$. 

We call this case \textbf{adaption to elastic state through microstructure evolution}. We observe in Fig.~\ref{Figs:ch7_fig2}a that due to the current microstructure state $\bar{m}^{x,j}_{n}$ (dashed line), the material goes into plastic state; however, the material adapts itself to fall back to the elastic state by updating the microstructure state to $\bar{m}^{x,j}_{n+1}$ (dotted line), while maximizing the total strain energy. 

Again, equation \eqref{eq:ch7_eq36} leads to $\boldsymbol{E}_{n+1}^{p} = \boldsymbol{E}_{n}^{p}$ and $\Delta \gamma_{n+1} = 0 $ following the discussion in Case~1. The solution $\bar{m}^{x,j}_{n+1} $ follows as
\begin{equation}
	\bar{m}^{x,j}_{n+1} = \argmax_{\mathfrak{F} (\boldsymbol{E}^{e}_{n+1},m^{x,j}_{n+1}) \leq  \mathfrak{F}^{tr(2)}_{n+1}   }
	\frac{1}{2}(\boldsymbol{E}_{n+1}   - \boldsymbol{E}^{p}_{n+1} ):\mathbb{C}(	m^{x,j}_{n+1}): (\boldsymbol{E}^{}_{n+1} - \boldsymbol{E}^{p}_{n+1}).
	\label{eq:ch7_eq44}
\end{equation}
The constraint in this problem ensures that the state remains elastic 
and can be interpreted as a restriction on the search space for $\bar{m}^{x,j}_{n+1} $ posed by the feasible solutions of \eqref{eq:ch7_eq36}. We write $ \mathfrak{F}$ in the strain space to emphasize that the strain state is known and problem \eqref{eq:ch7_eq44} is a function of $m^{x,j}_{n+1} $ only.\\ 

\noindent \textbf{Case 3:} If $\mathfrak{F}^{tr}_{n+1} > 0 $, and the problem $\mathfrak{F}^{tr(2)}_{n+1}:=\mathfrak{F}(\boldsymbol{\Sigma}^{tr}_{n+1},\hat{m}^{x,j}_{n+1}) = 0 $ does not have any solution. It implies that no microstructure state can solve \eqref{eq:ch7_eq36} with the chosen trial elastic strain $\boldsymbol{E}^{e,tr}_{n+1} $. Therefore, only a \textbf{plastic update} is feasible, and $ \boldsymbol{E}^{p}_{n+1} \ne \boldsymbol{E}^{p}_{n} $. This condition leads to $\Delta \gamma_{n+1} >0 $ from $\eqref{eq:ch7_eq36}_{1} $, 
and the microstructure evolution condition from $\eqref{eq:ch7_eq36}_{2} $ entails $\partial \mathfrak{F}^{}_{}/ \partial m^{x,j}_{n+1} =0$. It means that the microstructure configuration remains unchanged, that is $ \bar{m}^{x,j}_{n+1} = \bar{m}^{x,j}_{n} $. It implies that the solution $\hat{m}^{x,j}_{n+1}$ of \eqref{eq:ch7_eq36} restricts the search space for $\bar{m}^{x,j}_{n+1}$ to a single point, that is $\bar{m}^{x,j}_{n} $. 
With $ \bar{m}^{x,j}_{n+1} = \bar{m}^{x,j}_{n} $, the rest of the relations in \eqref{eq:ch7_eq36} reduces to the typical elastoplastic constitutive equations with known stiffness and yield criterion. 

This is illustrated in Fig.~\ref{Figs:ch7_fig2}b, where we see that no elastic update is possible for the current trial state; therefore, the microstructure state remains unchanged, and the next stress state $\boldsymbol{\Sigma}^{}_{n+1}$ is solved with a standard return map algorithm such as the closest point projection algorithm \cite{simo1985consistent}.


\begin{figure*}[h!]
	\centering
	\subfloat[Case 1: Purely elastic update ]{\includegraphics[width=0.40\textwidth]{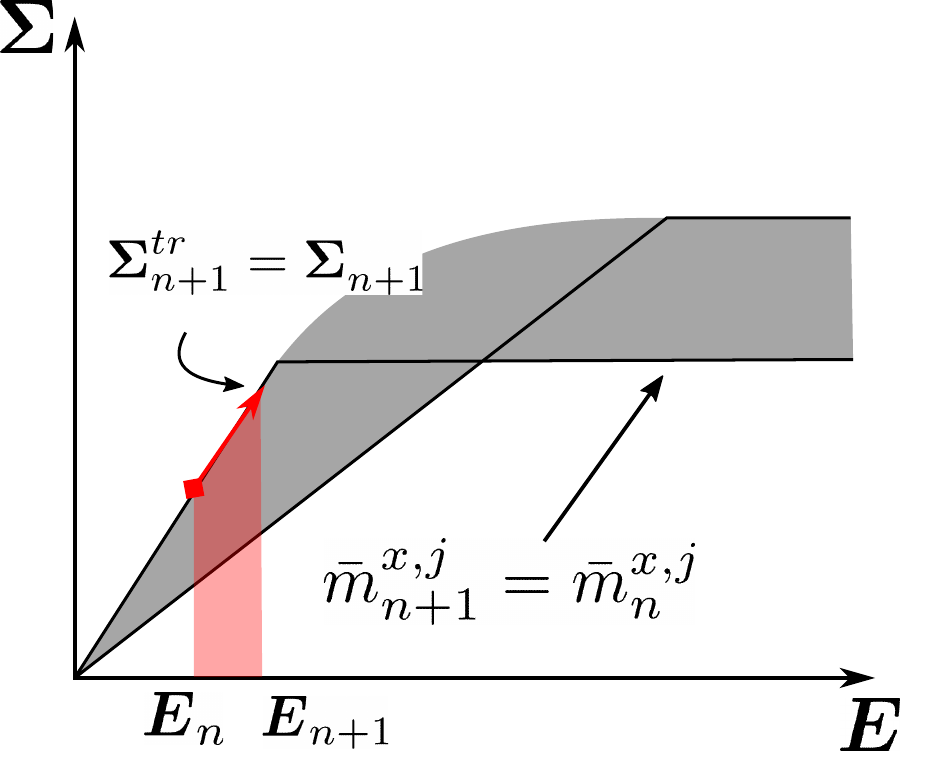}\label{Figs:ch7_fig3_a}}
	\hfil
	\subfloat[Case 2: Adaption to elastic state]{\includegraphics[width=0.40\textwidth]{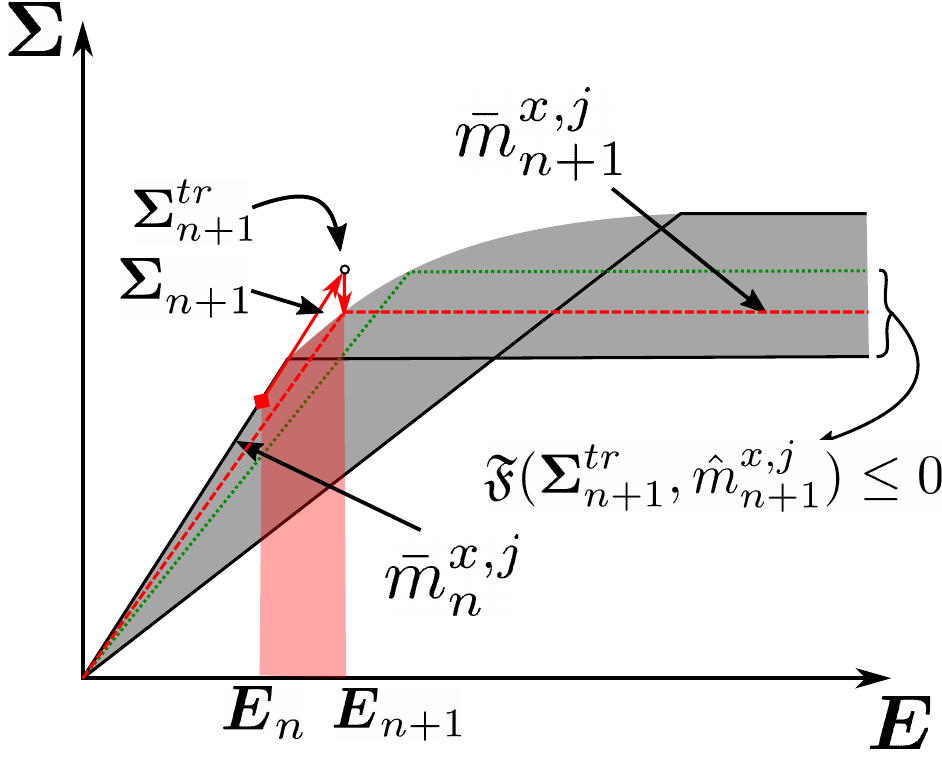}\label{Figs:ch7_fig3_b}}\\
	\subfloat[Case 3: Plastic update]{\includegraphics[width=0.54\textwidth]{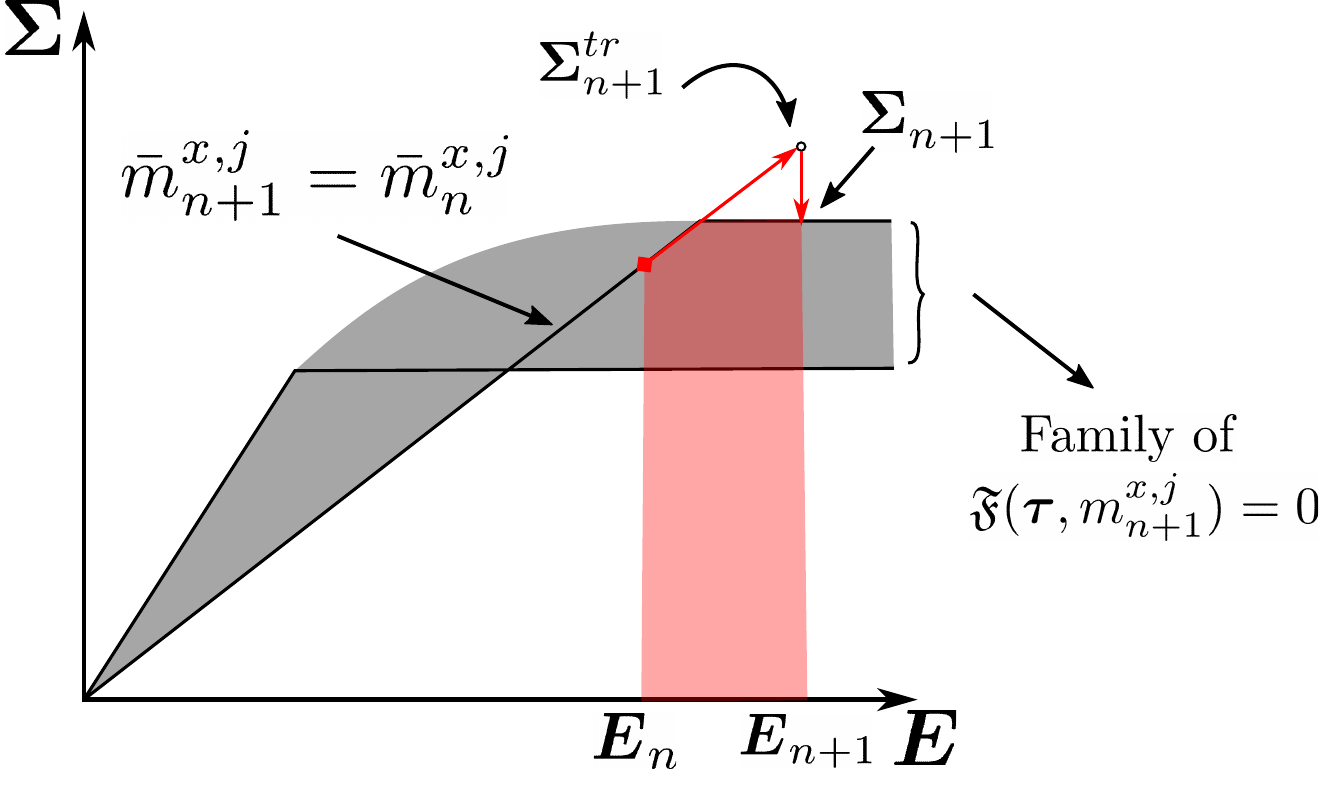}\label{Figs:ch7_fig3_c}}		
	\caption{Graphical solution of the material optimization problem in one dimension for the three possible cases.}
	\label{Figs:ch7_fig3}
\end{figure*}

\subsubsection{An analogue to the elastoplastic return map algorithm}

For an intuitive understanding, Figure~\ref{Figs:ch7_fig3} presents a graphical solution of the material optimization problem for the example of a one-dimensional linear elastic-perfectly plastic model. The gray region in these graphs represents the family of stress-strain curves for different microstructure design configurations $m^{x,j}_{n+1} $. The material state (stress, strain, microscale configuration) at load increment $n$ is known, and the macroscale strain $\boldsymbol{E}_{n+1} $ at load increment $(n+1)$ is given. Typical strain increments are infinitesimal, and increments are large for illustration only. The next material state from the material optimization problem warrants that the area increment (red shaded region in the graphs) is maximized. 

The trial elastic stress state $\boldsymbol{\Sigma}^{tr}_{n+1} $ assumes the microstructure state and the macroscale plastic strain in this load increment are equal to the previous load increment. Case~1 results in a purely elastic update as shown in Fig.~\ref{Figs:ch7_fig3}a. Except for the first load increment, this leads to a trivial solution for linear-elastic perfectly plastic models with the same microstructure configuration, that is $ \bar{m}^{x,j}_{n+1} = \bar{m}^{x,j}_{n} $. 
Case 2 in Fig.~\ref{Figs:ch7_fig3}b is of particular interest. The trial stress  $\boldsymbol{\Sigma}^{tr}_{n+1} $ predicts a plastic update. However, it is possible to find material configurations $ \hat{m}^{x,j}_{n+1} $ such that $\mathfrak{F}(\boldsymbol{\Sigma}^{tr}_{n+1},\hat{m}^{x,j}_{n+1}) \leq 0 $. The material adapts itself by falling back onto the elastic state via an appropriate update of the microscale configuration $\bar{m}_{n+1}^{x,j} $ (denoted with the red dashed line), maximizing the total strain energy following \eqref{eq:ch7_eq44}. In Case~3, no material configuration allows an elastic state for the trial stress $\boldsymbol{\Sigma}^{tr}_{n+1} $. Therefore, the material configuration remains unchanged, and the stress-strain state is updated through a return-mapping/closest point projection algorithm as shown in Fig.~\ref{Figs:ch7_fig3}c. 

We cast these cases into an algorithmic frame analogous to a standard elastoplastic return map algorithm. We summarize the result in the box below. 


\begin{mdframed}
	\begin{enumerate}
		\item Given: $\boldsymbol{E}_{n+1},\; \boldsymbol{E}_{n},\; \boldsymbol{E}^{p}_{n}, \;\bar{m}^{x,j}_{n},\;\rho_{j} $
		\item Compute elastic trial stress $\boldsymbol{\Sigma}^{tr}_{n+1}$
		\begin{equation*}
			\begin{split}
				& \boldsymbol{E}_{n+1}^{p} = \boldsymbol{E}_{n}^{p} \\
				& \boldsymbol{\Sigma}^{tr}_{n+1}:= \mathbb{C}(\bar{m}^{x,j}_{n}):(\boldsymbol{E}^{}_{n+1} - \boldsymbol{E}^{p}_{n})
			\end{split}
		\end{equation*}
		
		\item Check yield criterion $ \mathfrak{F}^{tr}_{n+1}:= \mathfrak{F}(\boldsymbol{\Sigma}^{tr}_{n+1},\bar{m}^{x,j}_{n}) $ 
		
		IF: $\mathfrak{F}^{tr}_{n+1} \leq 0 $ \\
		\textbf{CASE 1}: Elastic update $\Delta \gamma_{n+1} = 0;\boldsymbol{E}^{p}_{n+1} = \boldsymbol{E}^{p}_{n} $\\
		Microscale design $\bar{m}^{x,j}_{n+1} $ update through straightforward strain energy maximization
		\begin{equation*}
			\begin{split}
				& \argmax_{m^{x,j}_{n+1}\in E_{\textit{ad}}(\rho_j)}
				\frac{1}{2}(\boldsymbol{E}_{n+1}   - \boldsymbol{E}^{p}_{n+1} ):\mathbb{C}(	m^{x,j}_{n+1}): (\boldsymbol{E}^{}_{n+1} - \boldsymbol{E}^{p}_{n+1})\\
			\end{split}
		\end{equation*}
		
		ELSE IF: $\mathfrak{F}^{tr}_{n+1} > 0 $\\
		CHECK IF: $\mathfrak{F}^{tr(2)}_{n+1}:=\mathfrak{F}(\boldsymbol{\Sigma}^{tr}_{n+1},\hat{m}^{x,j}_{n+1}) \leq 0 ;\;\; \text{for} \;\; \hat{m}^{x,j}_{n+1} \in E_{ad}(\rho_{j})$\\
		\textbf{CASE 2}: Evolve microscale design ensuring elastic material behavior\\
		$ \Delta \gamma_{n+1} = 0;\; \boldsymbol{E}^{p}_{n+1} = \boldsymbol{E}^{p}_{n};\bar{m}^{x,j}_{n+1} \;\text{is solution of}$\\
		\begin{equation*}
			\argmax_{\mathfrak{F} (\boldsymbol{E}^{e}_{n+1},m^{x,j}_{n+1}) \leq  \mathfrak{F}^{tr(2)}_{n+1}   }
			\frac{1}{2}(\boldsymbol{E}_{n+1}   - \boldsymbol{E}^{p}_{n+1} ):\mathbb{C}(	m^{x,j}_{n+1}): (\boldsymbol{E}^{}_{n+1} - \boldsymbol{E}^{p}_{n+1})
		\end{equation*}
		ELSE: \\
		\textbf{CASE 3:} $\text{Only plastic update possible} \implies \Delta \gamma_{n+1} > 0;\; \bar{m}^{x,j}_{n+1} = \bar{m}^{x,j}_{n}$\\
		Update $\boldsymbol{E}^{p}_{n+1} $ through closest point projection algorithm. 
		
		\item Output: $\bar{m}_{n+1}^{x,j}, \boldsymbol{E}^{p}_{n+1}, \text{and}\; \boldsymbol{\Sigma}_{n+1}=\mathbb{C}(\bar{m}^{x,j}_{n+1}):(\boldsymbol{E}^{}_{n+1} - \boldsymbol{E}^{p}_{n+1}).  $	
	\end{enumerate}		
\end{mdframed}

\subsection{A special choice: continuum micromechanics based homogenization }
\label{sec:sec_43}

\begin{figure*}[b!]
	\centering
	\includegraphics[width=0.52\textwidth]{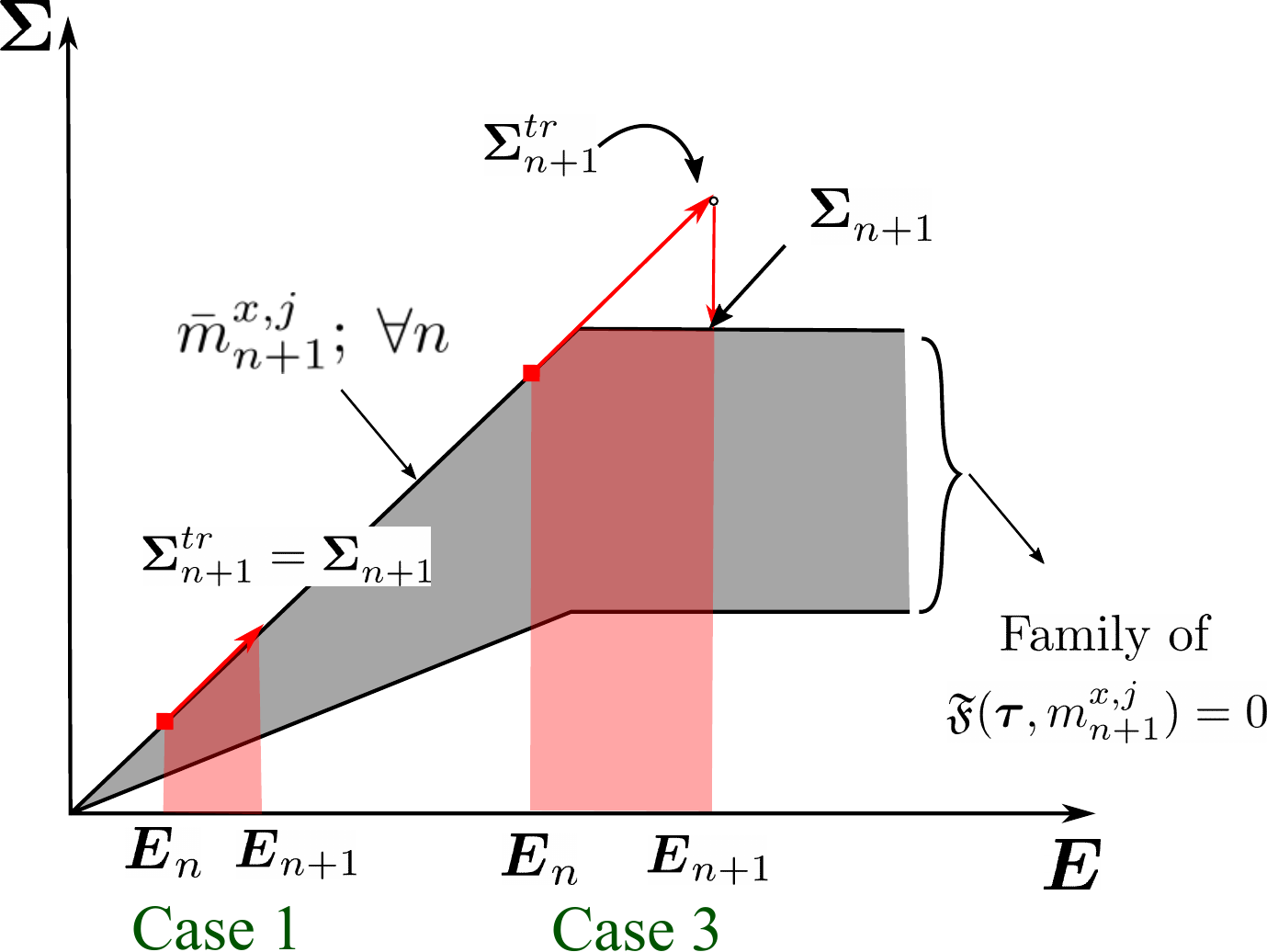}	
	\caption{Simplifications in the algorithmic procedure for material optimization induced by continuum micromechanics estimates. The optimized material configuration $ \bar{m}^{x,j}_{n+1}$ remains unchanged with loading history. }
	\label{Figs:ch7_fig3a}
\end{figure*}

In this article, we focus on homogenized yield criterion based on the quadratic stress average that can be obtained within a continuum micromechanics framework as reviewed in Section~\ref{sec:sec222}. To illustrate the simplifications that can be achieved with this choice, we consider again the one-dimensional example from Fig.~\ref{Figs:ch7_fig3}. Using the homogenized yield criterion based on continuum micromechanics, we arrive at Fig.~\ref{Figs:ch7_fig3a} that graphically illustrates the associated simplifications in the algorithmic procedure. In particular, the microscale configuration corresponding to the maximum stiffness (maximum strain energy density) also results in the maximum strength properties for the homogenized response, which implies that adaption to elastic state through microstructure evolution (Case~2) cannot occur here. In the following, we will provide a more detailed account of these simplifications. 


\subsubsection{Special properties induced through this choice}

For illustration purposes, we fall back to our initial example of a representative multiphase hierarchical system defined in Fig.~\ref{Figs:ch7_fig1}. We recall that Material C at the microscale is a perfectly elastoplastic material that follows the von Mises failure criterion with yield strength $\sigma^{Y}_{C}$ and bulk modulus $ \mu_{C} $.
For this example, we can then write the homogenized yield criterion $ \mathfrak{F} $ given in \eqref{eq:ch6_eq15} as a function of the stress $\boldsymbol{\tau}$ and microscale design variable $ m^{x,j}_{n+1} = [\phi^{x,j}_{A,n+1}, \theta^{x,j}_{A,n+1} ,  \zeta^{x,j}_{A,n+1}, \gamma^{x,j}_{C,n+1}] $:   
\begin{equation}
	\mathfrak{F} (\boldsymbol{\tau},m^{x,j}_{n+1}) = \sqrt{\boldsymbol{\tau}:[\mathbb{C}(m^{x,j}_{n+1})]^{-1}:\frac{\partial \; \mathbb{C}(m^{x,j}_{n+1})}{\partial \; \mu_{C}}:[\mathbb{C}(m^{x,j}_{n+1})]^{-1}:\boldsymbol{\tau}} \; - \;  \sqrt{\frac{\bar{\phi}_C}{3}} \; \frac{\sigma^{Y}_{C}}{\mu_{C}} \leq 0,
	\label{eq:ch7_eq45a}
\end{equation}
where $ \bar{\phi}_C $ is the equivalent volume fraction of Material C computed as $ \bar{\phi}_C = (1 - \phi^{x,j}_{A,n+1})\;\gamma^{x,j}_{C,n+1}  $. For a detailed derivation of the homogenized stiffness $\mathbb{C}(m^{x,j}_{n+1})$, we refer interested readers to Appendix~1 in \cite{gangwar2021concurrent}. 
These estimates 
hold the following three properties that form the basis for further simplifications in the algorithmic procedure for the material optimization:




\vspace{2mm}
\noindent \textbf{Property 1:} The microscale configuration $\bar{m}^{x,j}_{n+1} $ corresponding to the maximum stiffness (maximum strain energy density) from \eqref{eq:ch7_eq43} also maximizes the homogenized strength response. 
Figure~\ref{Figs:ch7_fig3a} graphically represents this property for the one-dimensional case with a family of possible stress-strain curves. Here, the microscale configuration corresponding to the higher linear-elastic slope leads to a higher limit strength for the homogenized response. Thus, the stress-strain curve for the configuration $\bar{m}^{x,j}_{n+1}$ acts as an envelope for all the stress-strain curves defined by the possible microscale configurations ${m}^{x,j}_{n+1}$. Utilizing the definitions $\boldsymbol{E}^{e,tr}_{n+1}:= \boldsymbol{E}^{}_{n+1} - \boldsymbol{E}^{p}_{n} $ and $\boldsymbol{\Sigma}^{tr}_{n+1}:= \mathbb{C}(\bar{m}^{x,j}_{n}):(\boldsymbol{E}^{}_{n+1} - \boldsymbol{E}^{p}_{n}) $, 
this property can be summarized as
\begin{equation}
	\begin{split}
		 \boldsymbol{E}^{e,tr}_{n+1}:\mathbb{C}(\bar{m}^{x,j}_{n+1}):\boldsymbol{E}^{e,tr}_{n+1} & \; \geq \; \boldsymbol{E}^{e,tr}_{n+1}:\mathbb{C}(m^{x,j}_{n+1}):\boldsymbol{E}^{e,tr}_{n+1} \\ \implies
		 \mathfrak{F} (\boldsymbol{\Sigma}^{tr}_{n+1}, \bar{m}^{x,j}_{n+1}) & \; \leq
		 \; \mathfrak{F} (\boldsymbol{\Sigma}^{tr}_{n+1}, m^{x,j}_{n+1}). 
	\end{split}
	\label{eq:ch7_eq51}
\end{equation} 
It is straightforward to see from \eqref{eq:ch7_eq51} that if $\mathfrak{F} (\boldsymbol{\Sigma}^{tr}_{n+1}, \bar{m}^{x,j}_{n+1}) > 0 $, then $\mathfrak{F} (\boldsymbol{\Sigma}^{tr}_{n+1}, m^{x,j}_{n+1}) \leq 0 $ is not possible for any microstructure configuration $m^{x,j}_{n+1} $. Therefore, following our discussion in Section~\ref{sec:sec_42}, we can conclude that an adaption to the elastic state through microstructure evolution (Case~2) is inconceivable for the continuum micromechanics schemes outlined in this paper. 


\vspace{2mm}
\noindent {\textbf{Property 2:}} An important conclusion from the previous section is that the microscale design update is possible in an elastic step only. The elastic part of macroscale strain $\boldsymbol{E}^{e}_{n+1}:= \boldsymbol{E}^{}_{n+1} - \boldsymbol{E}^{p}_{n+1}  $ at each Gauss point therefore entails the optimal material orientation $\bar{\theta}^{x,j}_{A,n+1} $ for load increment $(n+1)$. In the elastic step, the material optimization problem is essentially a strain energy maximization. The maximum strain energy is obtained for a general orthotropic material by aligning the material axis with the principal strain axes for the elastic strains \cite{jog1994topology,pedersen1989optimal}. 

\vspace{2mm}
\noindent \textbf{Property 3:} If the external loading is monotonically increasing, the optimal material orientation $\bar{\theta}^{x,j}_{A,n+1} $ is the only microscale variable that may change in each load increment. We denote the set of remaining  microscale design variables as  $m^{l(x,j)}_{n+1} = \; [\phi^{x,j}_{A,n+1},  \zeta^{x,j}_{A,n+1}, \gamma^{x,j}_{C,n+1} ]$. The optimal configuration $\bar{m}^{l(x,j)}_{n+1}$ for $m^{l(x,j)}_{n+1}$ remains unchanged throughout the loading history, that is $\bar{m}^{l(x,j)}_{n+1} = \bar{m}^{l(x,j)}_{n}\;\; \forall n = 1,2,...,n_{load}-1 $. We provide a proof of this property in \ref{App:App_A}. 

\subsubsection{Simplification of the material optimization problem}

The three special properties discussed above entail two important simplifications. First, adaption to the elastic state through microstructure evolution (Case~2) cannot occur. Second, except for the material orientation $\theta^{x,j}_{A,n+1}$, the optimized material configuration remains unchanged throughout the loading history. This implies that the material optimization problem is solved for the first load increment only via the strain energy maximization \eqref{eq:ch7_eq43} for the optimized configuration $\bar{m}^{x,j}_{n+1} $. Later, the optimized material orientation is updated for each load increment by aligning the material axis with the principal strain axes of the elastic part of the macroscale strain tensor $\boldsymbol{E}^{e}_{n+1}$. Furthermore, the continuum micromechanics-based estimates render the strain energy maximization \eqref{eq:ch7_eq43} a straightforward constraint optimization problem that seeks the solution in the microscale design variable space using fast gradient-based optimization methods \cite{boyd2004convex}, which we detailed in our previous work \cite{gangwar2021concurrent}. 



In conclusion, the total cost of solving all the material optimization problems is equivalent to the case of an end-compliance type optimization problem with a linear elastic response at the material scales. These simplifications result in an enormous reduction in computational effort, making our framework computationally tractable for the elastoplastic case.

\section{Comments on computer implementation}

\label{sec:sec5}
In this section, we provide an overview of our optimization framework with a focus on essential computer implementation details. First, we derive the essential sensitivity calculations of the objective function $ f_w $ with respect to the design variables $ \boldsymbol{\rho} $ for the structure optimization problem \eqref{eq:ch7_eq26}. We then briefly touch upon the optimality criteria method for updating the design variables in each structure optimization iteration, utilizing the computed sensitivities \cite{sigmund200199}. Finally, we consolidate all developments into a single algorithmic framework.

\subsection{Sensitivity analysis}
\label{sec:sec51}
 
 
 The format of the structure optimization problem \eqref{eq:ch7_eq26} is equivalent to the topology optimization formulation for elastoplastic structures presented by \textsc{Fritzen} and co-authors \cite{fritzen2016topology,xia2017evolutionary}. They derive the sensitivities using the path-dependent adjoint method \cite{buhl2000stiffness,cho2003design}. In the following, we provide a sketch of the derivation and highlight the important results. For a detailed derivation, we refer interested readers to \cite{fritzen2016topology,xia2017evolutionary}. 

The adjoint method begins with the construction of a Lagrangian function $ f^{*}_w$ that satisfies the zero residual constraints $ \boldsymbol{\bar{r}}_{n+1} $ and $ \boldsymbol{\bar{r}}_{n} $ at (quasi-)time $ t_{n+1} $ and $ t_{n} $ for each term of the trapezoidal rule stated in \eqref{eq:ch7_eq26}. With the Lagrange multipliers $ \boldsymbol{\lambda}_{n+1} $ and $ \boldsymbol{\mu}_{n+1} $ that are of the same dimensions as the vector of unknowns $ \boldsymbol{\bar{u}}_{n+1} $, the  Lagrangian function $ f^{*}_w$ follows as
\begin{equation}
	f^{*}_w = \frac{1}{2} \sum_{n=0}^{n_{load}-1} \Big\{  (\boldsymbol{f}_{n+1}^{\textit{ext}} + \boldsymbol{f}_{n}^{\textit{ext}})^{T} \;\Delta \boldsymbol{\bar{u}}_{n+1}   + (\boldsymbol{\lambda}^{}_{n+1})^{T} \boldsymbol{\bar{r}}_{n+1} +  (\boldsymbol{\mu}^{}_{n+1})^{T} \boldsymbol{\bar{r}}_{n} \Big\}. 
	\label{eq:ch6_eq31}
\end{equation}
Since $ \boldsymbol{\bar{r}}_{n+1} $ and $ \boldsymbol{\bar{r}}_{n}  $ vanish at the equilibrium solution, the sensitivity of $ f^{*}_w $ is same as that of $ f^{}_w $, implying that
\begin{equation}
	\frac{\partial f_w }{\partial \rho_{j}} = \frac{\partial f^{*}_w }{\partial \rho_{j}}\;. 
	\label{eq:ch6_eq33}
\end{equation}
The derivative of $ f^{*}_w $ with respect to the design variable $ \rho_{j} $ follows from (\ref{eq:ch6_eq31}) as
\begin{equation}
	\frac{\partial f^{*}_w }{\partial \rho_{j}} = \frac{1}{2} \sum_{n=0}^{n_{load}-1} \Bigg\{ \frac{\partial }{\partial \rho_{j}}  \Big((\boldsymbol{f}_{n+1}^{\textit{ext}} + \boldsymbol{f}_{n}^{\textit{ext}})^{T} \;\Delta \boldsymbol{\bar{u}}_{n+1}\Big)      + (\boldsymbol{\lambda}^{}_{n+1})^{T} \; \frac{\partial \boldsymbol{\bar{r}}_{n+1}}{\partial \rho_{j}}  +  (\boldsymbol{\mu}^{}_{n+1})^{T} \; \frac{\partial \boldsymbol{\bar{r}}_{n}}{\partial \rho_{j}} \Bigg\}. 
	\label{eq:ch6_eq34}
\end{equation}
The derivative of $ \boldsymbol{\bar{r}}_{n+1} $ with respect to $ \rho_{j} $ is evaluated following the residual definition in (\ref{eq:ch7_eq28}). Substituting the definition of $\boldsymbol{\Sigma}_{n+1}$ given in \eqref{eq:ch7_eq22} into (\ref{eq:ch7_eq28}), the derivative expression becomes
\begin{equation}
	\begin{split}
	\frac{\partial \boldsymbol{\bar{r}}_{n+1}}{\partial \rho_{j}} =  \frac{\boldsymbol{f}_{n+1}^{\textit{ext}}}{\partial \rho_{j}} -  \sum_{x=1}^{N_{gp}} \bigg[\boldsymbol{B}^{T} & \frac{\partial \mathbb{C}(\bar{m}^{x,j}_{n+1}) }{\partial \rho_{j}} (\boldsymbol{E}_{n+1}  - \boldsymbol{E}^{p}_{n+1}) w_x \bigg]  - \boldsymbol{K}^{\textit{tan}}_{n+1} \frac{\partial \Delta \boldsymbol{\bar{u}}_{n+1}}{\partial \rho_{j}}, \;\;  \\
	& \text{and} \;\; \boldsymbol{K}^{\textit{tan}}_{n+1} = - \frac{\partial \boldsymbol{\bar{r}}_{n+1}}{\partial \boldsymbol{\bar{u}}_{n+1}}\;.
	\end{split}
	\label{eq:ch6_eq40}
\end{equation}
$ \boldsymbol{K}^{\textit{tan}}_{n+1} $ is the global finite element stiffness matrix of the mechanical system at the equilibrium of load step $(n+1)$. We note that the quantities inside the square brackets in the second term of this equation are only computed for element $j$. For all remaining elements, $\boldsymbol{\Sigma}_{n+1}$ does not depend on $\rho_{j} $. The second term is zero for all corresponding entries, maintaining dimensional consistency with the vector $\boldsymbol{f}_{n+1}^{\textit{ext}}$.   

We observe that the sensitivities of $f_w $ as expressed in \eqref{eq:ch6_eq34} and \eqref{eq:ch6_eq40} require computationally extensive calculations of unknown derivatives. Therefore, our aim is to obtain the values of the Lagrange multipliers $ \boldsymbol{\lambda}_{n+1} $ and $ \boldsymbol{\mu}_{n+1} $ in such a way that these unknown derivatives can be eliminated from the sensitivity expression. 
To this end, we classify the degrees of freedom (DOF) into essential (index E; associated with the Dirichlet boundary conditions) and free (index F; remaining). According to this classification, we can partition vectors and matrices as shown for the following generic objects $\boldsymbol{v} $ and $\boldsymbol{M}$:
\begin{equation}
	\boldsymbol{v} \sim \begin{bmatrix}
		\boldsymbol{v}^E\\
		\boldsymbol{v}^F
	\end{bmatrix}\;\; \text{and} \;\;
	\boldsymbol{M} \sim \begin{bmatrix}
		\boldsymbol{M}^{EE} & \boldsymbol{M}^{EF} \\
		\boldsymbol{M}^{FE} & \boldsymbol{M}^{FF}
	\end{bmatrix}.
	\label{eq:ch6_eq35}
\end{equation}
Since the displacements $ \boldsymbol{\bar{u}}^{E} $ on the Dirichlet boundary $\Gamma_D $ are prescribed, they are independent of the current value of the optimization variable $ \boldsymbol{\rho} $. This observation leads to
\begin{equation}
	\frac{\partial \Delta \boldsymbol{\bar{u}}_q}{\partial \rho_{j}} = \frac{\partial}{\partial \rho_{j}} \begin{bmatrix}
		\Delta \boldsymbol{\bar{u}}_{q}^E\\
		\Delta \boldsymbol{\bar{u}}_{q}^F
	\end{bmatrix}\; =  \begin{bmatrix}
		\boldsymbol{0} \\
		\frac{\partial \Delta \boldsymbol{\bar{u}}_q^F }{\partial \rho_{j}}
	\end{bmatrix},
	\label{eq:ch6_eq36}
\end{equation}
at an arbitrary load step index $ q = 0,...,n_{load} -1$. With displacement-controlled loading, the only possible non-zero entries in the global force vector $\boldsymbol{f}_{q}^{\textit{ext}} $ are the reaction forces $\boldsymbol{f}_{p}^{\textit{ext,E}} $, that is
\begin{equation}
	\boldsymbol{f}_{q}^{\textit{ext}} = \begin{bmatrix}
		\boldsymbol{f}_{q}^{\textit{ext,E}} \\
		\boldsymbol{0}
	\end{bmatrix}.
	\label{eq:ch6_eq36a}
\end{equation}

The relations \eqref{eq:ch6_eq36} and \eqref{eq:ch6_eq36a} lead to an educated choices for the vectors $ \boldsymbol{\lambda}_{n+1} $ and $ \boldsymbol{\mu}_{n+1} $ such that the unknown derivatives with respect to the design variables in \eqref{eq:ch6_eq34} and \eqref{eq:ch6_eq40} can be eliminated (see \cite{fritzen2016topology,xia2017evolutionary} for details). The final expression for the sensitivity of the objective function $f_w $ with respect to the design variable $\rho_{j}$ is
\begin{equation}
	\begin{split}
		\frac{\partial f^{*}_w }{\partial \rho_{j}} = - \frac{1}{2} \sum_{n=0}^{n_{load}-1}  \Bigg\{    (\boldsymbol{\lambda}^{}_{n+1})^{T} \sum_{x=1}^{N_{gp}} \bigg[\boldsymbol{B}^{T} & \frac{\partial \mathbb{C}(\bar{m}^{x,j}_{n+1}) }{\partial \rho_{j}} (\boldsymbol{E}_{n+1}  - \boldsymbol{E}^{p}_{n+1}) w_x \bigg]\; \\  	
		& +\; (\boldsymbol{\mu}^{}_{n+1})^{T} \sum_{x=1}^{N_{gp}}\bigg[ \boldsymbol{B}^{T}  \frac{\partial \mathbb{C}(\bar{m}^{x,j}_{n}) }{\partial \rho_{j}} (\boldsymbol{E}_{n}  - \boldsymbol{E}^{p}_{n}) w_x \bigg]\;
		\Bigg\}.
	\end{split}
	\label{eq:ch7_eq39}
\end{equation}
With the prescribed displacement increments $\Delta \boldsymbol{\bar{u}}^{E}_{n+1} $, the choice of Lagrange multipliers $\boldsymbol{\lambda}^{}_{n+1} $ and $\boldsymbol{\mu}^{}_{n+1} $ that lead to the above expression is 
	\begin{align}
		\boldsymbol{\lambda}^E_{n+1} &= - \Delta \boldsymbol{\bar{u}}^{E}_{n+1} \;\; \text{and}\;\;  \boldsymbol{\lambda}^F_{n+1} =  \;[ \boldsymbol{K}^{\textit{tan},FF}_{n+1}]^{-1} \; \boldsymbol{K}^{\textit{tan},FE}_{n+1}\; \Delta \boldsymbol{\bar{u}}^{E}_{n+1}, \\
			\boldsymbol{\mu}^E_{n+1} &= - \Delta \boldsymbol{\bar{u}}^{E}_{n+1}\;\; \text{and}\;\; \boldsymbol{\mu}^F_{n+1} = \;[ \boldsymbol{K}^{\textit{tan},FF}_{n}]^{-1} \; \boldsymbol{K}^{\textit{tan},FE}_{n}\; \Delta \boldsymbol{\bar{u}}^{E}_{n+1}\;.
	\label{eq:ch7_eq38}
\end{align}
We note that the history of kinematic state variables $\boldsymbol{E} $ and $\boldsymbol{E}^{p} $ in \eqref{eq:ch7_eq39} are known from the solution of the global equilibrium equations at each load increment. 
For the representative multiscale configuration in Fig.~\ref{Figs:ch7_fig1}, the derivative of the homogenized stiffness $\mathbb{C} $ with respect to the element density $ \rho_{j}$ in \eqref{eq:ch7_eq39} can be evaluated by means of the chain rule. Following \cite{gangwar2021concurrent}, the expression is 
\begin{equation}
		\frac{\partial \mathbb{C}(\bar{m}^{x,j}_{n+1})}{\partial {\rho}_j} = 
		 \; \frac{\partial \mathbb{C}}{\partial \phi^{x,j}_{A,n+1}}  \frac{\partial  \phi^{x,j}_{A,n+1}}{\partial \rho_{j} } + \frac{\partial \mathbb{C}}{\partial \gamma^{x,j}_{C,n+1}}  \frac{\partial \gamma^{x,j}_{C,n+1} }{\partial \rho_{j} },
\end{equation}
where $\phi^{x,j}_{A,n+1} $ and $\gamma^{x,j}_{C,n+1} $ relate to $ {\rho}_j$ via \eqref{eq:ch7_eq27}. The partial derivatives of $\mathbb{C} $ with respect to $\phi^{x,j}_{A,n+1} $ and $\gamma^{x,j}_{C,n+1} $ are evaluated at the optimal microstructure configuration $\bar{m}^{x,j}_{n+1} $ by using finite difference approximations. 

\subsection{Structure optimization scheme}

We utilize the algorithmic procedure for the structure optimization that we outlined in our previous work \cite{gangwar2021concurrent} and that we briefly summarize in the following. First, we define sensitivity numbers to rank the element sensitivities that are used to update the macroscale design variables in each design iteration: 
\begin{equation}
	\alpha_j = - \frac{\partial f_c}{\partial {\rho}_j}. 
	\label{eq:eqs59f}
\end{equation}
To avoid mesh dependency and checkerboard patterns, the sensitivity numbers are first smoothed with a filtering scheme defined as
\begin{equation}
	\alpha_j = \frac{\sum_{j^{'}=1}^{N_j} g_{jj^{'}}\;\alpha_j}{\sum_{j^{'}=1}^{N_j} g_{jj^{'}}} \;\; \text{and} \;\; 	g_{jj^{'}} = \text{max}\;\{0, r_{\textit{min}} - \Delta(j,j^{'}) \},
	\label{eq:eqs59g}
\end{equation}
where $ N_j $ is the set of neighboring elements for which center-to-center distance $\Delta(j,j^{'})$ to element $ j^{'} $ is smaller than the filter radius $ r_{\textit{min}} $. 
To improve convergence, the sensitivity numbers are further averaged with the sensitivity numbers of the previous design iteration as
\begin{equation}
	\alpha_j^{i+1} \rightarrow (\alpha_j^{i+1} + \alpha_j^{i})/2.
	\label{eq:eqs59i}
\end{equation}

The ratio of sensitivity numbers and the mass constraint are combined to
\begin{equation}
	B_{j}^{i} = \Big(\frac{\alpha_j^{i}}{\Lambda^{i} |\Omega_{j} | }\Big)^{\eta},
	\label{eq:eqs59j}
\end{equation}
where $ \Lambda^{i} $ is the Lagrange multiplier corresponding to the total material mass constraint in design update $ i $, and $ \eta $ is a damping parameter. The macroscale density is updated by means of the well-known optimality criteria method \cite{sigmund200199}: 
\begin{equation}
	\scalebox{0.96}{%
		$\rho_{j}^{i+1} = 
		\begin{cases}
			\text{max} (\rho_{\textit{min}}, \rho_{j}^{i} - \mu )   & \text{if} \;\; \rho_{j}^{i} B_{j}^{i} \leq  \text{max} (\rho_{\textit{min}}, \rho_{j}^{i} - \mu )  \\
			\text{min} ( \rho_{j}^{i} + \mu, \rho_{\textit{max}})   & \text{if} \;\; \text{min} ( \rho_{j}^{i} + \mu, \rho_{\textit{max}}) \geq   \rho_{j}^{i} B_{j}^{i} \\
			\rho_{j}^{i} B_{j}^{i}                         & \text{otherwise}
		\end{cases}$}
	\label{eq:eqs59j1}
\end{equation}
To prevent a singular global stiffness matrix, the lower limit $ \rho_{\textit{min}} $ on $ \rho_{j} $ is limited by a small value, set in our case to 0.001. The maximum possible element density $\rho_{\textit{max}} $ depends on the density of the constituents at the microscales and the prescribed bounds in (\ref{eq:ch7_eq27}). $ \mu $ is a small move parameter that improves the stability, for instance by preventing multiple holes appearing and disappearing during optimization. The Lagrange multiplier $ \Lambda^{i} $ is updated using the bisection method to satisfy the mass constraint. The design iterations stop when the density convergence criteria are met.

\begin{algorithm}[t!]
	\SetAlgoLined
	\KwResult{Optimized solution vector $[\boldsymbol{\bar{\rho}}, \boldsymbol{\bar{m}}]^{T}$ }
	Initialize $ \boldsymbol{\rho}^0, \boldsymbol{m}^{0} $;  \\
	Set design iteration counter i = 0; \\
	\While { $|| \boldsymbol{\rho}^{i+1} - \boldsymbol{\rho}^{i} ||/ ||\boldsymbol{\rho}^{i}  || > \delta_{\textit{tol}} $   } {	
		Set up load increments $\Delta \boldsymbol{\bar{u}}_{n+1}^E$ for each loading step index $n = 0,..,n_{load} -1 $;\\
		Initialize load increment couter $ n = 0 $;\\
		Initialize $ \boldsymbol{\bar{u}}_{0} = \boldsymbol{0} \implies \boldsymbol{E}_{0} = \boldsymbol{0} $, and $\boldsymbol{E}^{p}_{0} = \boldsymbol{0}  $; \\
		\For{$n \leq n_{load} -1 $}{
			Increment load $ \boldsymbol{\bar{u}}^E_{n+1}  = \boldsymbol{\bar{u}}^E_{n} + 
			\Delta \boldsymbol{\bar{u}}^E_{n+1} $;\\
			Set Newton iteration counter k = 0;	\\
			$\boldsymbol{\bar{u}}^{(0)}_{n+1} =  \boldsymbol{\bar{u}}^{}_{n}$;\\
			\While{$||\boldsymbol{\bar{r}}^{(k)}_{n+1}|| < \epsilon_{tol} $}{
				
				\ForAll{macroscale Gauss points}{		
					Compute the macroscale strain $ \boldsymbol{E}^{(k)}_{n+1} =  \nabla^{s}(\boldsymbol{\bar{u}}^{(k)}_{n+1})  $;\\
					Update the state variables $ \boldsymbol{\Sigma}^{(k)}_{n+1}  $, $\boldsymbol{E}^{p,(k)}_{n+1}$, and $ \bar{m}^{x,j}_{n+1} $ by solving the material optimization problem; \\
				}
				Evaluate the residual force vector $\boldsymbol{\bar{r}}^{(k)}_{n+1}:= \boldsymbol{f}_{n+1}^{\textit{ext}} - \boldsymbol{f}_{}^{\textit{int}}(\boldsymbol{\Sigma}^{(k)}_{n+1}) $; \\
				Set up the linear system:
				$\boldsymbol{K}^{tan,(k)}_{n+1} \;\delta \boldsymbol{\bar{u}}^{(k+1)}_{} = \boldsymbol{\bar{r}}^{(k)}_{n+1}$, and solve for $ \delta \boldsymbol{\bar{u}}^{(k+1)}_{} $;\\
				Apply Newton correction to the displacements: $\boldsymbol{\bar{u}}^{(k+1)}_{n+1} = \boldsymbol{\bar{u}}^{(k)}_{n+1} + \delta \boldsymbol{\bar{u}}^{(k+1)}$; \\
				k++;		
			}
			Calculate and store Lagrange multipliers $\boldsymbol{\lambda}^{}_{n+1} $ and $\boldsymbol{\mu}^{}_{n+1} $ for senstivity calculations using converged state variables;\\
			n++;
		}
		Compute the objective function $ f_w (\boldsymbol{\rho}) $ and sensitivities $ \partial f_w / \partial\boldsymbol{\rho}  $;\\
		Update density $ \boldsymbol{\rho}^{i+1} $ using the optimality criteria algorithm; $ i$++;					 	
	}	
	\caption{Concurrent structure and material optimization framework for elastoplastic structures with multiphase hierarchical materials. }
	\label{Alg3}
\end{algorithm}

\subsection{General algorithm}


Algorithm~\ref{Alg3} consolidates all the developments into an algorithmic framework. It mainly consists of three blocks. The outer block represents macroscale structure optimization iterations with iteration index $ i $, using the optimality criteria method detailed in (\ref{eq:eqs59j1}). It stops when the relative change in macroscale density $ \boldsymbol{\rho} $ falls below the tolerance $\delta_{\textit{tol}} $, and the converged solution is the optimum macroscale density $\boldsymbol{\bar{\rho}}$. For a given macroscale density distribution, the middle block solves the initial boundary value problem with known load increments $\Delta \boldsymbol{\bar{u}}_{n+1}^E$ at each load increment $n $. The global equilibrium for each load increment is solved with the Newton-Raphson method that uses the linearization of \eqref{eq:ch7_eq28} \cite{simo2006computational}. Here, $ (\bullet)^{(k)}_{n+1} $ denotes the value of a particular variable $ (\bullet) $ at the $ k^{th} $ iteration at load step $(n+1)$. The Newton-Raphson scheme stops when the norm of the residual force vector drops below a tolerance threshold $\epsilon_{tol} $, and we adopt $\epsilon_{tol} = 10^{-5} $ in this article. 

The inner block solves the material optimization problem described in Section~\ref{sec:sec_4} at each Gauss point with prescribed state variables at this iteration stage for each load increment. For the schemes based on continuum micromechanics outlined in this paper, the material optimization problem is solved for the optimized configuration $\bar{m}^{x,j}_{n+1} $ maximizing the strain energy via \eqref{eq:ch7_eq43} at the first load increment only. Thereafter, the material orientation is updated for each load increment by aligning the material axis with the principal strain axes of the elastic part of the macroscale strain tensor according to our discussion in Section~\ref{sec:sec_43}. The equations due to strain energy maximization can be solved with standard gradient-based constraint optimization methods such as the quasi-Newton method of \textsc{Broyden}, \textsc{Fletcher},
\textsc{Goldfarb}, and \textsc{Shanno} (BFGS) or sequential least squares programming (SLSQP) methods. 

\section{Numerical examples}

\begin{figure*}[t!]
	\centering
	\includegraphics[width=0.94\textwidth]{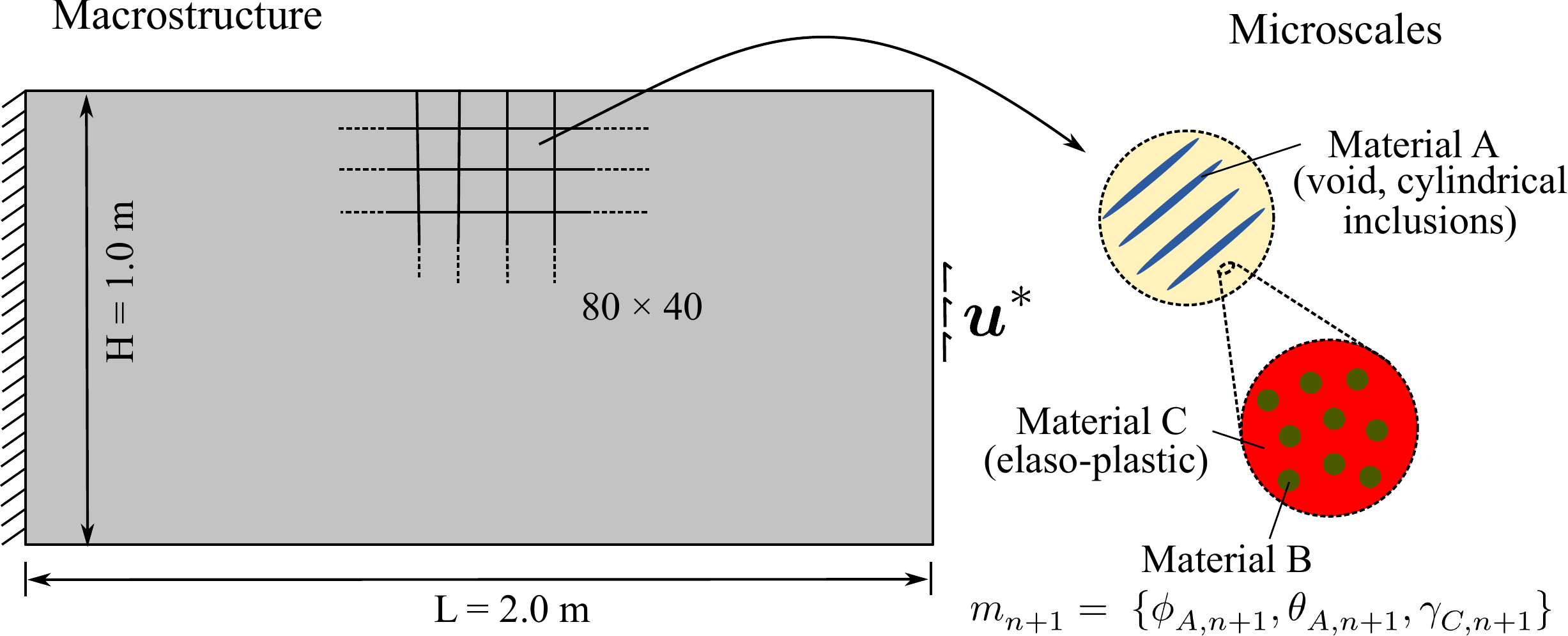}	
	\caption{Cantilever benchmark based on elastoplastic multiphase hierarchical materials.}
	\label{Fig:ch7_fig4}
\end{figure*}




In this section, we define two test examples with elastoplastic multiphase hierarchical material models that are suitable to illustrate the computational efficiency and validity of our path dependent concurrent material and structure optimization framework. First, we consider a standard cantilever type benchmark problem and modify its material definitions analogous to the multiscale configuration shown in Fig.~\ref{Figs:ch7_fig1}. Later, we demonstrate the potential of our framework for biotailoring applications by solving a prototype problem that integrates a hierarchical material model for cereal stems \cite{gangwar2020multiscale}.




\begin{figure*}[b!]
	\centering
	\includegraphics[width=0.60\textwidth]{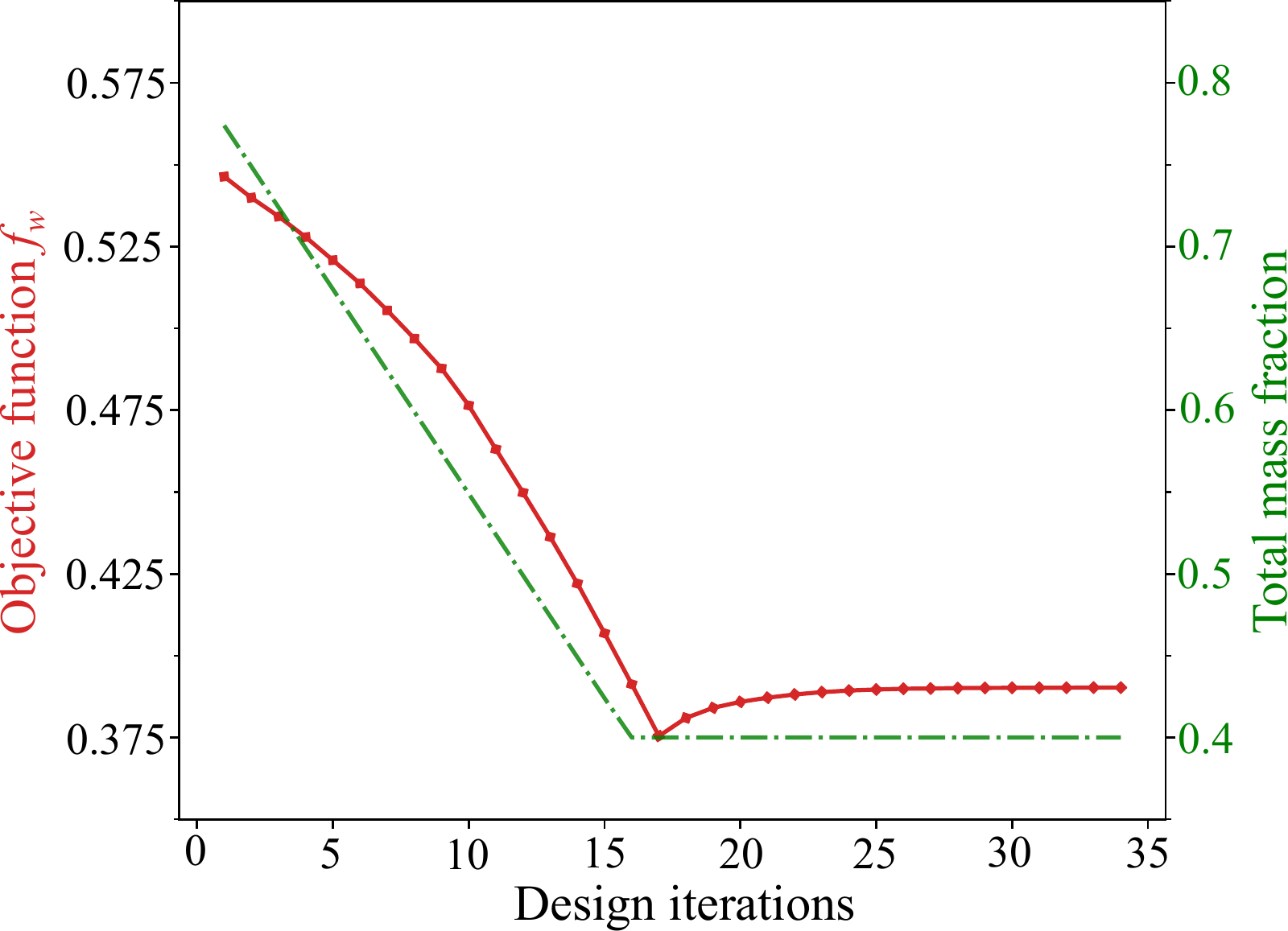}
	\caption{Convergence of objective function $f_w$ (in N-mm) and mass fraction with respect to number of macroscale design iterations.}
	\label{Fig:ch7_fig41}
\end{figure*}

\subsection{Cantilever benchmark problem}

Figure~\ref{Fig:ch7_fig4} modifies the definition of the standard cantilever design problem to demonstrate the developed concepts in this article. The length and height of the macrostructure are 2.0 m and 1.0 m, respectively. The left edge is fixed, and a displacement loading of $\boldsymbol{u}^{*} = 7.5$ mm is prescribed at the central 10\% of the right edge that we divide in six load steps with a constant load increment of $\Delta \boldsymbol{\bar{u}}^{E} = 1.25$ mm. We discretize the macroscale structure with an $ 80 \times 40 $ mesh of 4-node plane strain quadrilateral elements, resulting in a characteristic element size of $l_e = 25 $ mm and $ 3,200 $ macroscale design variables. Each element contains four Gauss points, resulting in $ 80 \times 40 \times 4 = 12,800 $ material optimization problems in each load step. 

As illustrated in Fig.~\ref{Fig:ch7_fig4}, we consider a hierarchical system that consists of Material A, B, and C at two different length scales. Their densities (in $ \text{Kg/m}^3 $) are $ \rho_A = 0 $, $ \rho_B = 0.5 $, and $\rho_C = 1.0$, their Young's moduli (in GPa) are $E_A = 0.0$, $E_B = 0.5$, and $E_C = 1.0$, and Poisson's ratio of all constituents is 0.3. Material C is elastoplastic with yield strength 1 MPa. We assume that Material A forms cylindrical inclusions in the homogenized matrix of Material B and C. 
At each Gauss point, the material microstructure is parametrized by the volume fraction $\phi^{x,j}_{A,n+1}$, the orientation $ \theta^{x,j}_{A,n+1} $, and the volume fraction $ \gamma^{x,j}_{C,n+1} $ for load step $(n+1)$, which results in $38,400$ microscale design variables in each load step.

\begin{figure*}[b!]
	\centering
	\subfloat[Optimum density distribution. ]{\includegraphics[width=0.48\textwidth]{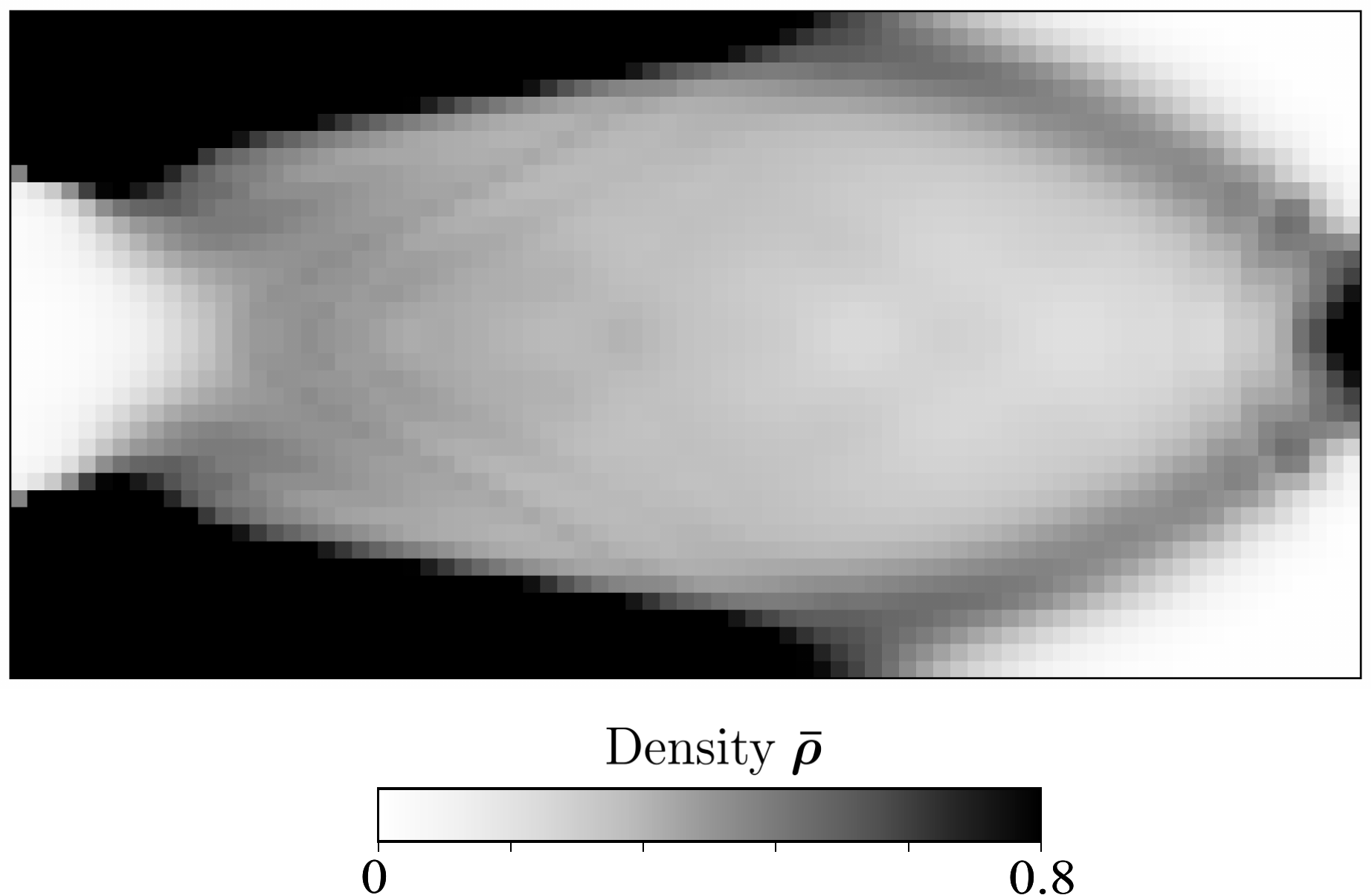}\label{Figs:ch7_fig5a}}
	\hspace{0.2cm}
	\subfloat[Equivalent plastic strain distribution.]{\includegraphics[width=0.48\textwidth]{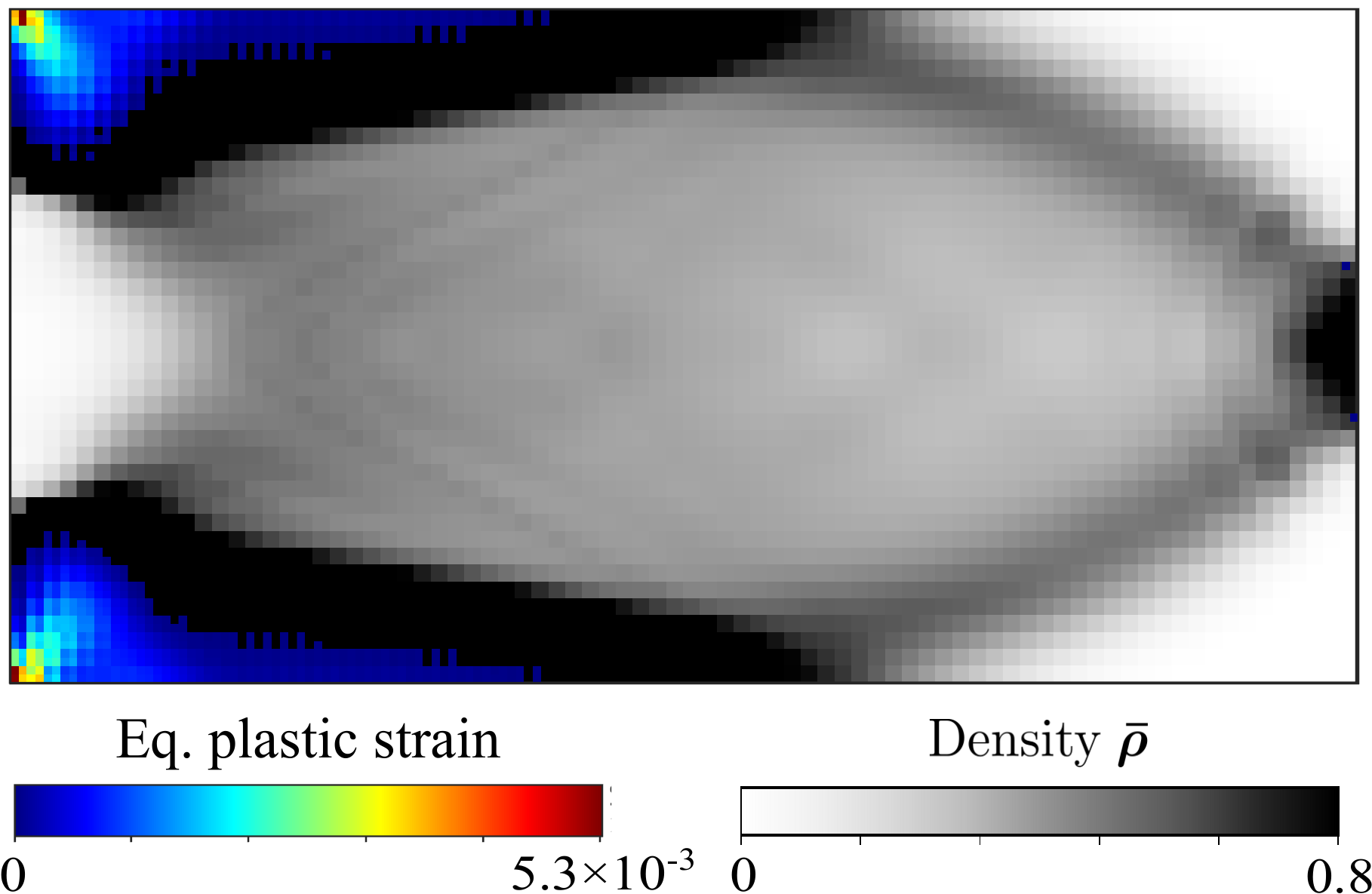}\label{Figs:ch7_fig5b}}	
	\caption{Macroscale density distribution and equivalent plastic strain distribution of the cantilever benchmark problem for a total prescribed displacement of $\boldsymbol{u}^{*} = 7.5 $ mm. }
	\label{Figs:ch7_fig5}
\end{figure*}

\begin{figure*}[t!]
	\centering
	\subfloat[Optimum microstructure at mesoscale. ]{\includegraphics[width=0.56\textwidth]{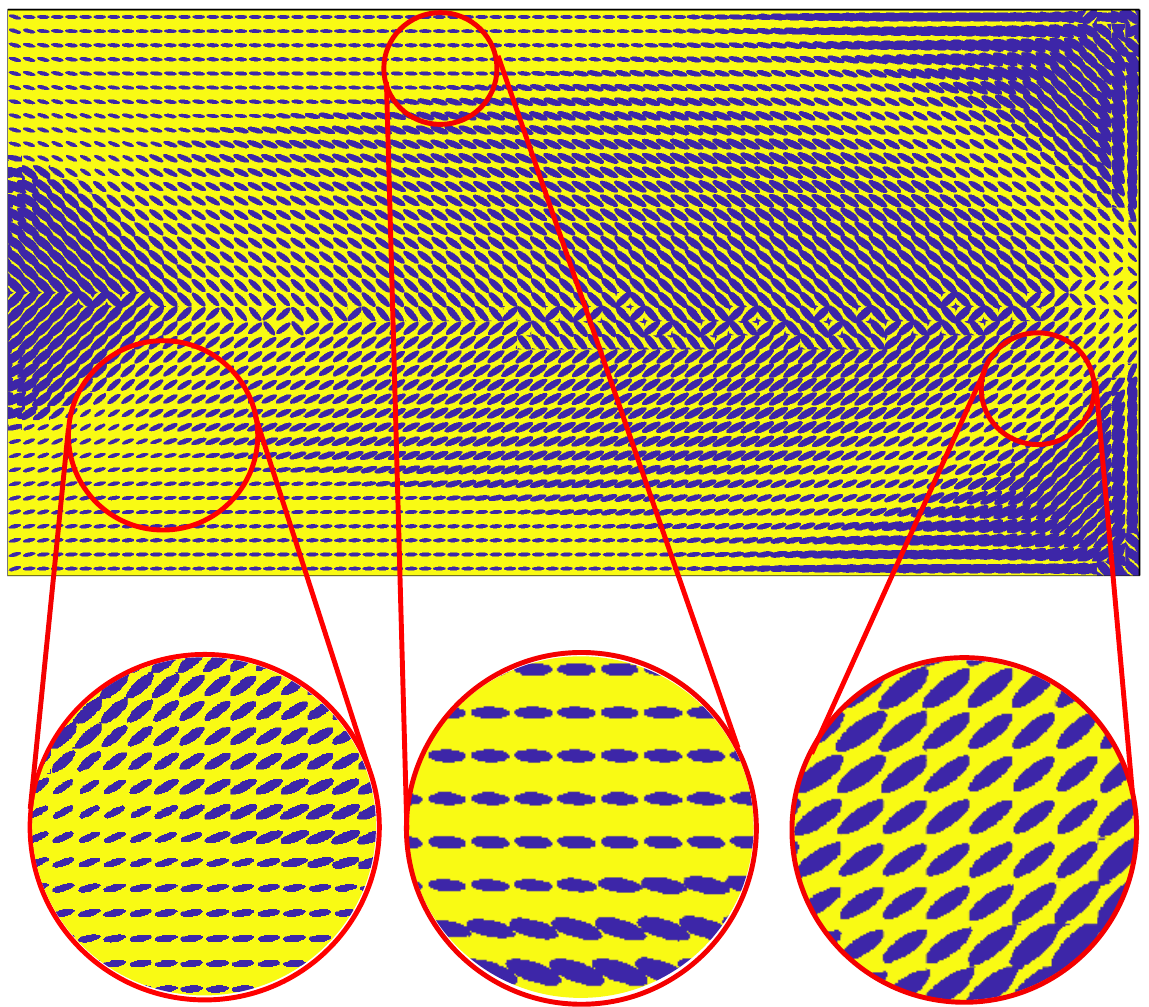}\label{Figs:ch7_fig6a}}\\
	\vspace{0.2cm}
	\subfloat[Optimized equivalent volume fractions of Material B and Material C.]{\includegraphics[width=0.62\textwidth]{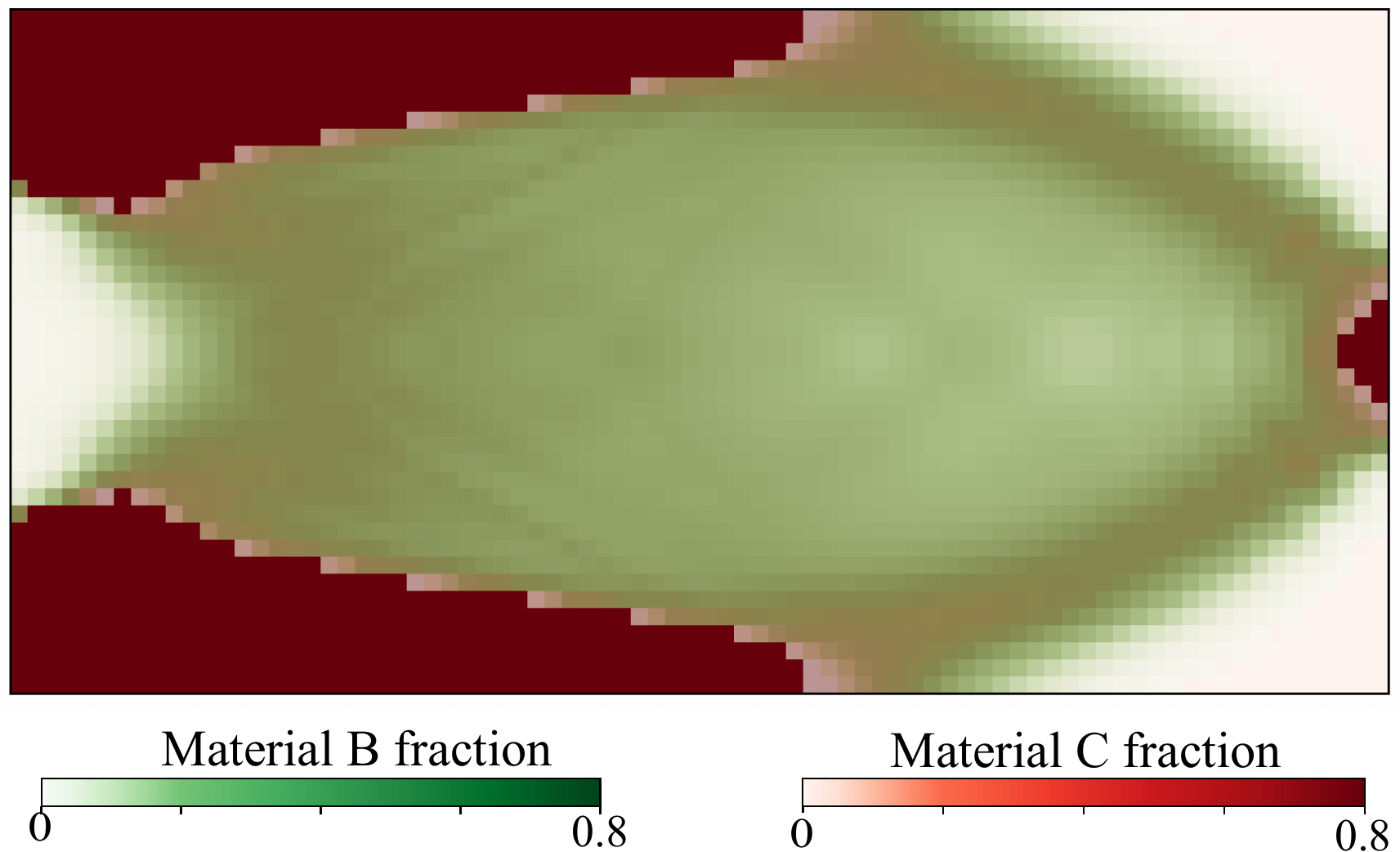}\label{Figs:ch7_fig6b}}	
	\caption{Optimal material configuration for total prescribed displacement of $\boldsymbol{u}^{*} = 7.5 $ mm. }
	\label{Figs:ch7_fig6}
\end{figure*}

The minimum volume fraction of Material A is set to $ \phi_{A}^{\textit{min}} = 0.2 $. The existence of the homogenized yield criterion $\mathfrak{F} $ in \eqref{eq:ch7_eq45a} requires $\bar{\phi}_C = (1 - \phi^{x,j}_{A,n+1})\;\gamma^{x,j}_{C,n+1} > 0$. It implies that the bounds $ \phi_{A}^{\textit{max}} < 1 - h $ and $ \gamma_{C}^{\textit{min}} > h $, where $h$ is a small positive number. We restrict $ \rho_{\textit{min}} $ to 0.001 and $ \rho_{\textit{max}} $ to 0.799 to satisfy these requirements. The total amount of material mass available is restricted to 40\% of the maximum possible mass.  As an initial condition at the macroscale, we assume the maximum possible density $ \rho_{\textit{max}} $ in each element. At the material level, we assume an initial microstructure configuration $ \phi^{}_{A} = 0.0 $, $ \theta^{}_{A} = 0.0  $, and $ \gamma^{}_{C} = 1.0 $ at each Gauss point. In each design update, we reduce the target mass fraction by 0.025 until we reach the specified mass fraction $ M_{\textit{frac}} = 0.4$. The move parameter $ \mu $ and the damping parameter $\eta$ are set to 0.05 and 0.5. The filter radius $r_{min} $ is reduced linearly from $r_{min} = 20 \, l_e $ to $r_{min} = 4 \, l_e $ with design iterations for improving the convergence of the structure optimization algorithm following \cite{xia2017evolutionary}.

The structure optimization algorithm stops when the relative change in the macroscale density field falls below the tolerance $\epsilon_{\textit{tol}} = 10^{-3} $. Figure~\ref{Fig:ch7_fig41} illustrate the convergence of the macroscale design update. We notice that the algorithm takes 34 density updates to converge to the final design with converged objective function value 0.39022 N-mm. Figure~\ref{Figs:ch7_fig5}a and b plot the optimized macroscale density and equivalent plastic strains overlaid on the density plot, respectively. The plastic strains are concentrated at the clamped end’s boundaries, consequently pushing the material towards these regions. As observed in Fig.~\ref{Figs:ch7_fig5}b, the sharp features near the clamped end in the optimal density distribution mimic the plastic front emphasizing its importance for the final design.


\begin{figure*}
	\centering
	\includegraphics[width=0.70\textwidth]{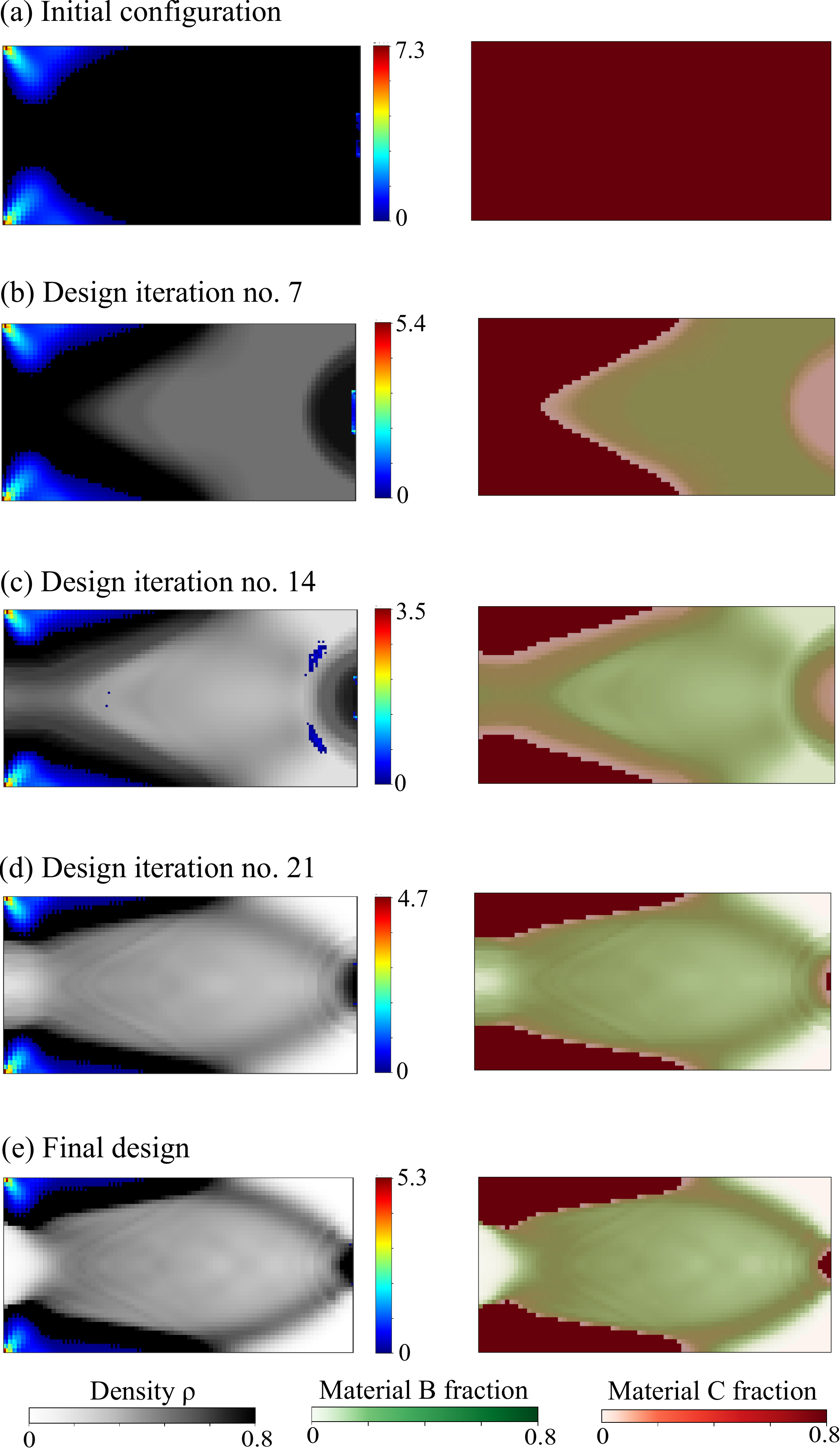}	
	\caption{Evolution of macroscale density configuration and equivalent plastic strains (rainbow colormap, $\times 10^{-3} $ units ) 
	and equivalent volume fractions of Material B and C.}  
	\label{Figs:ch7_fig7}
\end{figure*}

\begin{figure*}[t!]
	\centering
	\subfloat[Density distribution. ]{\includegraphics[width=0.49\textwidth]{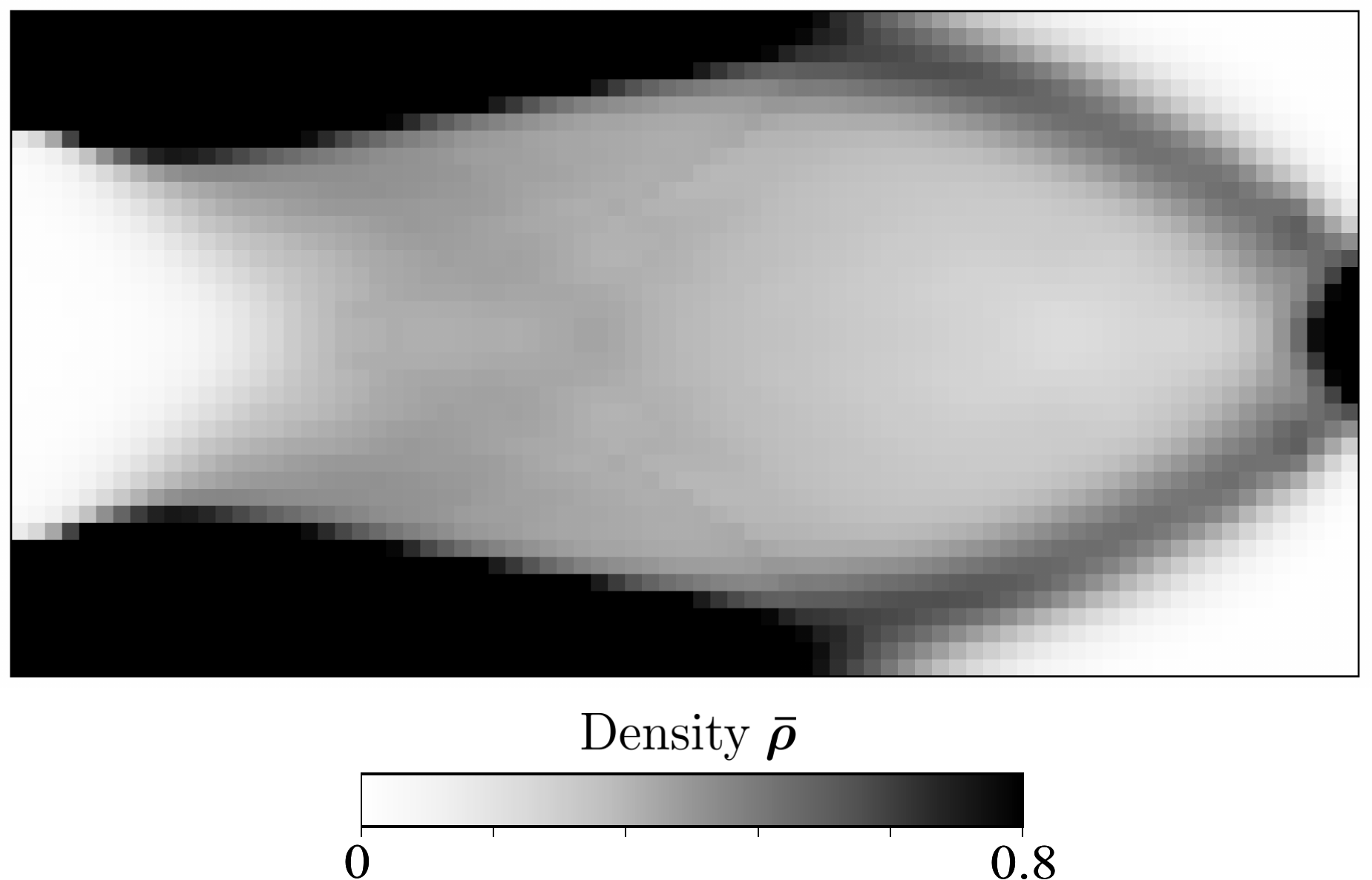}\label{Figs:ch7_fig8a}}
	\hspace{0.1cm}
	\subfloat[Equivalent volume fractions of Material B and C.]{\includegraphics[width=0.49\textwidth]{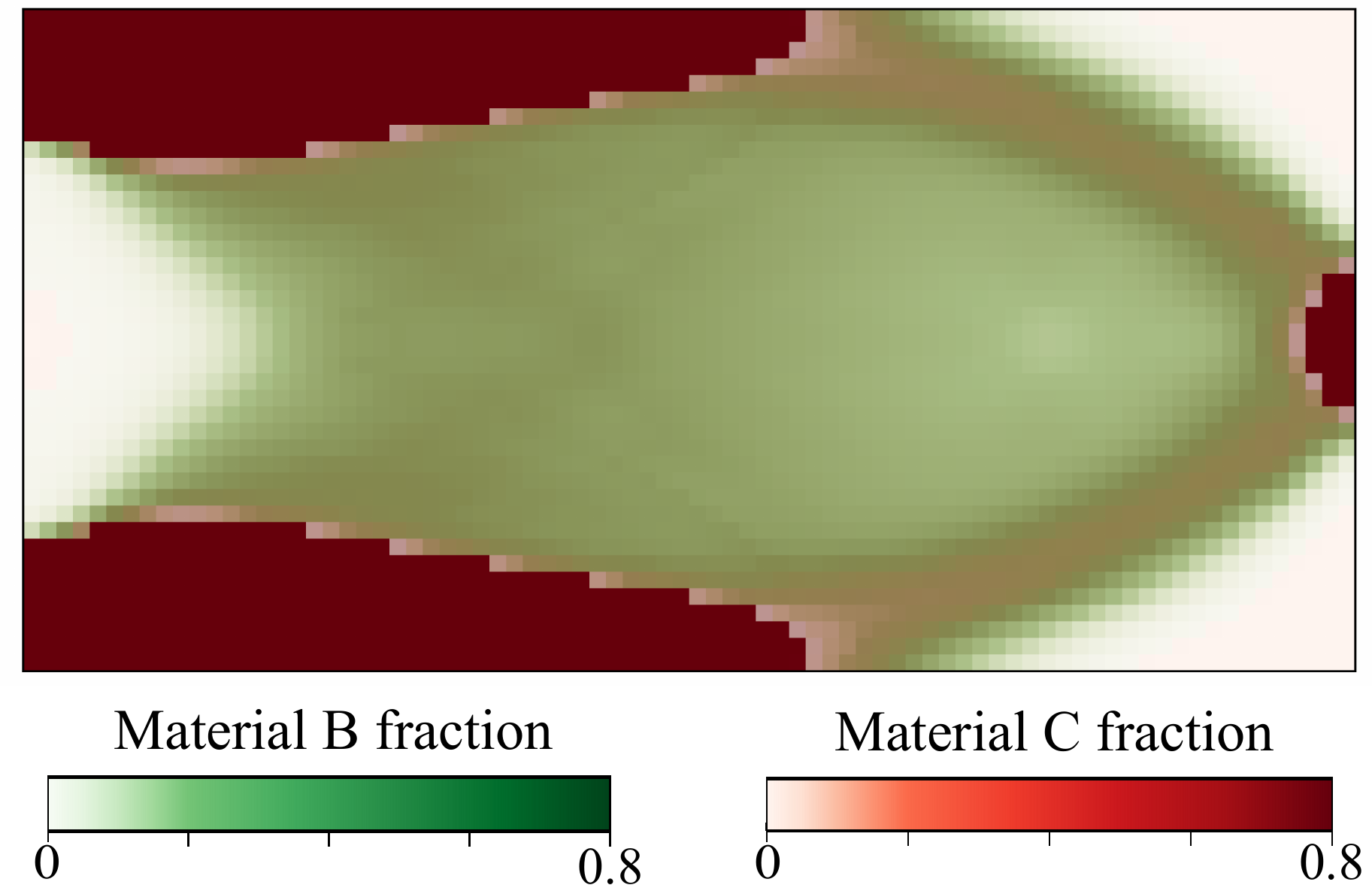}\label{Figs:ch7_fig8b}}	
	\caption{Final design of the cantilever benchmark in the linear elastic case (Material C purely elastic). }
	\label{Figs:ch7_fig8}
\end{figure*}

Figure~\ref{Figs:ch7_fig6} illustrates the optimized morphology at the mesoscale and the equivalent volume fraction of Material B and C from the lowermost scale. The yellow color in Fig.~\ref{Figs:ch7_fig6}a represents the matrix material that results from the homogenization of the lowermost scale, and the blue color displays the volume fraction and orientation of Material A inclusions. The inclusions follow the direction of the largest principal stress. The equivalent volume fractions of Material A, B, and C at the macroscale are defined as: $\bar{\phi}_{A} = {\phi}_{A}  $, $ \bar{\phi}_{B} = (1-{\phi}_{A}) (1 - \gamma_{C})  $ , and $ \bar{\phi}_{C} = (1-{\phi}_{A})  \gamma_{C}  $. Figure~\ref{Figs:ch7_fig6}b displays the equivalent volume fraction of Material B and C at the macroscale for the final design, where we use 60\% opacity for both. We can observe the regions dominated by Material B, C, and a mixing zone. The stiffer Material C is deposited in the regions anticipated to yield first, while Material B dominates the transition zone.

\begin{figure*}[b!]
	\centering
	\includegraphics[width=0.50\textwidth]{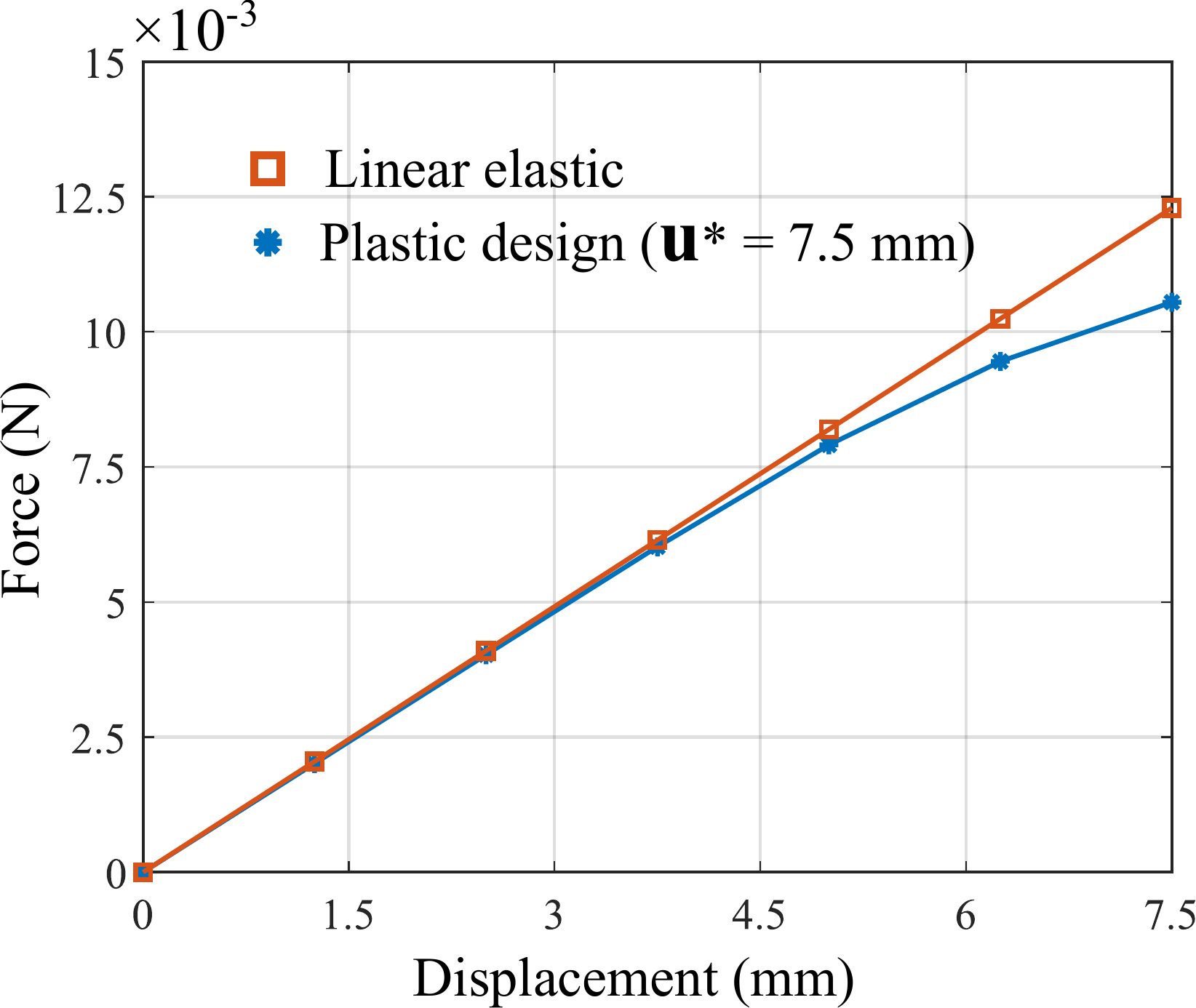}	
	\caption{Load vs displacement curves of the final designs in the linear elastic and elastoplastic cases. }
	\label{Figs:ch7_fig9}
\end{figure*}

Figure~\ref{Figs:ch7_fig7} illustrates the evolution of the optimization process by plotting equivalent plastic strains overlaid on the density distribution and the equivalent volume fraction of Material B and C, all at selected design iterations. The evolution of the macroscale density and equivalent plastic strains shows that the  design process attempts to attenuate the plastic front. In this process, the algorithm pushes more material towards the region close to the clamped boundary, delaying yielding in this region. 
Material C is the stiffest material among the constituents and exhibits elastoplastic behavior. Its evolution is heavily influenced by the plastic front, which leads for instance to sharp features in the macroscale density configuration.

We finally illustrate the impact of elastoplastic optimization with respect to optimization with linear elastic materials. To this end, Figure~\ref{Figs:ch7_fig8} plots the optimized density distribution and the equivalent volume fraction of Material B and C for the linear elastic design that assumes purely elastic properties of Material C at the lowermost scale. 
Comparing these plots with Fig.~\ref{Figs:ch7_fig5} and \ref{Figs:ch7_fig6b}, apparent differences can be observed in the optimized layouts. The plastic design places more material towards the clamped region with clear features imitating the plastic front, whereas these attributes are missing in the linear elastic design. Figure~\ref{Figs:ch7_fig9} plots the load-displacement curves for both cases. At the higher load levels, the load-displacement curves start to deviate from each other, when the plastic deformations play a crucial role. We conclude that the linear elastic design and the plastic design are functioning differently, and it is important to consider plastic effects at different scales in multiphase hierarchical systems that are expected to develop dissipation-based energy absorption mechanisms against external impacts.

\begin{figure*}[h!]
	\centering
	\includegraphics[width=0.99\textwidth]{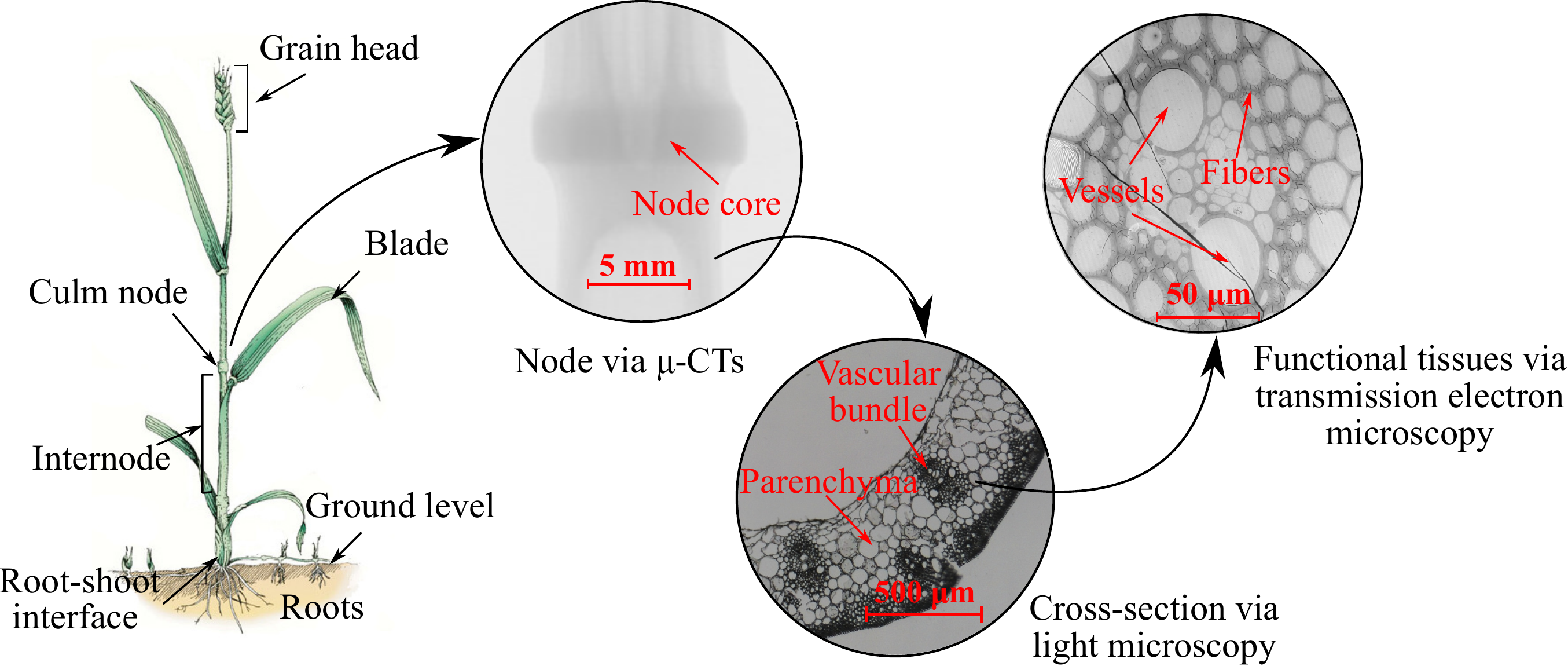}	
	\caption{Hierarchical structure of a cereal plant profiled through microimaging \cite{gangwar2020multiscale}. }
	\label{Figs:ch9_fig1a}
\end{figure*}

\subsection{Towards predicting self-adapting mechanisms in plants}
\label{sec:sec_62}
%
%
%



\begin{figure*}[t!]
	\centering
	\includegraphics[width=1.00\textwidth]{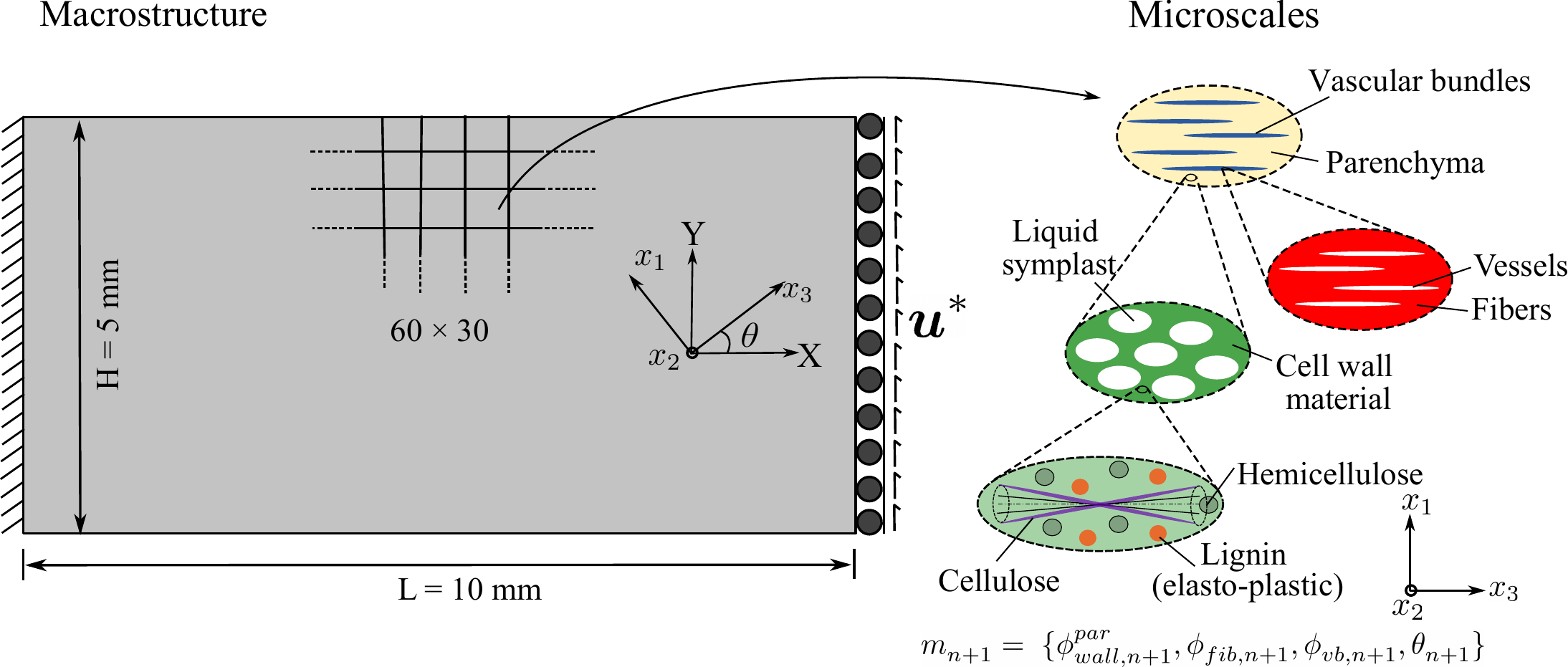}	
	\caption{Prototype model for the hierarchical optimization of a cereal node region. }
	\label{Figs:ch9_fig1}
\end{figure*}

Biomaterials exhibit multiscale inelastic behavior and develop dissipation-based energy absorption mechanisms optimizing its hierarchical composition across material microscales against the external biophysical stimuli through natural evolution 
\cite{wegst2015bioinspired,fratzl2007nature,bhushan2009biomimetics}. A rational understanding of  microstructure interdependencies with self-adapting mechanisms will pave the way towards many biotailoring applications with improved properties, for instance, in the context of the targeted breeding of agricultural crops \cite{brule2016hierarchies,berry2004understanding}. A few studies have attempted the multiscale optimization of biological systems such as bone-remodeling and bioinspired materials \cite{rodrigues1999global,coelho2008hierarchical,radman2013topology}. Several roadblocks, however, such as high computational cost and non-trivial problem decomposition in the case of elastoplastic behavior have limited these approaches to simple linear elastic problems with no more than two scales. With the following prototype model, we demonstrate the potential of our optimization framework in overcoming these roadblocks for the computationally efficient modeling of self-adaption of biomaterials.


Crop stem materials organize themselves hierarchically across multiple length scales. The hierarchical scales in crops range from base constituents such as cellulose, hemicellulose, and lignin, to cell wall, functional tissues, cross-section, and structure scale node morphology levels. In our previous work, we experimentally profiled this hierarchical organization through microimaging (micro-CT, light microscopy, transmission electron microscopy) and chemical analysis, focusing on cereal stems \cite{gangwar2020multiscale}, which we briefly summarize in Fig.~\ref{Figs:ch9_fig1a}. Exploiting this data, we developed and validated a continuum micromechanics model of cereal stem materials that accurately relates material composition with elastoplastic mechanical behavior across different scales. We provide all implementation information relevant in the scope of this article in \ref{App:App_oat}. 

\begin{figure*}[t!]
	\centering
	\includegraphics[width=0.75\textwidth]{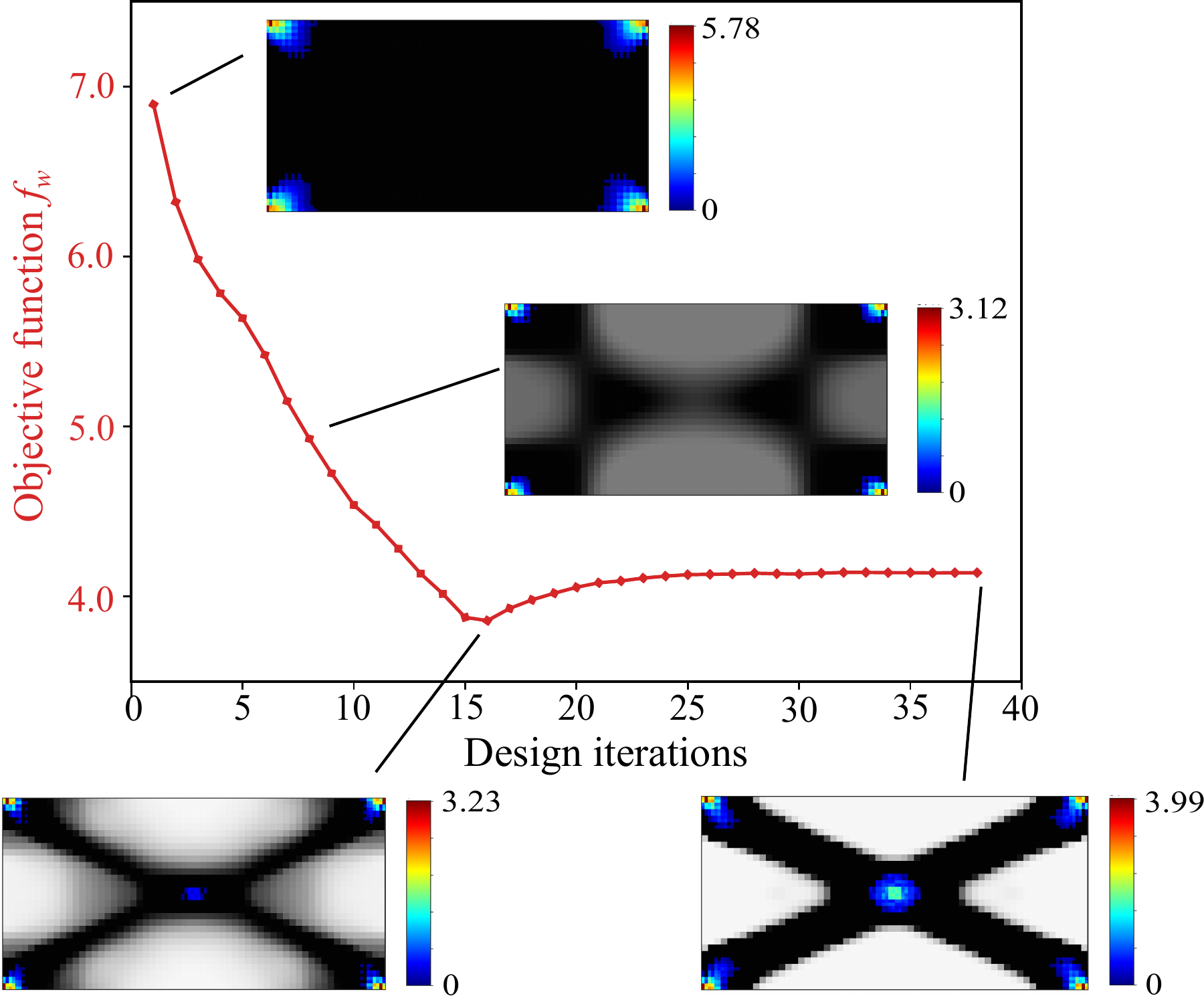}	
	\caption{Convergence of the objective function $f_w$ (in $10^{-3} $ N-mm)  with respect to the number of design iterations, plotted along with snapshots of the macroscale density configuration and the equivalent plastic strains (in $\times 10^{-3} $ units, rainbow colormap).}
	\label{Figs:ch9_fig4}
\end{figure*}



\begin{figure*}[t!]
	\centering
	\subfloat[Optimum desity distribution. ]{\includegraphics[width=0.48\textwidth]{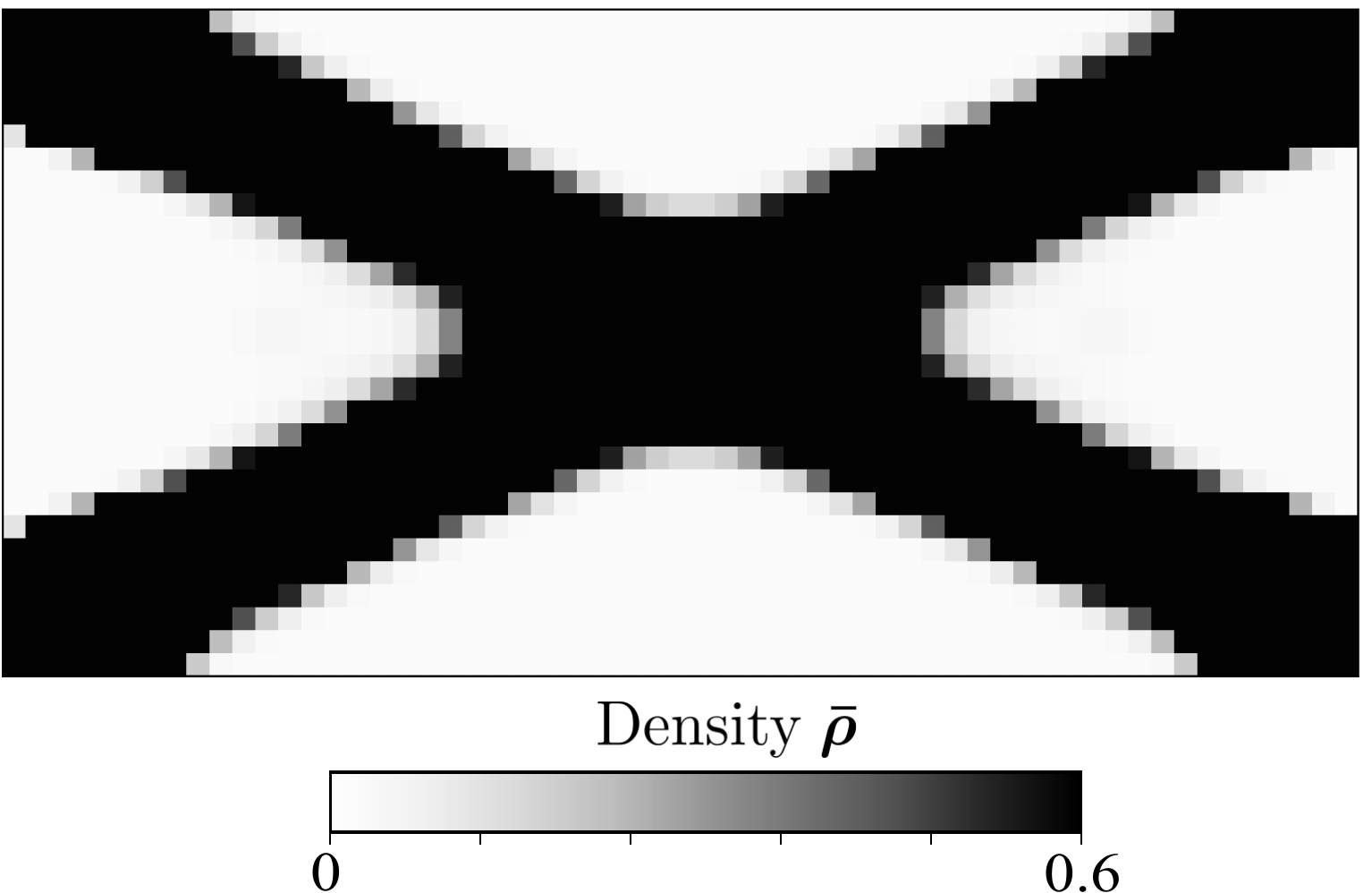}\label{Figs:ch9_fig2a}}
	\hspace{0.2cm}
	\subfloat[Equivalent plastic strain distribution.]{\includegraphics[width=0.48\textwidth]{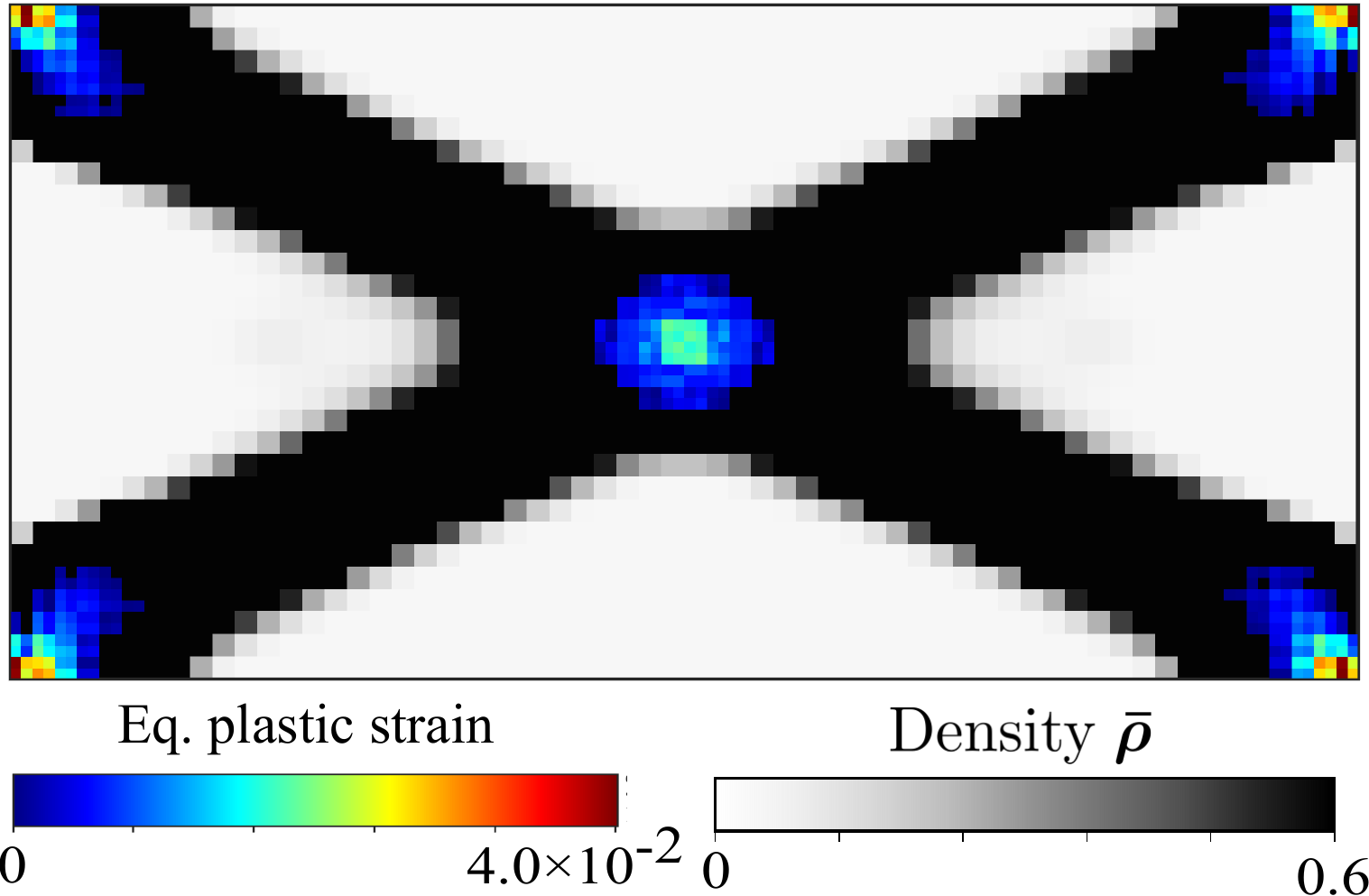}\label{Figs:ch9_fig2b}}	
	\caption{Final design of the cereal node region with equivalent plastic strain distribution for total prescribed displacement $\boldsymbol{u}^{*} = 0.4 $ mm. }
	\label{Figs:ch9_fig2}
\end{figure*}

\begin{figure*}[h!]
	\centering
	\includegraphics[width=1.00\textwidth]{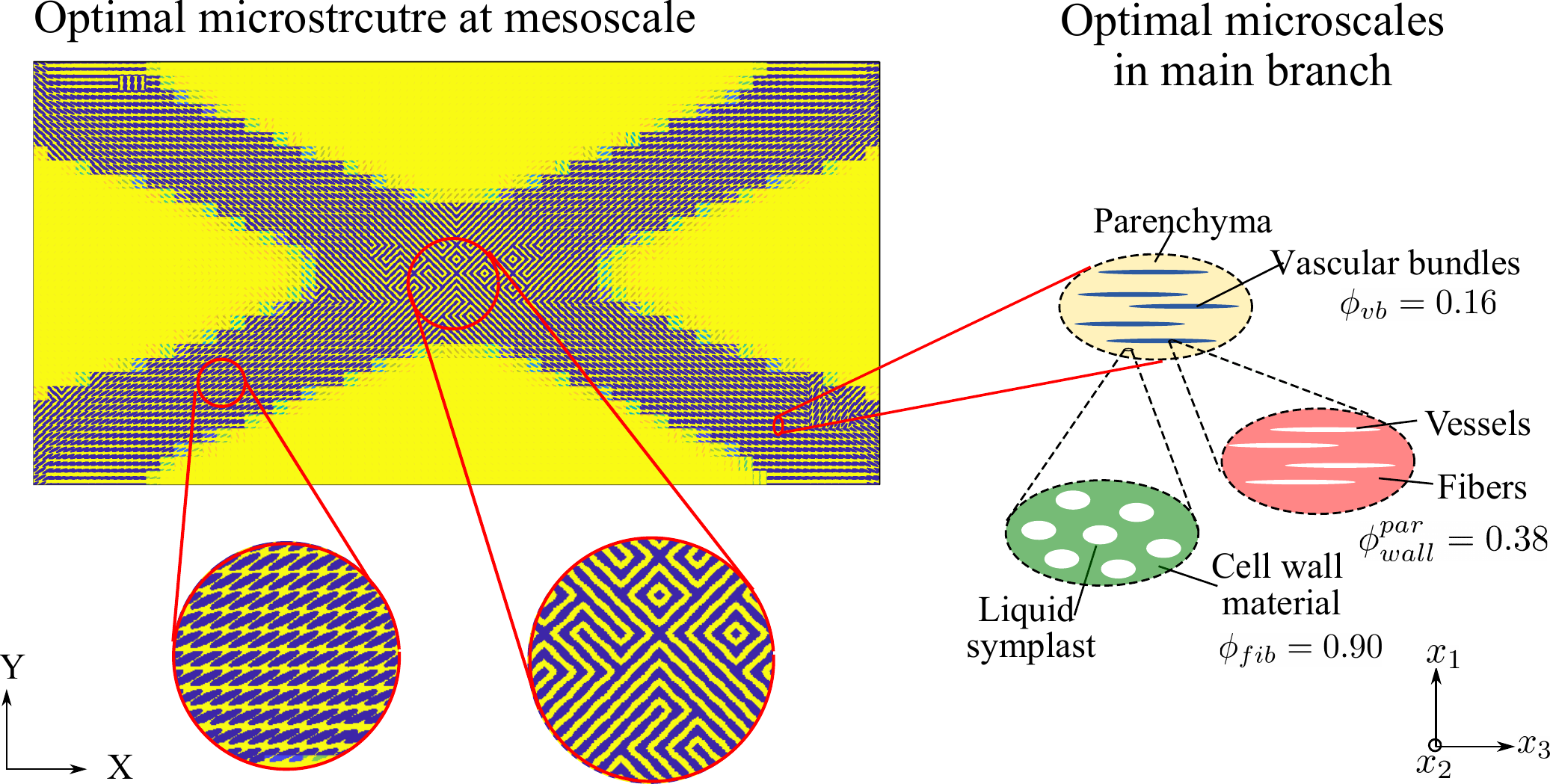}	
	\caption{Optimal microstructure configuration for total prescribed displacement $\boldsymbol{u}^{*} = 0.4 $ mm. }
	\label{Figs:ch9_fig3}
\end{figure*}

Figure~\ref{Figs:ch9_fig1} summarizes the prototype model for the hierarchical optimization of a cereal node region, given that stem failure has been generally observed in this region. We take the length and height of the macrostructure domain as 10 mm and 5 mm, which is typical of a node dimension. The left edge is completely fixed, and the displacement in $ X $ direction on the right edge is fixed. 
The prescribed displacement in the $Y $ direction on the right edge is $\boldsymbol{u}^{*} = 0.4$ mm. The displacement $\boldsymbol{u}^{*} = 0.4$ mm is divided in four load steps with increment $\Delta \boldsymbol{\bar{u}}^{E} = 0.1$ mm. We discretize the macroscale structure with a $ 60 \times 30 $ mesh of 4-node plane-strain quadrilateral elements. This macrostructure model definition is equivalent to a cereal node cross-section that undergoes combined shear and bending loads.


Following our multiscale material model, the stem cross-section consists of an outer-shell layer and a solid-pith region. The primary functions of the outer-shell are non-mechanical, such as protecting against insects and regulating gas exchange. Thus, we only consider the solid-pith region for hierarchical optimization. The microstructure design variables $ m^{x,j}_{n+1} $ consist of the cell wall fraction $\phi_{wall,n+1}^{par(x,j)} $ in the parenchyma, the fiber fraction $\phi^{x,j}_{fib,n+1}$ in the vascular bundles, the vascular bundle fraction $\phi^{x,j}_{vb,n+1}$, and the orientation $\theta^{x,j}_{n+1}$ of the anisotropy axis of the solid pith with respect to the global $X $ direction (see Fig.~\ref{Figs:ch9_fig1}). Lignin in the parenchyma cell wall material exhibits elastoplastic behavior. The parenchyma tissues and xylem-phloem vessels in the vascular bundles are also responsible for food storage and nutrient-water transport. We incorporate these biological constraints by adopting the bounds on the volume fractions that we measured through microimaging \cite{gangwar2020multiscale} in the material optimization problem. At the structure scale, the total amount of material is restricted by the reported average density, which can be interpreted as the limitation posed by the available biological energy required in the synthesis of biomass per unit of stem material.


Figures~\ref{Figs:ch9_fig4} and \ref{Figs:ch9_fig2} illustrate the design evolution history and the final macroscale density with the equivalent plastic strains. The macroscale design algorithm takes 38 density updates to converge to the final design. In the optimal layout, the branches from the left internode converge to the central node region, and the branches of the right internode emerges, a morphology that was also observed in real plants through micro-CT images \cite{ghaffar2015revealing}. 
The plastic strains are concentrated at the end and middle regions due to the anticipated high shear deformations. The optimal density layout puts material in areas to attenuate plastic fronts, which can also be observed in the design evolution history in Fig.~\ref{Figs:ch9_fig4}. These observations again emphasize the role of the plastic front in the optimal structural layout.

Figure~\ref{Figs:ch9_fig3} plots the optimal microstructure configuration at different scales at the final load level. At the mesoscale, the material orientation follows the stress flow direction in the main branches, while the morphology is more complex in the central node region. We also plot the optimal configurations at lower material scales in the main branches. These results indicate the choice of a stronger solid-pith material for the optimal mechanical response. The predicted morphology is in qualitative agreement with what our collaborators in plant science have found via field experiments for lodging-resistant cereals  \cite{gangwar2022wind}. Based on these promising results, we believe that our optimization framework can help pave the way towards efficient and sustained biotailoring applications, supported by modeling and simulation.



\section{Summary, conclusions, and outlook}

In this article, we established rigorous theoretical foundations for an efficient concurrent material and structure optimization framework for multiphase hierarchical systems with elastoplastic constituents at the material scales. In particular, we developed an efficient solution strategy for the material optimization problem based on the maximum plastic dissipation principle in the format of a typical return map algorithm for an elastoplastic constitutive law. Finally, we integrated analytical expressions of the homogenized stiffness and the yield criterion that are derived via continuum micromechanics, enabling a computationally tractable implementation for elastoplastic multiphase hierarchical systems.

We verified the validity and efficiency of our framework with newly defined benchmark problems that, for the first time, is computationally feasible via our framework. It consists of a macroscale configuration in the form of a standard cantilever, but involves hierarchical material definitions with elastoplastic constituents at the microscale. The optimized macroscale and microstructure configurations computed via our framework demonstrated the importance of plasticity effects that originate from the material microscales in developing dissipation-based energy absorbing mechanisms. In addition, we applied our framework for investigating self-adapting mechanisms in cereal plant structures, outlining its potential for biotailoring applications.

Our framework is a first attempt at a decomposed material and structure formulation that optimizes the path-dependent macroscale mechanical response of elastoplastic multiphase hierarchical systems. The formulation can potentially be extended for other path-dependent problems, where nonlinear effects such as viscoplasticity, fracture, and damage originate from the material microscales. Thus, our framework helps push forward path-dependent concurrent material and structure optimization to consider nonlinearities exhibited at the microscales, with a number of potential applications, including multiscale additive manufacturing and architecting metamaterials \cite{meza2015resilient,sanders2021optimal}. 




\noindent \textbf{Acknowledgments:} This work has received funding from the European Research Council (ERC) under the European Union's Horizon 2020 research and innovation programme (Grant agreement No.\ 759001). This support is gratefully acknowledged. In addition, the first author (T.\ Gangwar) was partially supported by a MnDRIVE Informatics Graduate Fellowship at the University of Minnesota, which is also gratefully acknowledged. The authors are also thankful to Pramesh Kumar (University of Minnesota) for helpful discussions and comments. 

\vspace{4mm}

\noindent \textbf{Declaration of competing interest:} The authors declare that they have no known competing financial interests or personal relationships that could have appeared to influence the work reported in this paper. 

\vspace{4mm}

\noindent \textbf{CRediT authorship contribution statement:} \underline{Tarun Gangwar}: Conceptualization, Methodology, Software, Validation, Visualization, Writing - original draft. \underline{Dominik Schillinger}: Writing - Review \& Editing, Supervision, Project administration, Funding acquisition.

\appendix
\normalsize

\section{Comments on continuum micromechanics-based simplifications}
\label{App:App_A}

We present a proof of Property~3 introduced in Section~\ref{sec:sec_43}. First, we write the strain energy maximization expressions in index notation. We denote the orthonormal basis that corresponds to the global coordinate system as $\{\boldsymbol{e}_{p}\} $. We drop superscript $ (x,j) $ from variables for conciseness. Problem \eqref{eq:ch7_eq43} at load increment $n$ with $ \boldsymbol{E}^{e}_{n}:= \boldsymbol{E}^{}_{n} - \boldsymbol{E}^{p}_{n}  $ can be rewritten in index notations as
	\begin{align}
			\max_{m_{n}}
			\frac{1}{2} \boldsymbol{E}^{e}_{n}:\mathbb{C}_{} (m_{n}):\boldsymbol{E}^{e}_{n} & =  \max_{m_{n}} \frac{1}{2}E^{e}_{pq(n)} \boldsymbol{e}_{p} \otimes \boldsymbol{e}_{q}: C_{pqrs}(m_{n})\boldsymbol{e}_{p} \otimes \boldsymbol{e}_{q}\otimes 
			\boldsymbol{e}_{r} \otimes \boldsymbol{e}_{s}:E^{e}_{rs(n)} \boldsymbol{e}_{r} \otimes \boldsymbol{e}_{s} \nonumber\\ 
			& = \max_{m_{n}} \frac{1}{2} E^{e}_{pq(n)}\;C_{pqrs}(m_{n})\;E^{e}_{rs(n)}.
		\label{eq:ch7_eq45}
\end{align}
The principle coordinate system, that is co-linear with the principal strain directions, uses the Roman indexed basis $\{\boldsymbol{\hat{e}}_{i}\} $. The transformation matrix $Q_{pi} $ between both systems depends on the optimal configuration $\bar{\theta}^{}_{A,n} $ from Property~2. Utilizing $Q_{pi} $ and its orthogonal property, we define the tensor component transformations as
\begin{equation}
	\boldsymbol{\hat{e}}^{}_{i} = Q_{pi} \boldsymbol{e}_{p};\;\;\hat{E}^{e}_{ij(n)} = Q_{pi}Q_{qj} E^{e}_{pq(n)};\;\; C_{pqrs}  = Q_{pi}Q_{qj}Q_{rk}Q_{sl} \hat{C}_{ijkl},
	\label{eq:ch7_eq46}
\end{equation}
where $\hat{E}^{e}_{ij(n)} $ are the components of the elastic part of the macroscale strain tensor 
in the principle coordinate system. Also, $\hat{E}^{e}_{ij(n)} = 0 \;\text{if}\; i \neq j$, and diagonal components ($i=j $) are the principle strain values. 
Similarly, $\hat{C}_{ijkl} $ are the components of the stiffness tensor in the principal coordinate system. Using (\ref{eq:ch7_eq46}), we reformulate the maximization problem (\ref{eq:ch7_eq45}) in terms of $m^{l}_{n} = \; [\phi^{}_{A,n},  \zeta^{}_{A,n}, \gamma^{}_{C,n} ]$ as
\begin{equation}
		\max_{m^{l}_{n}} \frac{1}{2} \hat{E}^{e}_{ij(n)} \; \hat{C}_{ijkl}(m^{l}_{n})\;\hat{E}^{e}_{kl(n)}.  
		\label{eq:ch7_eq47}
\end{equation}

With $\bar{m}^{l}_{n}$ as the solution to this maximization problem, we can rewrite \eqref{eq:ch7_eq47} as
\begin{equation}
	\begin{split}
		\hat{E}^{e}_{ij(n)}\; \hat{C}_{ijkl}(\bar{m}^{l}_{n})\;\hat{E}^{e}_{kl(n)} \geq \hat{E}^{e}_{ij(n)}\; \hat{C}_{ijkl}(m^{l}_{n})\;\hat{E}^{e}_{kl(n)}.
	\end{split}
	\label{eq:ch7_eq48}
\end{equation}
We assume that the macroscale loading increases monotonically. Therefore, the elastic part of the macroscale tensor $\hat{E}^{e}_{ij(n+1)} $ at load increment $ (n+1)$  in the principal coordinate system can be written in terms of the components $\hat{E}^{e}_{ij(n)} $ with an appropriate scaling. Exploiting the definition of the Kronecker delta $\delta $, we write $\hat{E}^{e}_{ij(n+1)} $ in terms of scaling components $a_{i\alpha} $ as
\begin{equation}
	\hat{E}^{e}_{ij(n+1)} = a_{i\alpha} \delta_{j\alpha} \hat{E}^{e}_{ij(n)}\;\; \text{and}\;\; a_{i\alpha} \geq 0.
	\label{eq:ch7_eq48a}
\end{equation}
As the scaling components are non-negative, we augment the expression \eqref{eq:ch7_eq48} and arrive at
\begin{equation}
	\begin{split}
		a_{i\alpha} \delta_{j\alpha}\hat{E}^{e}_{ij(n)}\; \hat{C}_{ijkl}(\bar{m}^{l}_{n})\;a_{k\beta} \delta_{l\beta}\hat{E}^{e}_{kl(n)} & \geq a_{i\alpha} \delta_{j\alpha}\hat{E}^{e}_{ij(n)}\; \hat{C}_{ijkl}(m^{l}_{n})\;a_{k\beta} \delta_{l\beta}\hat{E}^{e}_{kl(n)}\\
		\implies \hat{E}^{e}_{ij(n+1)}\; \hat{C}_{ijkl}(\bar{m}^{l}_{n})\;\hat{E}^{e}_{kl(n+1)} &\geq \hat{E}^{e}_{ij(n+1)}\; \hat{C}_{ijkl}(m^{l}_{n})\;\hat{E}^{e}_{kl(n+1)}. 
	\end{split}
	\label{eq:ch7_eq49}
\end{equation}
This is the expression for the strain energy maximization problem at load increment $(n+1)$ analogous to \eqref{eq:ch7_eq47} or \eqref{eq:ch7_eq48}.

We emphasize that our definition of the admissible set $E_{ad} $ depends only on the macroscale density distribution and, therefore, remains unchanged throughout the loading history. In addition, Case~2 is not conceivable as laid out in our discussion of Property~1 in Section~\ref{sec:sec_43}. Therefore, it implies that the possible admissible solutions of the right hand side expression at increment $(n+1)$ in \ref{eq:ch7_eq49} are the same as those at increment $n$. 
Replacing  $m^{l}_{n}$ with $m^{l}_{n+1} $, we arrive at
\begin{equation}
	\begin{split}
		\hat{E}^{e}_{ij(n+1)}\; \hat{C}_{ijkl}(\bar{m}^{l}_{n})\;\hat{E}^{e}_{kl(n+1)} &\geq \hat{E}^{e}_{ij(n+1)}\; \hat{C}_{ijkl}(m^{l}_{n+1})\;\hat{E}^{e}_{kl(n+1)} \implies \bar{m}^{l}_{n+1} = \bar{m}^{l}_{n}. 
	\end{split}
	\label{eq:ch7_eq50}
\end{equation}

\section{Details on the cereal prototype model}
\label{App:App_oat}
We briefly summarize the key components for implementing the hierarchical optimization of the cereal prototype model in Section~\ref{sec:sec_62}. For further details on multiscale modeling of stiffness and strength of crop stem material within continuum micromechanics, we refer the interested reader to Section~3 in \cite{gangwar2020multiscale}. In the current model, the microscale design variables defined at each Gauss point at load step $(n+1)$ are the cell wall fraction $\phi_{wall,n+1}^{par(x,j)} $ in the parenchyma, the fiber fraction $\phi^{x,j}_{fib,n+1}$ in the vascular bundles, the vascular bundle fraction $\phi^{x,j}_{vb,n+1}$, and the orientation $\theta^{x,j}_{n+1}$ of the anisotropy axis of the solid-pith material. The macroscale homogenized stiffness tensor $\mathbb{C} $ in the global coordinate system can be written as a function of the microscale design variables $m^{x,j}_{n+1} = [\phi_{wall,n+1}^{par(x,j)},\phi_{fib,n+1}^{x,j},\phi_{vb,n+1}^{x,j}, \theta^{x,j}_{n+1}]$ following Equation (27) in \cite{gangwar2020multiscale} as 
	\begin{align}
	\mathbb{C}(m^{x,j}_{n+1})  =  [\mathbf{T}( & \theta^{x,j}_{n+1})]^{-1}  \;\mathbb{C}_{pith}\;[\mathbf{T}(\theta^{x,j}_{n+1})]
		\label{eq:ch9_eq3} \\
		\text{with} \;\; \mathbb{C}_{pith}^{}  (\mathbb{C}_{\textit{par}} &  (\phi_{wall,n+1}^{par(x,j)}), \;  \mathbb{C}_{vb}(\phi_{fib,n+1}^{x,j}), \; \phi_{vb,n+1}^{x,j}) = \nonumber\\ & \Big \{(1 - \phi_{vb,n+1}^{x,j})\;  \mathbb{C}_{\textit{par}}^{} +  \phi_{vb,n+1}^{x,j} \mathbb{C}_{vb}:  [\mathbb{I} +  \mathbb{P}_{\textit{cyl}}^{\textit{par}}:(\mathbb{C}_{vb} - \mathbb{C}_{\textit{par}}^{})]^{-1} \Big \} : \\ 
& \nonumber\Big \{ (1 - \phi_{vb,n+1}^{x,j})\; \mathbb{I}\; +  \phi_{vb,n+1}^{x,j} [\mathbb{I} + \mathbb{P}_{\textit{cyl}}^{\textit{par}}:(\mathbb{C}_{vb} - \mathbb{C}_{\textit{par}}^{})]^{-1} \Big\}^{-1} .
	\end{align}
Here, $ \mathbb{C}_{\textit{par}} $ and $ \mathbb{C}_{vb} $ are the homogenized stiffness tensors of the parenchyma tissue and the vascular bundle tissue in the solid-pith region, respectively. For the analytical expression of these estimates, interested readers are referred to Section~3.3 in \cite{gangwar2020multiscale}. The composition of all other RVEs in the multiscale material model of cereal stems is considered constant, and model parameters corresponding to the Gopher oat variety are used (see Appendix~B in \cite{gangwar2020multiscale}). $\mathbf{T} $ is a standard rotation matrix for tensor transformations.

Lignin exhibits elastoplastic material behavior at the constituent level, and the macroscale limit state point corresponds to the yielding of lignin. 
In this prototype model, we assume that only lignin in parenchyma cell wall material is elastoplastic (see Fig.~\ref{Figs:ch9_fig1}). In this case, the macroscale homogenized yield criterion reads as
\begin{equation}
	\mathfrak{F} (\boldsymbol{\tau},m^{x,j}_{n+1}) = \sqrt{\boldsymbol{\tau}:[\mathbb{C}(m^{x,j}_{n+1})]^{-1}:\frac{\partial \; \mathbb{C}(m^{x,j}_{n+1})}{\partial \; \mu_{lig,par}}:[\mathbb{C}(m^{x,j}_{n+1})]^{-1}:\boldsymbol{\tau}} \; - \;  
	\sqrt{\frac{\overline{\phi}_{lig,par}}{3}} \; \frac{\sigma^{Y}_{lig}}{ \mu_{lig}},
	\label{eq:ch9_eq4}
\end{equation}
where lignin follows the von Mises failure criterion with yield strength $\sigma^{Y}_{lig}$ and the bulk modulus $ \mu_{lig} $. The equivalent volume fraction $\overline{\phi}_{lig,par}$ of lignin in parenchyma is computed as $ \overline{\phi}_{lig,par} =  \phi^{wall,par}_l \times \phi_{wall,n+1}^{par(x,j)} \times (1 - \phi_{vb,n+1}^{x,j}) $. The lignin volume fraction $\phi^{wall,par}_l $ in the parenchyma cell wall material is fixed as given in \cite{gangwar2020multiscale}.

With these definitions in hand, we write the material optimization problem for a Gauss point $ \boldsymbol{x} $ inside element $ j$ for load increment $(n+1)$ following \eqref{eq:ch7_eq27} as
\begin{equation}
	\begin{split}
		\bar{m}^{x,j}_{n+1} = \argmax_{	m^{x,j}_{n+1}(\rho_{j})} &  
		\Big\{   \boldsymbol{\Sigma}_{n+1}^{}: (\boldsymbol{E}^{p}_{n+1} - \boldsymbol{E}^{p}_{n}   )  +    \Psi(\boldsymbol{E}_{n+1} - \boldsymbol{E}^{p}_{n+1})
		- \Psi(   \boldsymbol{E}_{n} - \boldsymbol{E}^{p}_{n})\Big\}\\
		\text{s.t.}: & \;\; \boldsymbol{\Sigma}_{n+1} =  \mathbb{C}(m^{x,j}_{n+1}): (\boldsymbol{E}^{}_{n+1} - \boldsymbol{E}^{p}_{n+1})  \\		
		& \;\;\mathfrak{F}(\boldsymbol{\Sigma}_{n+1},m^{x,j}_{n+1}) \leq 0 \\
		& \;\; \Psi (\boldsymbol{E}_{n+1}   - \boldsymbol{E}^{p}_{n+1} ) =  \frac{1}{2}(\boldsymbol{E}_{n+1}   - \boldsymbol{E}^{p}_{n+1} ):\mathbb{C}(	m^{x,j}_{n+1}): (\boldsymbol{E}^{}_{n+1} - \boldsymbol{E}^{p}_{n+1}) \\
		& \;\; \rho_j = \rho_{wall} \; \phi_{wall,n+1}^{par (x,j)} (1-\phi_{vb,n+1}^{x,j}) + \rho_{fib}\; \phi_{fib,n+1}^{x,j} \; \phi_{vb, n+1}^{x,j} \\
		& \;\; \phi_{wall}^{\textit{par,min}} \leq \phi_{wall,n+1}^{par (x,j)} \leq \phi_{\textit{wall}}^{\textit{par,max}}; \;\; \phi_{\textit{fib}}^{\textit{min}} \leq \phi_{fib,n+1}^{x,j} \leq \phi_{\textit{fib}}^{\textit{max}} \\
		& \;\; \phi_{vb}^{\textit{min}} \leq \phi_{vb,n+1}^{x,j} \leq \phi_{\textit{vb}}^{\textit{max}}; \;\;-\pi/2 \leq   \theta^{x,j}_{n+1} \leq \pi/2,\\
	\end{split}
	\label{eq:ch9_eq2}
\end{equation}
where $ \rho_j $ is the given macroscale dry density for the finite element with index $j $. The first three lines in the constraints definition represent the microstructure dependent constitutive equations. The fourth statement connects $\rho_{j} $ with the microscale design variables via the rule of mixture. We adopt bounds for $ \phi_{wall,n+1}^{par(x,j)} $, $ \phi_{fib,n+1}^{x,j} $ and $ \phi_{vb,n+1}^{x,j} $ that are $ [0.01,0.38] $, $ [0.75,0.90] $, and $ [0.01,0.16] $, respectively. The upper bound reflects the measured microscale parameters reported in \cite{gangwar2020multiscale}. The strain energy maximization part in the algorithmic treatment of the material optimization problem is a constraint optimization problem with nonlinear equality constraint. We utilize the sequential least squares programming (SLSQP) method implemented in the SciPy library to solve this problem.



The setup of the structure optimization problem for this prototype model resembles \eqref{eq:ch7_eq26}. The sensitivity analysis for the macroscale design updates follows from Section~\ref{sec:sec51}. The derivative of the homogenized stiffness $\mathbb{C} $ with respect to the element density $ \rho_{j}$ follows via the chain rule as 
\begin{equation}
	\begin{split}
		\frac{\partial \mathbb{C}(m^{x,j}_{n+1})}{\partial {\rho}_j} =  \; \frac{\partial \mathbb{C}_{}}{\partial \phi_{wall,n+1}^{par(x,j)}} \frac{\partial  \phi_{wall,n+1}^{par(x,j)}}{\partial \rho_{j} } + \frac{\partial \mathbb{C}_{}}{\partial \phi_{fib,n+1}^{x,j}} \frac{\partial \phi_{fib,n+1}^{x,j} }{\partial \rho_{j} } +  \frac{\partial \mathbb{C}_{}}{\partial \phi_{vb,n+1}^{x,j}} \frac{\partial \phi_{vb,n+1}^{x,j}}{\partial \rho_{j} } \;.
	\end{split}
\end{equation}
The derivatives of $\mathbb{C} $ with respect to the microscale design variables at the material level are evaluated by finite difference approximations. The move parameter $ \mu $ and the damping parameter $\eta$ are set to 0.02 and 0.5. The filter radius $r_{min} $ is reduced linearly from $r_{min} = 15 \, l_e $ to $r_{min} = 4\,l_e $ with design iterations.

\bibliographystyle{elsarticle-num}
\bibliography{references}

\begin{thebibliography}{10}
\expandafter\ifx\csname url\endcsname\relax
  \def\url#1{\texttt{#1}}\fi
\expandafter\ifx\csname urlprefix\endcsname\relax\def\urlprefix{URL }\fi
\expandafter\ifx\csname href\endcsname\relax
  \def\href#1#2{#2} \def\path#1{#1}\fi

\bibitem{wegst2015bioinspired}
U.~G. Wegst, H.~Bai, E.~Saiz, A.~P. Tomsia, R.~O. Ritchie, Bioinspired
  structural materials, Nature Materials 14~(1) (2015) 23.

\bibitem{zheng2014ultralight}
X.~Zheng, H.~Lee, T.~H. Weisgraber, M.~Shusteff, J.~DeOtte, E.~B. Duoss, J.~D.
  Kuntz, M.~M. Biener, Q.~Ge, J.~A. Jackson, et~al., Ultralight, ultrastiff
  mechanical metamaterials, Science 344~(6190) (2014) 1373--1377.

\bibitem{fratzl2007nature}
P.~Fratzl, R.~Weinkamer, Nature’s hierarchical materials, Progress in
  Materials Science 52~(8) (2007) 1263--1334.

\bibitem{ritchie2009plasticity}
R.~O. Ritchie, M.~J. Buehler, P.~Hansma, Plasticity and toughness in bone,
  Physics Today 62~(6) (2009) 41.

\bibitem{bhushan2009biomimetics}
B.~Bhushan, Biomimetics: lessons from nature--an overview, Philosophical
  Transactions of the Royal Society A: Mathematical, Physical and Engineering
  Sciences 367~(1893) (2009) 1445--1486.

\bibitem{egan2015role}
P.~Egan, R.~Sinko, P.~R. LeDuc, S.~Keten, The role of mechanics in biological
  and bio-inspired systems, Nature Communications 6~(1) (2015) 1--12.

\bibitem{wolf1986law}
J.~Wolff, The law of bone remodelling (Das Gesetz der Transformation der
  Knocken), Springer, Berlin, 1986.

\bibitem{gibson2012hierarchical}
L.~J. Gibson, The hierarchical structure and mechanics of plant materials,
  Journal of The Royal Society Interface 9~(76) (2012) 2749--2766.

\bibitem{gao2003materials}
H.~Gao, B.~Ji, I.~L. J{\"a}ger, E.~Arzt, P.~Fratzl, Materials become
  insensitive to flaws at nanoscale: lessons from nature, Proceedings of the
  National Academy of Sciences 100~(10) (2003) 5597--5600.

\bibitem{brule2016hierarchies}
V.~Brul{\'e}, A.~Rafsanjani, D.~Pasini, T.~L. Western, Hierarchies of plant
  stiffness, Plant Science 250 (2016) 79--96.

\bibitem{mccann2014plants}
M.~C. McCann, M.~S. Buckeridge, N.~C. Carpita, Plants and bioenergy, Springer,
  2014.

\bibitem{rodrigues1999global}
H.~C. Rodrigues, C.~Jacobs, J.~M. Guedes, M.~P. Bends{\o}e, Global and local
  material optimization models applied to anisotropic bone adaptation, in:
  P.~Pedersen, M.~P. Bends{\o}e (Eds.), IUTAM Symposium on Synthesis in Bio
  Solid Mechanics, Springer Netherlands, Dordrecht, 1999, pp. 209--220.

\bibitem{blanchard2016patient}
R.~Blanchard, C.~Morin, A.~Malandrino, A.~Vella, Z.~Sant, C.~Hellmich,
  Patient-specific fracture risk assessment of vertebrae: A multiscale approach
  coupling x-ray physics and continuum micromechanics, International Journal
  for Numerical Methods in Biomedical Engineering 32~(9) (2016) e02760.

\bibitem{holstov2015hygromorphic}
A.~Holstov, B.~Bridgens, G.~Farmer, Hygromorphic materials for sustainable
  responsive architecture, Construction and Building Materials 98 (2015)
  570--582.

\bibitem{xia2014concurrent}
L.~Xia, P.~Breitkopf, Concurrent topology optimization design of material and
  structure within {FE2} nonlinear multiscale analysis framework, Computer
  Methods in Applied Mechanics and Engineering 278 (2014) 524--542.

\bibitem{xia2015multiscale}
L.~Xia, P.~Breitkopf, Multiscale structural topology optimization with an
  approximate constitutive model for local material microstructure, Computer
  Methods in Applied Mechanics and Engineering 286 (2015) 147--167.

\bibitem{rodrigues2002hierarchical}
H.~C. Rodrigues, J.~M. Guedes, M.~P. Bendsoe, Hierarchical optimization of
  material and structure, Structural and Multidisciplinary Optimization 24~(1)
  (2002) 1--10.

\bibitem{coelho2008hierarchical}
P.~G. Coelho, P.~R. Fernandes, J.~M. Guedes, H.~C. Rodrigues, A hierarchical
  model for concurrent material and topology optimisation of three-dimensional
  structures, Structural and Multidisciplinary Optimization 35~(2) (2008)
  107--115.

\bibitem{nakshatrala2013nonlinear}
P.~B. Nakshatrala, D.~A. Tortorelli, K.~Nakshatrala, Nonlinear structural
  design using multiscale topology optimization. {P}art {I}: Static
  formulation, Computer Methods in Applied Mechanics and Engineering 261 (2013)
  167--176.

\bibitem{da2017concurrent}
D.~Da, X.~Cui, K.~Long, G.~Li, Concurrent topological design of composite
  structures and the underlying multi-phase materials, Computers \& Structures
  179 (2017) 1--14.

\bibitem{zhang2018multiscale}
Y.~Zhang, M.~Xiao, H.~Li, L.~Gao, S.~Chu, Multiscale concurrent topology
  optimization for cellular structures with multiple microstructures based on
  ordered {SIMP} interpolation, Computational Materials Science 155 (2018)
  74--91.

\bibitem{feyel2000fe2}
F.~Feyel, J.-L. Chaboche, {FE2} multiscale approach for modelling the
  elastoviscoplastic behaviour of long fibre {S}i{C}/{T}i composite materials,
  Computer Methods in Applied Mechanics and Engineering 183~(3-4) (2000)
  309--330.

\bibitem{fish2013practical}
J.~Fish, Practical multiscaling, John Wiley \& Sons, 2013.

\bibitem{nguyen2019multiscale}
L.~H. Nguyen, D.~Schillinger, The multiscale finite element method for
  nonlinear continuum localization problems at full fine-scale fidelity,
  illustrated through phase-field fracture and plasticity, Journal of
  Computational Physics 396 (2019) 129--160.

\bibitem{da2019topology}
D.~Da, Topology optimization design of heterogeneous materials and structures,
  John Wiley \& Sons, 2019.

\bibitem{xia2018topology}
L.~Xia, D.~Da, J.~Yvonnet, Topology optimization for maximizing the fracture
  resistance of quasi-brittle composites, Computer Methods in Applied Mechanics
  and Engineering 332 (2018) 234--254.

\bibitem{li2021simp}
P.~Li, Y.~Wu, J.~Yvonnet, A {SIMP}-phase field topology optimization framework
  to maximize quasi-brittle fracture resistance of {2D} and {3D} composites,
  Theoretical and Applied Fracture Mechanics 114 (2021) 102919.

\bibitem{kato2010material}
J.~Kato, Material optimization of fiber reinforced composites applying a damage
  formulation, Ph.D. thesis, University of Stuttgart, Germany (2010).

\bibitem{kato2013multiphase}
J.~Kato, E.~Ramm, Multiphase layout optimization for fiber reinforced
  composites considering a damage model, Engineering Structures 49 (2013)
  202--220.

\bibitem{hilchenbach2015optimization}
C.~F. Hilchenbach, E.~Ramm, Optimization of multiphase structures considering
  damage, Structural and Multidisciplinary Optimization 51~(5) (2015)
  1083--1096.

\bibitem{lipton1994saddle}
R.~Lipton, A saddle-point theorem with application to structural optimization,
  Journal of Optimization Theory and Applications 81~(3) (1994) 549--568.

\bibitem{jog1994topology}
C.~S. Jog, R.~B. Haber, M.~P. Bends{\o}e, Topology design with optimized,
  self-adaptive materials, International Journal for Numerical Methods in
  Engineering 37~(8) (1994) 1323--1350.

\bibitem{yuan2009multiple}
Z.~Yuan, J.~Fish, Multiple scale eigendeformation-based reduced order
  homogenization, Computer Methods in Applied Mechanics and Engineering
  198~(21-26) (2009) 2016--2038.

\bibitem{le2015computational}
B.~Le, J.~Yvonnet, Q.-C. He, Computational homogenization of nonlinear elastic
  materials using neural networks, International Journal for Numerical Methods
  in Engineering 104~(12) (2015) 1061--1084.

\bibitem{liu2016self}
Z.~Liu, M.~Bessa, W.~K. Liu, Self-consistent clustering analysis: an efficient
  multi-scale scheme for inelastic heterogeneous materials, Computer Methods in
  Applied Mechanics and Engineering 306 (2016) 319--341.

\bibitem{bessa2017framework}
M.~Bessa, R.~Bostanabad, Z.~Liu, A.~Hu, D.~W. Apley, C.~Brinson, W.~Chen, W.~K.
  Liu, A framework for data-driven analysis of materials under uncertainty:
  Countering the curse of dimensionality, Computer Methods in Applied Mechanics
  and Engineering 320 (2017) 633--667.

\bibitem{zaoui2002continuum}
A.~Zaoui, Continuum micromechanics: {S}urvey, Journal of Engineering Mechanics
  128~(8) (2002) 808--816.

\bibitem{suquet2014continuum}
P.~Suquet, Continuum micromechanics, Vol. 377, Springer, 2014.

\bibitem{morin2017micromechanics}
C.~Morin, V.~Vass, C.~Hellmich, Micromechanics of elastoplastic porous
  polycrystals: theory, algorithm, and application to osteonal bone,
  International Journal of Plasticity 91 (2017) 238--267.

\bibitem{gangwar2019microimaging}
T.~Gangwar, D.~Schillinger, Microimaging-informed continuum micromechanics
  accurately predicts macroscopic stiffness and strength properties of
  hierarchical plant culm materials, Mechanics of Materials 130 (2019) 39--57.

\bibitem{gangwar2020multiscale}
T.~Gangwar, D.~J. Heuschele, G.~Annor, A.~Fok, K.~P. Smith, D.~Schillinger,
  Multiscale characterization and micromechanical modeling of crop stem
  materials, Biomechanics and Modeling in Mechanobiology
  https://doi.org/10.1007/s10237-020-01369-6 (2020).

\bibitem{hofstetter2005development}
K.~Hofstetter, C.~Hellmich, J.~Eberhardsteiner, Development and experimental
  validation of a continuum micromechanics model for the elasticity of wood,
  European Journal of Mechanics-A/Solids 24~(6) (2005) 1030--1053.

\bibitem{hellmich2004can}
C.~Hellmich, F.-J. Ulm, L.~Dormieux, Can the diverse elastic properties of
  trabecular and cortical bone be attributed to only a few tissue-independent
  phase properties and their interactions?, Biomechanics and Modeling in
  Mechanobiology 2~(4) (2004) 219--238.

\bibitem{fritsch2009ductile}
A.~Fritsch, C.~Hellmich, L.~Dormieux, Ductile sliding between mineral crystals
  followed by rupture of collagen crosslinks: experimentally supported
  micromechanical explanation of bone strength, Journal of Theoretical Biology
  260~(2) (2009) 230--252.

\bibitem{pichler2011upscaling}
B.~Pichler, C.~Hellmich, Upscaling quasi-brittle strength of cement paste and
  mortar: A multi-scale engineering mechanics model, Cement and Concrete
  Research 41~(5) (2011) 467--476.

\bibitem{gangwar2021concurrent}
T.~Gangwar, D.~Schillinger, Concurrent material and structure optimization of
  multiphase hierarchical systems within a continuum micromechanics framework,
  Structural and Multidisciplinary Optimization 64 (2021) 1175--1197.

\bibitem{simo2006computational}
J.~C. Simo, T.~J. Hughes, Computational inelasticity, Vol.~7, Springer Science
  \& Business Media, 2006.

\bibitem{truesdell2004non}
C.~Truesdell, W.~Noll, The non-linear field theories of mechanics, in: The
  non-linear field theories of mechanics, Springer, 2004, pp. 1--579.

\bibitem{tadmor2012continuum}
E.~B. Tadmor, R.~E. Miller, R.~S. Elliott, Continuum mechanics and
  thermodynamics: from fundamental concepts to governing equations, Cambridge
  University Press, 2012.

\bibitem{suquet1997effective}
P.~Suquet, Effective properties of nonlinear composites, in: Continuum
  micromechanics, Springer, 1997, pp. 197--264.

\bibitem{bendsoe1999material}
M.~P. Bends{\o}e, O.~Sigmund, Material interpolation schemes in topology
  optimization, Archive of Applied Mechanics 69~(9-10) (1999) 635--654.

\bibitem{allaire1999optimal}
G.~Allaire, S.~Aubry, On optimal microstructures for a plane shape optimization
  problem, Structural Optimization 17~(2-3) (1999) 86--94.

\bibitem{laws1977determination}
N.~Laws, The determination of stress and strain concentrations at an
  ellipsoidal inclusion in an anisotropic material, Journal of Elasticity 7~(1)
  (1977) 91--97.

\bibitem{laws1985note}
N.~Laws, A note on penny-shaped cracks in transversely isotropic materials,
  Mechanics of Materials 4~(2) (1985) 209--212.

\bibitem{masson2008new}
R.~Masson, New explicit expressions of the {H}ill polarization tensor for
  general anisotropic elastic solids, International Journal of Solids and
  Structures 45~(3-4) (2008) 757--769.

\bibitem{fritzen2016topology}
F.~Fritzen, L.~Xia, M.~Leuschner, P.~Breitkopf, Topology optimization of
  multiscale elastoviscoplastic structures, International Journal for Numerical
  Methods in Engineering 106~(6) (2016) 430--453.

\bibitem{hughes2012finite}
T.~J. Hughes, The finite element method: linear static and dynamic finite
  element analysis, Dover Publications, 2000.

\bibitem{de2011computational}
E.~A. de~Souza~Neto, D.~Peric, D.~R. Owen, Computational methods for
  plasticity: theory and applications, John Wiley \& Sons, 2011.

\bibitem{swan1997voigtnon}
C.~C. Swan, I.~Kosaka, Voigt--{R}euss topology optimization for structures with
  nonlinear material behaviors, International Journal for Numerical Methods in
  Engineering 40~(20) (1997) 3785--3814.

\bibitem{huang2008optimal}
X.~Huang, Y.~Xie, Optimal design of periodic structures using evolutionary
  topology optimization, Structural and Multidisciplinary Optimization 36~(6)
  (2008) 597--606.

\bibitem{schwarz2001topology}
S.~Schwarz, K.~Maute, E.~Ramm, Topology and shape optimization for
  elastoplastic structural response, Computer Methods in Applied Mechanics and
  Engineering 190~(15-17) (2001) 2135--2155.

\bibitem{maute1998adaptive}
K.~Maute, S.~Schwarz, E.~Ramm, Adaptive topology optimization of elastoplastic
  structures, Structural optimization 15~(2) (1998) 81--91.

\bibitem{cho2003design}
S.~Cho, H.-S. Jung, Design sensitivity analysis and topology optimization of
  displacement--loaded non-linear structures, Computer Methods in Applied
  Mechanics and Engineering 192~(22-24) (2003) 2539--2553.

\bibitem{simo1985consistent}
J.~C. Simo, R.~L. Taylor, Consistent tangent operators for rate-independent
  elastoplasticity, Computer Methods in Applied Mechanics and Engineering
  48~(1) (1985) 101--118.

\bibitem{pedersen1989optimal}
P.~Pedersen, On optimal orientation of orthotropic materials, Structural
  Optimization 1~(2) (1989) 101--106.

\bibitem{boyd2004convex}
S.~Boyd, S.~P. Boyd, L.~Vandenberghe, Convex optimization, Cambridge University
  Press, 2004.

\bibitem{sigmund200199}
O.~Sigmund, A 99 line topology optimization code written in {M}atlab,
  Structural and Multidisciplinary Optimization 21~(2) (2001) 120--127.

\bibitem{xia2017evolutionary}
L.~Xia, F.~Fritzen, P.~Breitkopf, Evolutionary topology optimization of
  elastoplastic structures, Structural and Multidisciplinary Optimization
  55~(2) (2017) 569--581.

\bibitem{buhl2000stiffness}
T.~Buhl, C.~B. Pedersen, O.~Sigmund, Stiffness design of geometrically
  nonlinear structures using topology optimization, Structural and
  Multidisciplinary Optimization 19~(2) (2000) 93--104.

\bibitem{berry2004understanding}
P.~Berry, M.~Sterling, J.~Spink, C.~Baker, R.~Sylvester-Bradley, S.~Mooney,
  A.~Tams, A.~Ennos, Understanding and reducing lodging in cereals, Advances in
  Agronomy 84~(04) (2004) 215--269.

\bibitem{radman2013topology}
A.~Radman, X.~Huang, Y.~Xie, Topology optimization of functionally graded
  cellular materials, Journal of Materials Science 48~(4) (2013) 1503--1510.

\bibitem{ghaffar2015revealing}
S.~H. Ghaffar, M.~Fan, Revealing the morphology and chemical distribution of
  nodes in wheat straw, Biomass and Bioenergy 77 (2015) 123--134.

\bibitem{gangwar2022wind}
T.~Gangwar, A.~Susko, S.~Baranova, M.~Guala, K.~Smith, D.~Heuschele, Multiscale
  modeling predicts plant stem bending behavior in response to wind to inform
  lodging resistance, Royal Society Open Science Submitted (2022).

\bibitem{meza2015resilient}
L.~R. Meza, A.~J. Zelhofer, N.~Clarke, A.~J. Mateos, D.~M. Kochmann, J.~R.
  Greer, Resilient 3{D} hierarchical architected metamaterials, Proceedings of
  the National Academy of Sciences 112~(37) (2015) 11502--11507.

\bibitem{sanders2021optimal}
E.~Sanders, A.~Pereira, G.~Paulino, Optimal and continuous multilattice
  embedding, Science Advances 7~(16) (2021) eabf4838.

\end{thebibliography}

\end{document}